\documentclass[12pt,letterpaper]{article}

\usepackage{amssymb,amsthm,amsmath,amsfonts,bbm}
\usepackage[colorlinks=true,linkcolor=red,urlcolor=blue,citecolor=blue,anchorcolor=blue,filecolor=red,
pdftex,breaklinks,pdfencoding=auto,psdextra,hypertexnames=false,
bookmarksopenlevel=1,bookmarksopen=true]{hyperref}

\usepackage{amssymb,amsthm,amsmath,amsfonts,float}
\usepackage[T1]{fontenc}
\usepackage{lmodern}
\usepackage[margin=1in]{geometry}
\usepackage[compact,small]{titlesec}
\usepackage[onehalfspacing]{setspace}
\usepackage{microtype}

\numberwithin{equation}{section}

\usepackage[round,authoryear,comma,sort]{natbib}
\usepackage{booktabs,dcolumn}
\newcolumntype{d}[1]{D{.}{.}{#1}} 
\usepackage{graphicx,epstopdf}
\graphicspath{{graphics/}}
\usepackage{subcaption}
\usepackage{tikz}
\usetikzlibrary{decorations.pathreplacing}

\makeatletter
\DeclareRobustCommand\citepos
{\begingroup\def\NAT@nmfmt##1{{\NAT@up##1's}}%
	\NAT@swafalse\let\NAT@ctype\z@\NAT@partrue
	\@ifstar{\NAT@fulltrue\NAT@citetp}{\NAT@fullfalse\NAT@citetp}}
\makeatother

\makeatletter
\renewcommand*{\eqref}[1]{\hyperref[{#1}]{\textup{\tagform@{\ref*{#1}}}}}
\makeatother

\usepackage[inline,shortlabels]{enumitem}
\setlist[enumerate,1]{itemsep=0pt, topsep=2pt, partopsep=0pt}
\setlist[enumerate,2]{nosep}
\setlist[itemize,1]{itemsep=0pt, topsep=2pt, partopsep=0pt}
\setlist[itemize,2]{nosep}

\expandafter
\def \expandafter \normalsize \expandafter{\normalsize \setlength \abovedisplayskip{5pt plus 2pt minus 3pt}}
\expandafter
\def \expandafter \normalsize \expandafter{\normalsize \setlength \abovedisplayshortskip{0pt plus 2pt}}
\expandafter
\def \expandafter \normalsize \expandafter{\normalsize \setlength \belowdisplayskip{5pt plus 2pt minus 3pt}}
\expandafter
\def \expandafter \normalsize \expandafter{\normalsize \setlength \belowdisplayshortskip{2pt plus 2pt}}

\makeatletter
\def\thm@space@setup{%
	\thm@preskip=4pt plus 2pt minus 4pt
	\thm@postskip=4pt plus 2pt minus 4pt
}
\makeatother
\theoremstyle{plain}
\newtheorem{theorem}{Theorem}[section]
\newtheorem{lemma}{Lemma}[section]
\newtheorem{corollary}{Corollary}[section]
\theoremstyle{definition}
\newtheorem{assumption}{Assumption}
\newtheorem{remark}{Remark}[section]

\DeclareMathOperator{\ran}{ran}
\DeclareMathOperator{\op}{op}
\DeclareMathOperator{\spn}{span}
\DeclareMathOperator{\tr}{trace}
\DeclareMathOperator{\rank}{rank}
\DeclareMathOperator{\diag}{diag}
\DeclareMathOperator{\E}{\mathbb{E}}

\def\dto{\overset{\mathrm{d}}\rightarrow}
\def\pto{\overset{\mathrm{P}}\rightarrow}
\def\dlowto{\rightarrow_{\mathrm{d}}}
\def\plowto{\rightarrow_{\mathrm{P}}}
\def\bigOp{\mathcal{O}_{\mathrm{P}}}
\def\smallop{\scalebox{0.7}{$\mathcal{O}$}_{\mathrm{P}}}
\def\smallo{\scalebox{0.7}{$\mathcal{O}$}}
\def\bigO{\mathcal{O}}
\newcommand{\RO}{\Lambda_{\mathsf{S},R}}
\newcommand{\ROL}{\Lambda_{\mathsf{S},L}}
\newcommand{\Fmax}{\mathcal{F}_{\max}}
\newcommand{\Ftr}{\mathcal{F}_{\tr}}
\newcommand{\NNN}{\mathsf{N}}
\newcommand{\SSS}{\mathsf{S}}
\newcommand{\nd}{{\mathbbm{d}_{\NNN}}}
\newcommand{\sd}{{\mathbbm{d}_{\SSS}}}
\newcommand{\km}{\mathrm{k}_\mathit{m}}
\newcommand{\KK}{\mathrm{K}}
\newcommand{\smalls}{\mathbbm{d}}
\newcommand{\bigs}{\mathbbm{D}}
\newcommand{\px}{p_X}
\newcommand{\ROK}{\Lambda_{\SSS,R}^{\KK}}
\newcommand{\ROKL}{\Lambda_{\SSS,L}^{\KK}}
\newcommand{\revlab}[1]{\phantomsection\label{#1}} 

\begin{document}
\title{\vspace*{-1cm}
Inference on common trends in functional time series\thanks{We are grateful to the co-editor, Yixiao Sun, two anonymous referees, Brendan Beare, James Duffy, Massimo Franchi, Bent Nielsen, Hanlin Shang, and seminar participants at Australian National University, University of Oxford, University of Queensland, University of Sydney, University of Zurich, University of Amsterdam, University of Warwick, Rome Econometrics, as well as the 2022~NBER-NSF time series conference, SETA~2023, ESAM~2023, EcoSta~2023, IAER~2024, IAAE~2024, and ESCW~2025 for helpful comments. Nielsen is grateful for financial support from the Aarhus Center for Econometrics (ACE) funded by the Danish National Research Foundation grant number DNRF186 and from the DNRF Chair program grant number DNRF154. A replication package for \textsf{R} is available at \url{https://github.com/sdkseong/Inference-on-common-trends-in-functional-time-series}.}}

\author{Morten {\O}rregaard Nielsen\thanks{Corresponding author.}\\ 
{\small Aarhus Center for Econometrics}\\[-0.2cm]
{\small Aarhus University}\\ [-0.2cm]
{\footnotesize \texttt{mon@econ.au.dk}}
\and 
Won-Ki Seo\\
{\small School of Economics}\\[-0.2cm]
{\small University of Sydney}\\[-0.2cm] 
{\footnotesize \texttt{won-ki.seo@sydney.edu.au}}
\and 
Dakyung Seong\\ 
{\small School of Economics}\\[-0.2cm]
{\small University of Sydney}\\[-0.2cm] 
{\footnotesize \texttt{dakyung.seong@sydney.edu.au}}
}
	
\maketitle
	
\begin{abstract}
We study statistical inference on unit roots and cointegration for time series in a Hilbert space. We develop statistical inference on the number of common stochastic trends embedded in the time series, i.e., the dimension of the nonstationary subspace. We also consider tests of hypotheses on the nonstationary and stationary subspaces themselves. The Hilbert space can be of an arbitrarily large dimension, and our methods remain asymptotically valid even when the time series of interest takes values in a subspace of possibly unknown dimension. This has wide applicability in practice; for example, to cointegrated vector time series that are either high-dimensional or of finite dimension, to high-dimensional factor models that include a finite number of nonstationary factors, to cointegrated curve-valued (or function-valued) time series, and to nonstationary dynamic functional factor models. To illustrate our methods, we include two empirical examples.
		
\medskip \noindent \textbf{JEL codes}: C32.
		
\medskip \noindent \textbf{MSC 2010}: primary 62G99, 62H99; secondary 62H25, 62M10, 91B84.
		
\medskip \noindent \textbf{Keywords}: cointegration, common trends, functional data, high-dimensional data, nonstationarity, stochastic trends, variance ratio.
\end{abstract}	
	
\newpage

\section{Introduction}
\label{sintro}

We consider statistical inference on unit roots and common stochastic trends for time series taking values in a Hilbert space of arbitrary dimension or a subspace of possibly unknown dimension. An important first step in the analysis of such time series is the determination of the dimension of the nonstationary subspace, i.e., the subspace in which the time series behaves like a unit root process (see Section~\ref{sec_cointeg}). This dimension is the number of common stochastic trends. Our objectives are to test hypotheses on the dimension of the nonstationary subspace as well as hypotheses on the stationary and nonstationary subspaces themselves.

Since we want to examine nonstationary time series in a possibly unknown-dimensional subspace of a Hilbert space~$\mathcal H$, our tests need to be statistically valid regardless of (i)~whether the dimension of the space in which the time series takes values is finite or not and (ii)~whether the dimension is known in advance or not. That is, the tests need to be (at least asymptotically) invariant to the dimensionality of the time series, and due to this property they will be called Asymptotically Dimension Invariant (ADI) tests.

This is especially relevant in the recent literature on functional time series, where it is both empirically and theoretically supported that nonstationarity tends to be driven by a finite-dimensional process \citep[e.g.,][]{Chang2016,BS2018,Franchi2017b,LRS2020,LRS2020nonst,NSS}. Even when a function-valued random element can be accommodated in a finite-dimensional space, and thus can be represented by a finite number of basis functions, its dimension is generally unknown and large. In such cases, most existing cointegration rank tests are not applicable since they require either finite (preferably small) dimensionality or a priori information on the dimensionality of the time series; this is true even for recently developed cointegration tests for high-dimensional time series \citep[e.g.,][]{Onatski2018,BG2022,BG2023}.

Our testing procedures have wide applicability in practice. For example, they can be used with (i)~cointegrated vector time series of finite dimension \citep[e.g.,][]{SW1988,Johansen1991}, (ii)~high-dimensional factor models with a finite number of  nonstationary factors \citep[e.g.,][]{NelsonSiegel87,PENA2004291,PENA20061237}, (iii)~cointegrated curve-valued (or function-valued) time series \citep[e.g.,][]{Chang2016,NSS}, and (iv)~nonstationary dynamic functional factor models \citep[e.g.,][]{martinez2020nonparametric}. 

\revlab{ce_2nd_major1}Our approach to developing ADI tests utilizes a dimension-reduction method that projects a high-dimensional or functional time series onto a fixed $\KK$-dimensional subspace. This projection is constructed such that, asymptotically, the $\KK$-dimensional projected component includes all stochastic trends, while the residual component (of unknown or possibly infinite dimension) contains no information regarding these trends. By constructing test statistics solely from the projected series, we effectively bypass interference from the residual component, ensuring that the resulting inference is inherently ADI. The analysis proceeds by examining generalized eigenvalues of two variance operators associated with the projected time series. Thus, the proposed tests will be classified as variance ratio-type tests, and some are generalizations of existing tests (Remark~\ref{revremadd1} and Appendix~\ref{secexisting}). Variance ratio tests have desirable properties for the study of nonstationary time series. First, they can avoid estimation of the long-run variance, which implies a consistency property not shared by other tests \citep{muller2007,muller2008}. Second, the limiting behavior of the tests does not depend on a parametric assumption, such as a vector autoregression. As shown in the finite-dimensional case, finite-sample properties of parametric cointegration rank tests, such as those proposed by \citet{SW1988,Ahn1990,Johansen1991,Bewley1995,Ahn1997}, depend crucially on the model specification; see \citet{Toda1995,Haug1996,Bewley1998}. \revlab{r2_minor1}Third, many parametric assumptions, e.g.\ autoregressive structures, are not preserved under projection, and tests relying on such assumptions cannot generally be ADI (Remark~\ref{rem1}). In these regards, nonparametric variance ratio tests are appealing.

Our theoretical results can be summarized by the following four points. First, we provide limit theory for general variance ratio-type statistics based on partial summation and/or differencing, and this limit theory is applied to obtain our ADI inferential methods. Second, we apply our ADI tests sequentially to determine the dimension of the nonstationary subspace via either a top-down, bottom-up, or hybrid approach. The bottom-up does not require the choice of an initial hypothesis, but the top-down has better finite-sample properties, while the hybrid combines these advantages. Third, we consider also a direct estimator of the dimension of the nonstationary subspace based on ratios of generalized eigenvalues. This is similar to estimation of the dimension of the ``dominant subspace'' in \citet{LRS2020,LRS2020nonst} and may be considered complementary to their estimator for curve-valued time series and to those in \citet{Zhang2018}, \citet{Zhang_et_al2019}, and \citet{Franchi2023} for vector-valued time series. Fourth, we consider hypothesis testing on the nonstationary or stationary subspaces.


\subsection*{Comparisons with existing work}

First, we view the time series of interest, $\{X_t\}_{t\geq 1}$, as a sequence in a known Hilbert space~$\mathcal H$, but allow for the possibility that $\{X_t\}_{t \geq 1}$ takes values only in a subspace of~$\mathcal H$. In the latter case, the variance operator of $X_t$ is singular on~$\mathcal H$ and may allow only finitely many nonzero eigenvalues even in an infinite-dimensional setting. This violates the common assumption in the literature of having sufficiently many (often infinitely many) nonzero eigenvalues. Of course, the eigenstructure of such a population variance operator is not known in practice, and assuming that infinitely many eigenvalues are nonzero may not only be unrealistic but is also not testable. For example, each functional observation of interest could be constructed by only a finite number of discrete and regularly spaced points on a grid of the domain, and such empirical examples can easily be found in the literature; see e.g., \citet{NSS} (age-specific employment rates), \citet{LRS2020nonst} (US treasury yield curves, in the working paper version), and Section~\ref{sec_empirical_1}. This poses an obstacle to the use of existing methods, and this is precisely where our ADI inference methods have a distinctive advantage over existing ones. As will be detailed, our methods are designed to be asymptotically valid under a more general scenario where the (long-run) variance operator of the stationary part of $X_t$ permits only a few nonzero eigenvalues (we also provide analysis on the exact number of nonzero eigenvalues required for the proposed tests). In comparison, existing methods rule out this possibility by assumption (including our own recent work, \citealp{NSS}). Thus, our methodology can accommodate more general and realistic functional time series.

Second, the common requirement of infinitely many nonzero eigenvalues for existing methods distinguishes them from cointegration rank tests developed in a finite-dimensional setting, where any variance operator necessarily permits only finitely many nonzero eigenvalues. We show a natural connection between our ADI tests and well-known tests for cointegration (or stationarity) developed in a finite-dimensional setting, and we find that our ADI tests generalize those tests. Furthermore, we demonstrate that, in specific cases, our ADI tests reduce to some recently developed tests for nonstationary functional time series. These results imply that some well-known tests, developed in both conventional and functional setups, can be understood as special cases of our ADI tests, thereby bridging the gap between them.

Third, the advantages of our methods lie not only in their applicability to a more general setting but also in their practical usefulness for real datasets. Many existing methods require a reasonable conjecture on the maximum possible number of stochastic trends \citep[e.g.,][]{Chang2016,LRS2020nonst,NSS}. Although some rule-of-thumb choices can be used in practice, such as those based on the scree plot, a more formal procedure may be preferable. The current study complements this aspect by providing a testing procedure based on a generalized version of the functional KPSS test \citep{Horvath2014} that can be used in practice to construct an upper bound.

Fourth, we provide statistical inference on the space spanned by stochastic trends based on our ADI tests. Specifically, we can test whether a particular subspace of interest is contained in the nonstationary subspace, spans the nonstationary subspace, or is contained in the stationary subspace. We illustrate the empirical usefulness of these tests in Section~\ref{sec:NSM} in the context of the well-known \citet{NelsonSiegel87} model. 

Compared with our earlier work in \citet{NSS}, all four preceding points apply (see Appendix~\ref{sec_NSS}). Also, in \citet{NSS} we only considered a particular variance ratio statistic (VR(2,1) under more restrictive assumptions than in Section~\ref{sec_vr21} below), whereas in this paper we consider a family of statistics as well as eigenvalue ratio estimators.

Compared with recent work on cointegration testing in high-dimensional vector autoregressive (VAR) processes \citep{Onatski2018,BG2022,BG2023}, our setup is fundamentally different. They allow only a finite number of cointegrating vectors, whereas our setup requires that the number of stochastic trends is finite and hence is more suitable when nonstationarity is driven by a small number of factors. Interestingly, under the assumption of compact autoregressive operators widely adopted in the literature, a function-valued VAR implies that the number of stochastic trends is finite (Remark~\ref{rem2}). Moreover, our setup does not rely on a VAR structure, and it covers intrinsically infinite-dimensional time series which theirs does not. Finally, our tests can be naturally applied to determine the number of stochastic trends while their tests are for the existence of cointegration and cannot immediately be applied as cointegration rank tests. In general, therefore, their tests serve a different purpose and may be viewed as complementary; see also Remark~\ref{rem3}.

\revlab{r1_2nd_major2}In concurrent work, \citet{LiLiPhillips2025} investigated common stochastic trends within a more complex framework involving a growing number of curve-valued time series. Such data, often referred to as high-dimensional functional time series (e.g., \citealp{Tang2025}), can be understood as an integration of high-dimensional time series analysis and functional data analysis. They proposed an approximate factor model for such data and developed inferential methods, including a consistent estimator for the low-dimensional stochastic trends that drive the nonstationarity of the high-dimensional functional series. While the present setup does not accommodate their framework, the exploration of a possible extension of our methodology to their setting may be an interesting avenue for future research.

The remainder is organized as follows. Section~\ref{sec_cointeg} introduces I(1) time series in Hilbert space, and Section~\ref{sec_dfvrtest} presents our ADI variance ratio tests. Section~\ref{sec_determine} discusses determination of the dimension of the nonstationary subspace via sequential testing or via eigenvalue ratio estimation. In Section~\ref{sec:tests} we discuss testing hypotheses about the subspaces. Section~\ref{sec:sim} presents Monte Carlo simulations, and Section~\ref{sec_empirical} presents two empirical applications to the term structure of interest rates and to labor market indices. Finally, Section~\ref{sec:conc} concludes. The appendix contains mathematical notation, details, and proofs, and a supplementary appendix includes a practical implementation guide, additional discussion, simulations, and proofs.

\section{I(1) time series and stochastic trends in Hilbert space}
\label{sec_cointeg}

\revlab{ce_2nd_minor1}We consider a cointegrated linear I(1) process $X_t$ taking values in a separable Hilbert space, $\mathcal H$, equipped with an inner product $\langle v_1, v_2 \rangle$ and the induced norm $\|v_1\| = \langle v_1,v_1 \rangle^{1/2}$ for $v_1,v_2\in \mathcal H$, as defined by \citet{Beare2017}. For stationary time series in $\mathcal H$, the real-valued sequence $\langle X_t,v \rangle$ must be stationary for all possible choices of~$v \in \mathcal H$ (see Proposition~3.1 and its proof in \citealp{Beare2017}). For example, if $\mathcal H$ is the standard $L^2[a,b]$ Hilbert space and hence $X_t(r)$ is a function defined on $[a,b]$, then $\langle X_t,v \rangle = \int_a^b X_t(r)v(r)dr$ may be viewed as a continuous linear combination of~$X_t(r)$. For many empirical applications involving economic functional time series, stationarity is too restrictive \citep[e.g.,][]{Chang2016,SEO2019,NSS}. Thus, we consider a sequence of $X_t$ that allows $\langle X_t, v\rangle$ to be either I(1) nonstationary or I(0) stationary. It is then important to characterize under what choices of $v\in \mathcal H$ that $\langle X_t, v\rangle$ becomes stationary or nonstationary, which leads to notions of stationary and nonstationary subspaces. We next discuss this in more detail.

Let $\{ X_t \}_{t \geq 1}$ be a sequence whose first difference, denoted $\Delta X_t$, satisfies
\begin{equation}
	\label{eqlinear}
	\Delta X_t = \sum_{j=0} ^\infty \Phi_j \epsilon_{t-j}, \quad t \geq 1,
\end{equation}
where $\{ \epsilon_t \}_{t \in \mathbb{Z}}$ is an independent and identically distributed (iid) sequence with $\E [\epsilon_t] = 0$, $\E [\|\epsilon_t\|^{4}] < \infty$, and positive definite variance~$C_\epsilon$. We further assume that $\{\Phi_j \}_{j \geq 0}$ is a sequence of bounded, linear operators satisfying
\begin{equation}
	\label{eqlinear2}
	\sum_{j=0} ^\infty j^2 \Vert \Phi_j \Vert_{\op} <\infty \quad \text{and} \quad \Phi(1) = \sum_{j=0} ^\infty \Phi_j \neq 0,
\end{equation}
where $\Vert \cdot\Vert_{\op}$ denotes the operator norm. Then the long-run variance of $\Delta X_t$ is well defined as $\Lambda_{\Delta X} = \Phi(1) C_\epsilon \Phi  (1)^\ast \neq 0$; see \citet{Beare2017}. Any stationary sequence with nonzero long-run variance is~I(0). Hence, because $\{\Delta X_t\}_{t\geq1}$ satisfying \eqref{eqlinear} and \eqref{eqlinear2} is necessarily stationary and $\Lambda_{\Delta X}\neq 0$, it is I(0) by construction, so that $\{ X_t \}_{t \geq 1}$ is~I(1). The summability condition in \eqref{eqlinear2} is common in the unit root literature \citep{PS1992}.

Let $\mathcal H_{\SSS} = \ker \Lambda_{\Delta X}$ denote the kernel of $\Lambda_{\Delta X}$ and let $\mathcal H_{\NNN}=\mathcal H_{\SSS}^\perp$ be its orthogonal complement. We call $\mathcal H_{\SSS}$ and $\mathcal H_{\NNN}$ the stationary (cointegrating) subspace and nonstationary (attractor) subspace, respectively. These names are related to some distinctive properties possessed by elements in those spaces.

First, it follows from \citet[Remark~3.1]{Beare2017} that $\langle X_t , v \rangle$ is stationary for $v \in \mathcal H_\SSS$ and nonstationary for $v \notin \mathcal H_\SSS = \mathcal H_{\NNN}^{\perp}$. In fact, we will allow $X_t$ to take values only in a strict subspace of unknown dimension. For all $v$ that are in the orthogonal complement of the strict subspace it holds that $\langle X_t,v\rangle =0$, and hence also $\langle \Delta X_t,v\rangle =0$ with long-run variance $\langle v, \Lambda_{\Delta X} v \rangle =0$. Because $\Lambda_{\Delta X}$ is self-adjoint and nonnegative, this implies that $v \in \ker \Lambda_{\Delta X}$ and thus~$v \in \mathcal H_{\SSS}$. In general, it therefore follows that if $v \in \mathcal H_{\SSS}$ then $\{\langle X_t, v\rangle\}_{t \geq 1}$ is either a stationary random sequence or equal to zero almost surely (in which case it is also stationary).

Second, regarding $\mathcal H_{\NNN}$, we will assume throughout that
\begin{equation}
\label{eqfinite}
\dim (\mathcal H_{\NNN})  = \nd < \infty.
\end{equation}
If \eqref{eqfinite} is true, we say that $\{X_t\}_{t\geq1}$ contains $\nd$ stochastic trends. To see this in detail, we note that $\{X_t\}_{t\geq 1}$ satisfying \eqref{eqlinear} and \eqref{eqlinear2} allows the decomposition
\begin{equation}
\label{eqbn}
X_t = \Phi(1)\sum_{s=1} ^t \epsilon_s + \nu_t+ X_0-\nu_0 , \quad t \geq 1,
\end{equation}
where $\nu_t = \sum_{j=0} ^\infty \tilde \Phi_j \epsilon_{t-j} $ and $\tilde \Phi_j = -\sum_{s=j+1} ^\infty \Phi_s$; see \citet{PS1992}. From the summability condition in \eqref{eqlinear2}, we also find that $\sum_{j=0}^\infty j \|\tilde \Phi_j\|_{\op}<\infty$.
It is known \citep[Proposition~3.2]{Beare2017} that $\mathcal H_{\NNN}$ is the closure of~$\ran \Phi(1)$. Thus, under \eqref{eqbn} we have 
\begin{align*}
\mathcal H = \mathcal H_{\SSS} \oplus \mathcal H_{\NNN}, \quad \mathcal H_{\SSS} = \{v \in \mathcal H: \Phi(1)C_{\epsilon}\Phi(1)^\ast v = 0\}, \quad	\mathcal H_{\NNN} = \{\Phi(1) v : v \in \mathcal H\}.
\end{align*}
Since the random walk component in \eqref{eqbn} clearly takes values in $\mathcal H_{\NNN}$ and $\dim (\mathcal H_{\NNN})  = \nd$, we say that $\{X_t\}_{t\geq1}$ contains $\nd$ stochastic trends.

Since $\dim (\mathcal H_{\NNN}) = \dim (\ran\Phi(1)) = \rank (\Phi(1))$, the projection onto $\mathcal H_{\NNN}$, denoted $P_{\NNN}$, is a finite-rank operator of rank $\nd$ under~\eqref{eqfinite}, regardless of whether $\dim (\mathcal H)$ is finite or not. 
However, the rank of the projection onto~$\mathcal H_{\SSS}$, denoted $P_{\SSS} = I - P_\NNN$ with $I$ denoting the identity operator on~$\mathcal H$, depends on~$\dim (\mathcal H)$. Specifically, the rank is $p-\nd$ if $\mathcal H = \mathbb R^p$, while it is $\infty$ if $\mathcal H$ is infinite-dimensional. Importantly, even in the latter case, if $X_t$ only takes values in a strict subspace of dimension~$\px$, then the (long-run) variance of $\{P_{\SSS}X_t\}_{t \geq 1}$ is not injective on $\mathcal H_{\SSS}$ and allows at most $\px-\nd$ nonzero eigenvalues, which may be only finitely many.

\begin{remark}
\label{remfac}
The representation in \eqref{eqlinear} and \eqref{eqbn} is that of a typical cointegrated vector- or curve-valued time series. However, it is also closely related to nonstationary vector- or curve-valued dynamic factor models, where $X_t$ allows the representation
\begin{equation}
\label{eqfactor}
X_t = \sum_{j=1}^{\nd} \beta^{\NNN}_{j,t} f^{\NNN}_j +\sum_{j=1}^{\sd} \beta^{\SSS}_{j,t} f^{\SSS}_j + e_t, 
\quad \spn\{f_j^{\NNN}\}_{j=1}^{\nd} \subseteq \mathcal H, 
\quad \spn\{f_j^{\SSS}\}_{j=1}^{\sd} \subseteq \mathcal H .
\end{equation}
Here, $\nd$ and $\sd$ are nonnegative integers that can be infinite if $\dim(\mathcal H)=\infty$, $\{\beta_{j,t}^{\NNN}\}_{j\geq 1}$ is a real-valued I(1) sequence, $\{\beta^{\SSS}_{j,t}\}_{j \geq 1}$ is a real-valued stationary sequence, and $\{e_t\}_{t\geq 1}$ is a $\mathcal H$-valued centered iid sequence with bounded variance operator. In fact, $\{X_t\}_{t\geq 1}$ satisfying \eqref{eqlinear} with compact, self-adjoint $\Phi_j$ can always be written as \eqref{eqfactor} \citep{martinez2020nonparametric}. If $\nd < \infty$, $\sd < \infty$, and $e_t$ is a curve-valued random element, \eqref{eqfactor} is the functional nonstationary dynamic factor model considered by \citet{martinez2020nonparametric}, and in that case, $X_t$ is a nonstationary time series in $\mathcal H$, but $\langle X_t,v \rangle$ is stationary if and only if $v$ is orthogonal to every $f^{\NNN}_j$ (i.e, $\mathcal H_{\NNN}=\spn\{f^{\NNN}_j\}_{j=1}^{\nd}$ and $\mathcal H_{\SSS}$ is its orthogonal complement). See also \citet{PENA2004291,PENA20061237} for related discussion in a Euclidean space setting.
\end{remark}
	
\begin{remark}
\label{rem3}
In the context of a high-dimensional cointegrated time series, the existing literature considers vector time series of dimension $p \to \infty$, but $\nd /p \to 1$ as $p \to \infty$ and $T \to \infty$, while this paper considers the case where $\nd$ is fixed regardless of the dimension~$p$. Specifically, \citet{Onatski2018} and \citet{BG2022,BG2023} studied tests for existence of cointegration in a high-dimensional setup, but not for cointegration rank or the number of stochastic trends,~$\nd$. In contrast with our condition \eqref{eqfinite} that $\nd<\infty$, they assume that $\nd$ increases such that $(p-\nd)/p \to 0$ and $\nd/p \to 1$; that is, the cointegration rank ($p-\nd$) is negligible relative to~$\nd$ (e.g., (12) of \citealp{Onatski2018}, or Section~3.2.2 of \citealp{BG2022}). Thus, their methods are not applicable in our setup, and should be considered complementary to our methods since they apply in a different context. Despite this distinction, our methodology is in fact applicable to high-dimensional vector time series in some cases. For example, a vector-valued time series $X_t$ of arbitrary dimension can be understood as a Hilbert-valued I(1) time series if the eigenvalues of the variance matrix of $\Delta X_t$ are square-summable \citep[Lemmas~1.3 and~1.4]{Bosq2000}. Finally, our methodology can be applied to time series of intrinsically infinite or large unknown dimension, and this is not considered or explored in the above-mentioned papers.
\end{remark}

\begin{remark}
\label{rem2}
The condition in \eqref{eqfinite} seems reasonable in many empirical examples of functional time series (e.g., in \citealp{Chang2016} and \citealp{NSS}). From a theoretical point of view, suppose that $\{X_t\}_{t\geq 1}$ is generated by the functional ARMA law of motion,
\begin{equation*}
\Theta(L) X_t = \Psi(L) \varepsilon_t,
\end{equation*}
where $\{\varepsilon_t\}_{t \in \mathbb{Z}}$ is an $\mathcal H$-valued white noise, $\Theta (z) = I - \sum_{j=1}^{q_1} \Theta_j z^j$, $\Psi(z) = I + \sum_{j=1}^{q_2} \Psi_j z^j$, and $q_1,q_2$ are allowed to be infinite. In this case, if $\Theta (z)$ is a Fredholm operator-valued function with a unit root, then it can be shown from the Granger-Johansen representation theorem in Hilbert space \citep{BS2018,Franchi2017b,seo_2023} that \eqref{eqfinite} is always satisfied. A sufficient (but not necessary) condition for $\Theta(z)$ to be a Fredholm operator is that $\Theta_1,\ldots , \Theta_{q_1}$ are compact and $q_1<\infty$. \revlab{ce_2nd_minor2} It is common to assume compactness of autoregressive operators in statistical analysis of the functional ARMA model, and thus finiteness of $\nd$ arises as a natural consequence. \revlab{ce_2nd_major2}Of course, for vector-valued time series, where the dimension $p$ is fixed, $\nd$ is inherently finite. However, in high-dimensional settings, where $p$ grows with the sample size as discussed in Remark~\ref{rem3}, the condition $\nd < \infty$ may be less natural, and there is no obvious counterpart to the assumption of compact autoregressive operators. Nonetheless, even in high-dimensional settings the condition \eqref{eqfinite} does encompass some relevant examples like factor structures (such as the nonstationary dynamic factor models discussed in Remark~\ref{remfac}) where nonstationarity is driven by a limited number of factors.
\end{remark}

\section{ADI variance ratio tests}
\label{sec_dfvrtest}

%

We let $\mathcal H$ be a separable Hilbert space and let $\{X_t\}_{t\geq 1}$ take values in $\mathcal H$ or any subspace. $\mathcal H$ can be an infinite-dimensional space of square-integrable functions or sequences depending on whether $X_t$ is a random function or a random vector, but we do not exclude the possibility that $\mathcal H$ is finite-dimensional. We assume that $\{X_t\}_{t\geq 1}$ satisfies the following assumption.
\begin{assumption}
	\label{assum1}
	$\{X_t\}_{t\geq1}$ satisfies the conditions in Section~\ref{sec_cointeg}; in particular \eqref{eqlinear}--\eqref{eqfinite}.
\end{assumption}
A crucial input to estimation and inference on unit roots and cointegration \citep{Chang2016,NSS,seo2024functional} is $\nd=\dim(\mathcal H_{\NNN})$, i.e., the number of (linearly independent) stochastic trends. To determine this quantity, we consider the testing problem
\begin{equation}
	\label{eqhypo}
	H_0: \nd = \smalls_0 \quad\text{vs}\quad H_1: \nd \leq  \smalls_0 -1, 
\end{equation}
for $\smalls_0 \in \bigs_0 = \{1,2,\ldots,\smalls_{\max}\}$, where $\smalls_{\max}<\infty$ is an upper bound on the number of stochastic trends so that $\bigs_0$ is a finite set. In fact, $\smalls_{\max}$ can be reasonably chosen from data in such a way that $\smalls_{\max} \geq \nd$, but $\smalls_{\max}$ does not greatly exceed~$\nd$ (Remark~\ref{remslack3}, Section~\ref{sec:sequential}).

As mentioned above, our tests can accommodate the case where $\{X_t\}_{t\geq1}$ only takes values in a strict subspace of possibly unknown dimension~$\px$, and the tests do not require knowledge of~$\px$. For this reason, they are called Asymptotically Dimension Invariant (ADI) tests. However, it should be noted that this dimension invariance property does not mean that our tests are not affected by $\px$ at all. Rather, it means that we do not explicitly require \emph{a priori} information on $\px$ in order to implement the tests.

Our approach to developing ADI tests is based on a dimension-reduction technique that preserves stochastic trends. We call this \emph{slack extraction} of stochastic trends. To elaborate on this, suppose that $\{X_t\}_{t\geq1}$ is a $\mathcal H$-valued cointegrated I(1) process as described in Section~\ref{sec_cointeg} and only takes values in a possibly strict subspace of dimension $\px \leq p_{\mathcal H} = \dim(\mathcal H) \leq \infty$, where $\px$ may be much larger than~$\nd$. In addition, assume that we can find a finite integer $\KK \in [ \nd , \px ]$ and an orthonormal set $\{ f_j\}_{j=1}^{\KK}$ in $\mathcal H$ such that $\spn\{ f_j\}_{j=1}^{\nd}=\mathcal H_{\NNN}$ and $\{\langle X_t, f_j \rangle\}_{t\geq 1}$ is a stationary process for $j=\nd + 1 , \ldots , \KK$. Define $P_{\KK} = \sum_{j=1} ^{\KK} f_j \otimes f_j$, where $\otimes$ denotes the tensor product and thus $f_j \otimes f_j  (\cdot) = \langle f_j, \cdot \rangle f_j$; that is, $P_{\KK}$ is the unique orthogonal projection onto $\mathcal H_{\KK}=\spn\{ f_j\}_{j=1}^{\KK}$. In this case, we have $\mathcal H = \mathcal H_{\KK} \oplus \mathcal H_{\KK}^\perp$ and $ \mathcal H_{\KK}  =  \mathcal H_{\NNN} \oplus  \mathcal H_{\KK}^{\SSS}$, where $\mathcal H_{\KK}^{\SSS} = \spn\{ f_j\}_{j=\nd+1}^{\KK}$. Then, from the fact that $\{\langle X_t, f_j \rangle\}_{t\geq 1}$ is an I$(1)$ sequence for $j \leq \nd$, we may deduce that (i)~the projected time series $\{  P_{\KK} X_t \}_{t\geq 1}$ contains $\nd$ stochastic trends and a $(\KK-\nd)$-dimensional stationary component, and (ii)~the nonstationary subspace of $\{  P_{\KK} X_t  \}_{t\geq 1}$ is the same as that of~$\{ X_t  \}_{t\geq 1}$. Due to these properties, we call $P_{\KK}$ a \emph{slack extractor} of stochastic trends. In other words, we do not lose any information about stochastic trends by using the projected time series $\{ P_{\KK} X_t \}_{t\geq 1}$ in the statistical analysis and ignoring the residual time series $\{ (I-P_{\KK})X_t\}_{t\geq 1}$ (whose dimension generally depends on~$\px$). Therefore, a statistical test, which is not dependent on the residual part $\{ (I-P_{\KK}) X_t  \}_{t\geq 1}$ but is constructed only from $\{ P_{\KK} X_t  \}_{t\geq 1}$, naturally becomes~ADI. Based on this idea, one may readily develop ADI tests by, for example, extending existing tests developed in a finite-dimensional setting (because $\{ P_{\KK} X_t  \}_{t\geq 1}$ itself may be viewed as a $\KK$-dimensional cointegrated time series, see Remark~\ref{remisomorphism}). 

Of course, a slack extractor $P_{\KK}$ satisfying the required conditions is not observable, but it can be replaced with a suitable estimator~$\widehat{P}_{\KK}$. For now, we impose some high-level conditions on $\widehat{P}_{\KK}$ and its rank~$\KK$ (the latter is specified by the practitioner). Define the operator
\begin{equation}
	\label{eqlongrun}
	\RO = \RO (a_R) = \E  [ P_{\SSS}X_{t} \otimes  P_{\SSS} X_{t}] + 1_{\{a_R>0\}}\sum_{s=1} ^\infty (\E [P_{\SSS} X_{t} \otimes P_{\SSS}X_{t-s}] + \E [P_{\SSS} X_{t-s} \otimes P_{\SSS} X_{t}]),
\end{equation}
where $1_{\{\cdot \}}$ is the indicator function, and $a_R$ is a parameter that will be naturally specified (in Assumption~\ref{assumkernel} below) for each of the proposed tests. Note that $\RO$ is the long-run variance (if $a_R > 0$) or variance (if $a_R = 0$) of the stationary component of~$\{X_{t}\}_{t\geq 1}$. 

\newcounter{foo}
\setcounter{foo}{\value{assumption}}
\begin{assumption}
	\label{assumvr1}
	(i)~$\KK \geq \nd$, (ii)~$\|\widehat{P}_{\KK}P_{\NNN} - P_{\NNN}\|_{\op} = \smallop (1)$, and (iii)~$\|\widehat{P}_{\KK}P_{\SSS} - P_{\SSS}^{\KK}\|_{\op} = \smallop (1)$ for some orthogonal projection $P_{\SSS}^{\KK}$ satisfying $\ran P_{\SSS}^{\KK} \subseteq \mathcal H_{\SSS}$ and $\rank (P_{\SSS}^{\KK}\RO P_{\SSS}^{\KK}) = \KK-\nd$.
\end{assumption}	

Assumption~\ref{assumvr1}(i) implies that we need an appropriate choice of $\KK$ depending on the value of~$\nd$. As will be discussed in Section~\ref{sec_determine}, see in particular Remarks~\ref{remadd},~\ref{remadd2}, and the subsequent discussion, it is possible to obtain a reasonable upper bound $\smalls_{\max}$ for~$\nd$. By choosing $\KK$ so that $\KK \geq \smalls_{\max}$, the condition on $\KK$ required by Assumption~\ref{assumvr1}(i) does not raise any issues in the practical implementation of our tests. For a given choice of $\KK$, feasible choices of $\widehat{P}_{\KK}$ satisfying Assumption~\ref{assumvr1}(ii)--(iii) are discussed in Section~\ref{secprojec} under mild low-level conditions.

\begin{remark}
\label{rem1} 
It seems reasonable to ask if one can apply standard (cointegration rank) tests of \eqref{eqhypo} from a Euclidean space setting to the projected time series~$\{ \widehat{P}_{\KK} X_t \}_{t \geq 1}$. This is in fact not always the case, because some properties of the time series are not preserved under projection. In particular, parametric tests based on an AR($p$) law of motion, such as those in \citet{Johansen1991}, cannot be directly applied since parametric relationships are not generally preserved under projection. To see this in detail, suppose that $\dim(\mathcal H)=p_{\mathcal H}$, but $X_t$ follows an AR(1) law of motion in a subspace ${\mathcal H}_X$ of (unknown) dimension $\px \leq p_{\mathcal H}$ such that $\Delta X_t = \Theta X_{t-1} +\epsilon_t$ for some bounded linear operator $\Theta$ and~$t \geq 1$. Then, for $\nd \leq {\KK} \leq \px$, we have $ \Delta \widehat P_{\KK} X_t = \widehat P_{\KK} \Theta \widehat P_{\KK} X_{t-1} + \widehat P_{\KK} \Theta (I -\widehat P_{\KK}) X_{t-1} +\widehat P_{\KK}\epsilon_t$. The term $\widehat  P_{\KK} \Theta (I -\widehat  P_{\KK}) X_{t-1}$ is dependent on $\px$ and nonzero in general. Thus, the parametric relationship between $\Delta \widehat P_{\KK} X_t$ and $\widehat P_{\KK} X_{t-1}$ is generally different from that between $\Delta X_t$ and $X_{t-1}$, and also dependent on $\KK$, $\widehat P_{\KK}$, and~$\px$. Even if $\widehat{P}_{\KK}$ is a consistent estimator of a slack extractor~$P_{\KK}$, a parametric test applied to $\{\widehat{P}_{\KK} X_t\}_{t\geq 1}$ not only depends on $\px $, but is in fact misspecified. \revlab{r2_main2}This point is illustrated in Appendix~\ref{appsim}, where we provide simulation evidence that \citepos{Johansen1991} trace test does not generally work with a consistent estimator of a slack extractor. 
\end{remark}

A consequence of Remark~\ref{rem1} is that, even if we may have a reasonable parametric assumption such as an AR structure for the original time series~$\{X_t\}_{t\geq 1}$, such an assumption is not generally preserved under the projection~$\widehat{P}_{\KK}$. Therefore, when we use the projected time series $\{\widehat{P}_{\KK} X_t\}_{t\geq 1}$ for our statistical analysis to obtain ADI tests, these need to be developed without parametric assumptions. To achieve this goal, we consider tests based on various sample (long-run) variance operators that can be computed from the projected time series without any parametric assumptions.

Thus, define $X_{d,t}$ by
\begin{equation*}
	X_{0,t} =\Delta X_t, \quad \quad	X_{1,t} = X_t, 
	\quad\quad X_{d,t} = \sum_{s=1}^t X_{d-1,s} \text{ for } d\geq 2 ,
\end{equation*}
so that $d$ denotes the integration order of~$X_{d,t}$. Further define the unnormalized sample long-run variance operator of $\{X_{d,t}\}_{t\geq 1}$, denoted $\widehat{\Lambda}_d (h,\mathrm{k})$, as
\begin{align}
	\label{eqslrv}
	\widehat{\Lambda}_d (h,\mathrm{k}) = \sum_{s=-T+1}^{T-1} \mathrm{k} \left( \frac{s}{h} \right) \widehat{\Gamma}_{d,s}, \quad\quad \widehat{\Gamma}_{d,s} = 
	\begin{cases}
		\sum_{t=s+1}^T X_{d,t-s} \otimes X_{d,t}, \quad &\text{if } s \geq 0,\\
		\sum_{t=-s+1}^T X_{d,t} \otimes X_{d,t+s}, \quad &\text{if } s < 0,
	\end{cases} 
\end{align}
where $\mathrm{k}(\cdot)$ is a kernel function and $h$ is the associated bandwidth parameter. From \eqref{eqslrv} and the fact that $\widehat P_{\KK}$ is self-adjoint, it is readily found that $\widehat P_{\KK} \widehat{\Lambda}_d (h,\mathrm{k}) \widehat P_{\KK}$ is the unnormalized sample (long-run) variance operator of the projected time series~$\{\widehat P_{\KK} X_{d,t}\}_{t\geq 1}$. As a special case, $\widehat P_{\KK} \widehat{\Lambda}_d (0,\mathrm{k}) \widehat P_{\KK}$ is the unnormalized sample variance operator of~$\{\widehat P_{\KK} X_{d,t}\}_{t\geq 1}$. 

\begin{remark} \label{remisomorphism}
Throughout, we will use the well-known result that any $\KK$-dimensional subspace of $\mathcal H$ with orthonormal basis $\{g_j\}_{j=1}^{\KK}$ is isomorphic to $\mathbb{R}^{\KK}$ via the isomorphism $v \mapsto (\langle v,g_1\rangle,\ldots,\langle v,g_{\KK}\rangle)'$; e.g., Proposition~5.2 and Theorem~5.4 of \citet{Conway1990}. Moreover, any linear operator $A$ acting on the $\KK$-dimensional subspace can be represented as the $\KK \times \KK$ matrix $[A]_{ij}=\langle g_i,Ag_j \rangle$ \citep[see also][p.~32]{NSS}.
Suppose $\widehat{P}_{\KK}=\sum_{j=1}^{\KK} \hat{f}_j \otimes \hat{f}_j$ for an orthonormal set~$\{\hat{f}_j\}_{j=1}^{\KK}$. Then $\widehat{P}_{\KK} X_{d,t} = \sum_{j=1}^{\KK} \langle X_{d,t},\hat{f}_j\rangle \hat{f}_j$ such that $\widehat{P}_{\KK} X_{d,t}$ may be identified as the $\KK$-dimensional vector $x_{\KK,d,t} = (\langle X_{d,t},\hat{f}_1 \rangle,\ldots, \langle X_{d,t},\hat{f}_{\KK} \rangle)'$. From these results we find that if $[\widehat{P}_{\KK}\widehat{\Lambda}_d (h,\mathrm{k})\widehat{P}_{\KK}]$ is the $\KK \times \KK$ sample long-run variance matrix of~$x_{\KK,d,t}$, then $\widehat P_{\KK} \widehat{\Lambda}_d (h,\mathrm{k}) \widehat P_{\KK}$ can be found as $[\widehat{P}_{\KK}\widehat{\Lambda}_d (h,\mathrm{k})\widehat{P}_{\KK}]$. That is, $\widehat P_{\KK} \widehat{\Lambda}_d (h,\mathrm{k}) \widehat P_{\KK}$ can also be computed from the vector-valued time series~$\{x_{\KK,d,t}\}_{t\geq 1}$.
Finally, for any $v \in \mathcal{H}$, $\widehat P_{\KK} \widehat{\Lambda}_d (h,\mathrm{k}) \widehat P_{\KK} v$ can be computed as $\sum_{j=1}^{\KK} a_j \hat{f}_j$, where $a_j$ is the $j$-th element of the $\KK$-dimensional vector $[\widehat{P}_{\KK}\widehat{\Lambda}_d (h,\mathrm{k})\widehat{P}_{\KK}] v_{\KK}$ with $v_{\KK} = (\langle v, \hat{f}_1 \rangle, \ldots, \langle v, \hat{f}_{\KK} \rangle)'$. Additional details and notation are given in Appendix~\ref{sec:app-notation} and in Section~\ref{revremadd2} of the supplement.
\end{remark}

Our tests are based on the generalized eigenvalue problem
\begin{equation}
	\label{eqgev}
	\mu_j \widehat P_{\KK}\widehat{\Lambda}_{d_L,L}\widehat P_{\KK} \nu_j = \widehat P_{\KK}\widehat{\Lambda}_{d_R,R}\widehat P_{\KK} \nu_j ,
\end{equation} 
where $\widehat{\Lambda}_{d_m,m} = \widehat{\Lambda}_{d_m} (h_m,\km)$ for $m \in \{ L,R \}$ (left and right) and choices of integration orders, $d_L$ and~$d_R$. Here, $\mu_j$ are the ordered (smallest to largest) eigenvalues of the operators $\widehat P_{\KK}\widehat{\Lambda}_{d_L,L}\widehat P_{\KK}$ and $\widehat P_{\KK}\widehat{\Lambda}_{d_R,R}\widehat P_{\KK}$, and $\nu_j$ are the corresponding eigenvectors. These can easily be computed as detailed in Section~\ref{revremadd2} in the supplement. Because $\widehat P_{\KK}\widehat{\Lambda}_{d_m,m} \widehat P_{\KK}$ are sample (long-run) variance operators, we call \eqref{eqgev} a generalized variance ratio (VR) eigenvalue problem, and any test based on it is called a VR($d_L, d_R$)-based test. We assume $d_L > d_R$ without loss of generality because $d_L <d_R$ is obtained by simply redefining $\mu_j$ as its inverse.

In the sequel, $\km(\cdot)$ and $h_m$ are assumed to satisfy the following assumption.

\begin{assumption}
\label{assumkernel}
$\km(\cdot)$ and $h_m$, for $m \in \{L,R\}$, satisfy:
\begin{enumerate}[label=(\roman*)]
\item $\km(\cdot)$ is a twice continuously differentiable even function from $\mathbb{R}$ to $[-1,1]$ such that $\km(x)= 0$ for $|x| \geq 1$, $\km(0)=1$, $\km '(0) = 0$, $\km ''(0) \neq 0$, and $\lim_{|x|\to 1} \km(x)/(1-|x|)^2$ = constant.
\item $h_m = [a_m T^{b_m}]$ for some $a_m\geq 0$ and $b_m \in (0,1/2)$, where $[x]$ is the nearest integer to~$x$. 
\end{enumerate}
\end{assumption} 

Assumption~\ref{assumkernel}(i) is adopted from \citet[Assumption~KL]{phillips1995fully}. The requirements on $\km(\cdot)$ in Assumption~\ref{assumkernel}(i) are not restrictive in practice and are satisfied by many widely used kernel functions, including the Epanechnikov, Parzen, Tukey-Hanning, and quartic kernels. The functional form of $h_m$ depending on $T$ in Assumption~\ref{assumkernel}(ii) is quite standard in practice.

\begin{remark}\label{remvrext}
Our theory covers all pairs of $(d_L, d_R)$ with $d_R \in \{ 0,1 \}$, but not $d_R \geq 2$. Our theoretical approach requires a normalizing diagonal operator $D_T$ (acting on $\ran \widehat P_{\KK}$) and a scaling factor $m_T$, both depending on $T$, such that $m_T D_T \widehat P_{\KK}\widehat{\Lambda}_{d_R,R}\widehat P_{\KK} D_T$ converges to a nonzero block-diagonal finite-rank operator (see \eqref{DRlimit} and \eqref{eqpf002}). For $d_R \geq 2$, however, this is not generally possible. As a simple illustration with $d_R =2$, consider a bivariate I(1) dynamic system $z_t=(z_{\NNN,t},z_{\SSS,t})'$, where $z_{\NNN,t}$ is I(1) and $z_{\SSS,t}$ is~I(0), and define $\tilde{z}_t = \sum_{s=1}^t z_s$. 
Then the asymptotic orders of elements of $\widehat{C}_{2,R}=\sum_{t=1}^T \tilde{z}_t \tilde{z}_t'$ are 
$\left[\begin{smallmatrix}
\bigOp (T^4) & \bigOp (T^3) \\ \bigOp (T^3) &  \bigOp (T^2) 
\end{smallmatrix}\right]$, 
and they are generally exact. Thus, for $m_T D_T \widehat{C}_{2,R} D_T$ to have nonzero diagnal blocks, we must have 
$D_T = \left[\begin{smallmatrix}
T^{-1} & 0 \\ 0 & 1 
\end{smallmatrix}\right]$ 
and $m_T = T^{-2}$, but in this case off-diagonal elements are not generally negligible. 
\end{remark}

\begin{remark}\label{revremadd1}
To motivate some choices of $(d_L , d_R )$, consider a finite-dimensional time series $X_t=(x_{1,t},\ldots,x_{\KK,t})'$ for fixed~$\KK$. Then, for example, \eqref{eqgev} with $(d_L , d_R ) = (2,1)$ is similar to the generalized eigenvalue problem studied in \citet{Breitung2002} with $a_L=a_R=0$, and \eqref{eqgev} with $(d_L , d_R ) = (1,0)$ is similar to that studied in \citet{shintani2001simple}. The locally best invariant tests studied in \citet{KPSS92} and \citet{nyblom2000tests} can also be seen as special cases of our inverse tests in Section~\ref{sec_inversevr}. Intuitively, these tests are based on the fact that the $\nd$ largest eigenvalues of (the finite-dimensional version of) \eqref{eqgev} exhibit different limiting behavior compared to the remaining $\KK-\nd$ ones when there are $\nd (\leq \KK)$ stochastic trends in~$X_t$. In view of the isomorphism in Remark~\ref{remisomorphism}, it is reasonable to conjecture that similar results hold in a functional time series setup if we choose $\widehat P_{\KK}$ appropriately. Additional motivation based on comparisons with existing tests is given in Appendix~\ref{secexisting}.
\end{remark}

\subsection{Tests based on VR($d_L$,1) with $d_L \geq 2$}
\label{sec_vr21}

In this section we consider tests based on the VR($d_L$,1) eigenvalue problem in \eqref{eqgev} with $d_L \geq 2$ and $d_R = 1$. We first obtain the asymptotic properties of the eigenvalues of this problem.

To this end, we let $\{W_{1,q}(r)\}_{r\in[0,1]} $ denote a $q$-dimensional standard Brownian motion and recursively define the $(d-1)$-fold integrated Brownian motion $W_{d,q}(r)  = \int_0^r  W_{d-1,q} (u)du$ for $d \geq 2$. For any matrix or compact operator~$A$, we let $\lambda_j \{ A \}$ be the $j$-th smallest eigenvalue of~$A$, i.e.\ $\lambda_1\{A\}\leq \lambda_2\{A\} \leq \ldots$. Given a kernel $\km$ satisfying Assumption~\ref{assumkernel}, we define
\begin{align}
	\label{eqcm}
	c_m = \int_{-1}^1 \km (u) du = 2 \int_0^1 \km (u) du.
\end{align} 
Moreover, when there is no risk of confusion, we write $\int f$ to denote $\int_0^1 f(u)du$.

\begin{theorem}
	\label{thmv2}
	Suppose that Assumptions~\ref{assum1}, \ref{assumvr1}, and~\ref{assumkernel} hold and define $\widetilde{\mu}_j = n_T \mu_j$, where $\{\mu_j\}_{j=1}^{\KK}$ are the eigenvalues from \eqref{eqgev} with $d_L \geq 2$, $d_R = 1$, and
	\begin{equation}
		\label{nT21}
		n_T = T^{2d_L-2} (h_Lc_L + 1_{\{ a_L=0 \}})/(h_Rc_R  + 1_{\{ a_R = 0\}}) ,
	\end{equation}
	where $h_m,a_m,c_m$ are given in Assumption~\ref{assumkernel}(ii) and~\eqref{eqcm}. Then
	\begin{align}
		\label{eqthm1}
		\widetilde{\mu}_j & \dto \widetilde{\lambda}_j =  \lambda_j \left\{ \left(\int W_{d_L,\nd} W_{d_L,\nd}'\right)^{-1} \int W_{1,\nd} W_{1,\nd}' \right\} \text{ jointly for $j \leq \nd$ (if $\nd \geq 1$),}\\
		\label{eqthm2}
		\widetilde{\mu}_j & \pto \infty \quad \text{for $j \geq \nd +1$}.
	\end{align}
\end{theorem}

Note that the normalization factor \eqref{nT21} of the eigenvalues generally depends on the choice of bandwidth parameters $h_m$ associated with the kernel $\km ( \cdot )$ for $m \in \{L,R\}$. However, in the case where $a_L=a_R=0$ (i.e., $h_L=h_R=0$), $\widehat{P}_{\KK} \widehat{\Lambda}_{d_L,L} \widehat{P}_{\KK}$ and $\widehat{P}_{\KK} \widehat{\Lambda}_{1,R} \widehat{P}_{\KK}$ reduce to the variances of $\{ \widehat{P}_{\KK}X_{d_L,t}\}_{t\geq 1}$ and $\{ \widehat{P}_{\KK}X_{1,t}\}_{t\geq 1}$, and the convergence rate is simply $n_T^{-1} = T^{2-2d_L}$.

Theorem~\ref{thmv2} shows that the eigenvalues in \eqref{eqgev} have distinct asymptotic properties depending on whether $j$ is greater than the dimension of $\mathcal H_{\NNN}$ or not, as motivated in Remark~\ref{revremadd1}. That is, if we consider the vector $(\widetilde\mu_1 ,\ldots, \widetilde\mu_{\KK})$, the first $\nd$ elements converge jointly in distribution to the vector $(\tilde{\lambda}_1 , \ldots, \tilde{\lambda}_{\nd})$, while the last $\KK - \nd$ elements are divergent in probability. If $\KK \geq \smalls_0$, for any continuous map $\mathcal{F} : \mathbb{R}^{\smalls_0} \to \mathbb{R}$ we have 
\begin{equation*}
	\mathcal{F}(\{ \widetilde{\mu}_j\}_{j=1}^{\smalls_0}) \dto \mathcal{F}(\{\widetilde{\lambda}_j\}_{j=1}^{\smalls_0}) \quad \text{under $H_0$ of \eqref{eqhypo}}.
\end{equation*}
If $\mathcal{F}$ additionally satisfies $\mathcal{F}(\{ \widetilde{\mu}_j\}_{j=1}^{\smalls_0}) \plowto \infty$ under $H_1$ of \eqref{eqhypo}, we can consistently test the hypothesis of interest in an obvious way; hereafter such an $\mathcal{F}$ is called a proper \emph{test functional}. Among many possible choices of proper test functionals, we focus on the two most common choices, $\Fmax$ and $\Ftr$, which are defined by
\begin{equation*}
	\Fmax(\{x_j\}_{j=1}^{\smalls_0}) = \max_{1\leq j\leq \smalls_0}\{x_j\}  \quad\text{and}\quad
	\Ftr (\{x_j\}_{j=1}^{\smalls_0}) =  \sum_{j=1}^{\smalls_0} x_j .
\end{equation*}

The following corollary delivers consistent tests of \eqref{eqhypo} based on~VR($d_L$,1).

\begin{corollary}
	\label{corthmv2}
	Consider the setup of Theorem~\ref{thmv2} with $\KK \geq \smalls_0$. Under $H_0 : \nd = \smalls_0$,
	\begin{equation}
		\label{eqcor1}
		\Fmax(\{{\widetilde{\mu}_j}\}_{j=1}^{\smalls_0}) \dto \max_{1\leq j\leq \smalls_0}\{\lambda_{j}\{ \mathcal A \}  \} , \quad \Ftr(\{ {\widetilde{\mu}_j}\}_{j=1}^{\smalls_0}) \dto \sum_{j=1}^{\smalls_0}\lambda_{j}\{ \mathcal A \} ,
	\end{equation} 
	where $\mathcal A= (\int W_{d_L,\smalls_0} W_{d_L,\smalls_0}' )^{-1} \int W_{1,\smalls_0} W_{1,\smalls_0}'$. Under $H_1 : \nd <\smalls_0$,
	\begin{equation}
		\label{eqcor2}
		\Fmax(\{\widetilde{\mu}_j\}_{j=1}^{\smalls_0}) \pto \infty , \quad \Ftr(\{\widetilde{\mu}_j\}_{j=1}^{\smalls_0}) \pto \infty.	
	\end{equation}
\end{corollary}

The test statistics in Corollary~\ref{corthmv2} are only functions of~$\{\widehat{P}_{\KK}X_t\}_{t\geq1}$, which may be understood as a $\KK$-dimensional vector-valued time series (Remark~\ref{remisomorphism}), and the associated limiting distributions are functionals of $\smalls_0$-dimensional standard Brownian motion and do not depend on $p_X$, $\KK$, or any nuisance parameters. These results imply that the proposed tests have the ADI property and that critical values depend only on $\smalls_0$ and the test functional.

\begin{remark}
\label{remvr21}
The VR($d_L$,1) tests allow $a_L=a_R = 0$, and hence they can be completely free from the choice of bandwidth parameters. This seems desirable in practice, since practitioners typically do not want test results to be dependent on the choice of bandwidth parameters. 
\end{remark}

\subsection{Tests based on VR($d_L$,0) with $d_L \geq 1$}

We now consider VR($d_L$,0)-based tests, which are established from the eigenvalue problem \eqref{eqgev} with $d_L \geq 1$ and $d_R =0$. As with VR($d_L$,1)-based tests, the tests in this section are based on the asymptotic properties of the sample operators, $\widehat{P}_{\KK}\widehat{\Lambda}_{d_L,L}\widehat{P}_{\KK}$ and $\widehat{P}_{\KK}\widehat{\Lambda}_{0,R}\widehat{P}_{\KK}$ for $d_L \geq 1$. The important difference relative to tests based on VR($d_L$,1) is that VR($d_L$,0)-based tests generally require~$a_R> 0$; that is, $\widehat{\Lambda}_{0,R}$ must be the sample \emph{long-run} variance of~$X_{0,t}$. This is in contrast with the case in Section~\ref{sec_vr21}, where $a_R = 0$ can be chosen; see Remarks~\ref{remvr21} and~\ref{remLRV}.

The asymptotic properties of the VR($d_L$,0) eigenvalues with $d_L \geq 1$ are as follows.

\begin{theorem}
	\label{thmvd0}
	Suppose that Assumptions~\ref{assum1}, \ref{assumvr1}, and~\ref{assumkernel} hold with $a_R>0$ and define $\widetilde{\mu}_j = n_T \mu_j$, where $\{\mu_j\}_{j=1}^{\KK}$ are the eigenvalues from \eqref{eqgev} with $d_L \geq 1 , d_R=0$, and $n_T = T^{2d_L-1}(h_Lc_L  + 1_{\{a_L=0\}})$, where $h_L,a_L,c_L$ are given in Assumption~\ref{assumkernel}(ii) and~\eqref{eqcm}. Then
	\begin{align*}
		\widetilde{\mu}_j &\dto \lambda_j \left\{ \left( \int W_{d_L,\nd} W_{d_L,\nd}' \right)^{-1} \right\} \quad \text{jointly for $j \leq \nd$ (if $\nd \geq 1$),} \\
		\widetilde{\mu}_j &\pto \infty \quad \text{for $j \geq \nd +1$}.
	\end{align*}
\end{theorem}

The next corollary delivers consistent tests of \eqref{eqhypo} based on VR($d_L$,0) with $d_L \geq 1$.

\begin{corollary}
	\label{corthmv12}
	Consider the setup of Theorem~\ref{thmvd0} with $\KK \geq \smalls_0$. Under $H_0 : \nd = \smalls_0$,
	\begin{equation}
		\label{eq1corthmv12}
		\Fmax(\{{\widetilde{\mu}_j}\}_{j=1}^{\smalls_0}) \dto \max_{1\leq j\leq \smalls_0} \{\lambda_{j}\{ \mathcal A \} \}, \quad \Ftr(\{ {\widetilde{\mu}_j}\}_{j=1}^{\smalls_0}) \dto \sum_{j=1}^{\smalls_0}\lambda_j \{ \mathcal A \} ,
	\end{equation} 
	where $\mathcal A = ( \int W_{d_L,\smalls_0} W_{d_L,\smalls_0}' )^{-1}$. Under $H_1 : \nd <  \smalls_0$,
	\begin{equation} 
		\label{eq2corthmv12}
		\Fmax(\{\widetilde{\mu}_j\}_{j=1}^{\smalls_0}) \pto \infty , \quad \Ftr(\{\widetilde{\mu}_j\}_{j=1}^{\smalls_0})  \pto \infty .
	\end{equation}
\end{corollary}

\begin{remark}
\label{remLRV}
For the VR($d_L$,0) tests, $a_L = 0$ is allowed, which may seem desirable as in Remark~\ref{remvr21}. However, $h_L>0$ is supported by simulation results for the corresponding finite-dimensional VR(1,0) test in \citet{shintani2001simple} (see Appendix~\ref{sec_shintani}), where it was found that using a long-run variance estimator on the left was superior to using a variance estimator. For nonzero~$h_L$, we also note that a larger $h_L$ helps certain asymptotically negligible quantities decay at faster rates; see, e.g., the paragraph following~\eqref{eqppgeq2a}. On the other hand, $a_R=0$ is generally not allowed because $\widehat{\Lambda}_{0,R}$ needs to be a consistent estimator of the long-run variance of~$\Delta X_t$, which requires a nonzero bandwidth~$h_R$. Given that $\Delta X_t$ is stationary, we may use existing data-dependent choices of $h_R$ that achieve certain mean-square optimality properties \citep[e.g.,][]{Rice_Shang}. Since $\widehat{P}_{\KK}\widehat{\Lambda}_{0,R}\widehat{P}_{\KK}$ is the sample long-run variance of the projected time series $\{\widehat{P}_{\KK} \Delta X_t\}_{t \geq 1}$, which may be understood as a $\KK$-dimensional vector-valued time series (Remark~\ref{remisomorphism}), we may also apply more conventional methods \citep[e.g.,][]{andrews1991}. We follow the latter approach in our simulations in Section~\ref{sec:sim}.
\end{remark}

\subsection{Inverse VR tests}
\label{sec_inversevr}

We showed in Theorem~\ref{thmv2} that the first $\nd$ eigenvalues $\{\mu_j\}_{j=1}^{\nd}$ from the VR($d_L$,1) problem with $d_L \geq 2$, properly normalized, converge jointly to a well-defined limit. As we now show, it is also possible to obtain the limiting behavior of the remaining $\KK-\nd$ eigenvalues, $\{\mu_j\}_{j=\nd +1}^{\KK}$. This enables us to examine the inverse testing problem,
\begin{equation}
	\label{eqhypo2}
	H_0: \nd = \smalls_0 \quad\text{vs}\quad H_1: \nd \geq \smalls_0 + 1,
\end{equation}
for some $\smalls_0 \geq 0$. Unlike in the testing problem \eqref{eqhypo}, we do not require any prior information on a reasonable upper bound for~$\nd$.

Related to testing~\eqref{eqhypo2}, we will replace Assumption~\ref{assumvr1} with the following high-level assumption, which allows the possibility that $\KK < \nd$.

\newcounter{foobar}
\setcounter{foobar}{\value{assumption}}
\setcounter{assumption}{\value{foo}}
\renewcommand\theassumption{\arabic{assumption}'}
\begin{assumption}
	\label{assumvr1add}
	If $\KK \geq \nd$ then Assumption~\ref{assumvr1} holds. If $\KK < \nd$ then $\|\widehat{P}_{\KK}P_{\SSS}\|_{\op} = \smallop (1)$.
\end{assumption}
\setcounter{assumption}{\value{foobar}}
\renewcommand\theassumption{\arabic{assumption}}

\begin{theorem}
	\label{thmvr2ab}
	Suppose that Assumptions~\ref{assum1}, \ref{assumvr1add}, and \ref{assumkernel} hold with $a_R>0$ and define $\widetilde{\mu}_j = n_T \mu_j$, where $\{\mu_j\}_{j=1}^{\KK}$ are the eigenvalues from \eqref{eqgev} with $d_L \geq 2$, $d_R = 1$, and $n_T = T^{2d_L -3}( h_Lc_L   + 1_{\{a_L=0\}})$, where $h_L,a_L,c_L$ are given in Assumption~\ref{assumkernel}(ii) and~\eqref{eqcm}. Then
	\begin{align}
		\label{thmeqaa1}
		(\widetilde{\mu}_j)^{-1} &\dto \lambda_{\KK -j+1} \{ \mathcal A \} \quad \text{jointly for $j = \nd +1 , \ldots , \KK$ (if $\nd < \KK$),} \\
		\label{thmeqaa2}
		(\widetilde{\mu}_j)^{-1} &\pto \infty \quad \text{for $j\leq \nd$ (if $\nd \geq 1$),}
	\end{align}
	where $\mathcal A = \int B_{\KK-\nd} B_{\KK-\nd}' - 1_{\{ \nd \geq 1 \}} \int B_{\KK-\nd} W_{d_L,\nd}' (\int W_{d_L,\nd} W_{d_L,\nd}' )^{-1} \int W_{d_L,\nd}B_{\KK-\nd}'$ and $B_{\KK-\nd}$ is a $(\KK-\nd)$-dimensional standard Brownian motion which is independent of~$W_{d_L,\nd}$.
\end{theorem}

Theorem~\ref{thmvr2ab} complements Theorem~\ref{thmv2}, where the asymptotic properties of the first $\nd$ eigenvalues are presented under similar assumptions. Specifically, the asymptotic results given in Theorem~\ref{thmvr2ab} are for $\nd \geq 0$ under the requirement $a_R>0$, which differs from Theorem~\ref{thmv2}, where the limiting distributions are given under the existence of unit roots ensuring $\nd \geq 1$ but allowing $a_R = 0$. This means that Theorem~\ref{thmvr2ab} is not simply a byproduct of Theorem~\ref{thmv2} that can be obtained with slight modifications, but in fact has its own theoretical and practical justification as we now illustrate.

When $\KK > \smalls_0$, and given a proper test functional~$\mathcal{F}$, the asymptotic results in Theorem~\ref{thmvr2ab} can be used to deliver consistent tests of \eqref{eqhypo2} as follows.

\begin{corollary}
	\label{corthmv2a}
	Consider the setup of Theorem~\ref{thmvr2ab} with $\KK > \smalls_0$. Under $H_0 : \nd=\smalls_0$,
	\begin{equation}
		\label{thmeqaa3}
		\Fmax ( \{  \widetilde{\mu}_j^{-1} \}_{j=\smalls_0+1}^{\KK} ) \dto \max_{1\leq j\leq \KK-\smalls_0} \{\lambda_j\{ \mathcal A \} \}, \quad \Ftr ( \{ \widetilde{\mu}_j ^{-1} \}_{j=\smalls_0+1}^{\KK}) \dto \sum_{j=1}^{\KK-\smalls_0} \lambda_j \{ \mathcal A \},   
	\end{equation} 
	where $\mathcal A = \int B_{\KK-\smalls_0} B_{\KK-\smalls_0}' - 1_{\{ \smalls_0\geq 1 \}} \int B_{\KK-\smalls_0} W_{d_L,\smalls_0}' (\int W_{d_L,\smalls_0} W_{d_L,\smalls_0}' )^{-1} \int W_{d_L,\smalls_0}B_{\KK-\smalls_0}'$. Under $H_1 : \nd >\smalls_0$,
	\begin{equation}
		\label{eqcorthmv2a}
		\Fmax ( \{ \widetilde{\mu}_j ^{-1} \}_{j=\smalls_0+1}^{\KK}) \pto \infty , \quad \Ftr ( \{ \widetilde{\mu}_j ^{-1} \}_{j=\smalls_0+1}^{\KK}) \pto \infty .
	\end{equation}
\end{corollary}

The test functionals described in Corollary~\ref{corthmv2a} are based on the limiting behavior of the inverse eigenvalues from the VR($d_L$,1) problem, so we call them inverse VR tests. Compared to Corollary~\ref{corthmv2} concerning the testing problem \eqref{eqhypo}, we allow $\smalls_0=0$ in Corollary~\ref{corthmv2a}. This means that we can, for example, test the null of stationarity of the time series $\{X_t\}_{t\geq 1}$ against an alternative of unit root nonstationarity. Hence, this test generalizes some existing, and widely used, KPSS-type tests of stationarity that are essentially obtained when $\smalls_0=0$; see Appendix~\ref{sec_nyblom} for a detailed discussion of related procedures. 

\begin{remark}
\label{remInvVR}
Note that $a_R>0$ is generally required in Theorem~\ref{thmvr2ab} (and Corollary~\ref{corthmv2a}), whereas $a_R = 0$ is allowed in Theorem~\ref{thmv2} for the same VR($d_L$,1) problem. This is because Theorem~\ref{thmvr2ab} requires $P_{\SSS}\widehat{\Lambda}_{1,R} P_{\SSS}$ to converge in probability in the sense of the operator norm to the true long-run variance of $\{P_{\SSS}X_t\}_{t\geq 1}$. Also, with $d_R=1$, $X_{1,t}$ is nonstationary, such that the optimal bandwidths in Remark~\ref{remLRV} do not appear immediately applicable.
\end{remark}

\subsection{Estimation of slack extractor}
\label{secprojec}

To implement the VR-based tests in practice, a slack extractor $\widehat{P}_{\KK}$ satisfying either Assumption~\ref{assumvr1} or~\ref{assumvr1add} is needed. As shown in Sections~\ref{sec_vr21}--\ref{sec_inversevr}, both the VR and inverse VR tests require Assumption~\ref{assumvr1} to hold for $\KK \geq \nd$, and when $\KK < \nd$ the inverse VR tests require instead that Assumption~\ref{assumvr1add} holds. In this section, we construct such estimators $\widehat{P}_{\KK}$ from observations $\{X_t\}_{t\geq 1}$, and we propose low-level conditions on the eigenstructure of the operator $\RO$ defined in \eqref{eqlongrun}, under which the estimators satisfy the high-level conditions.

We let $\{\tau_{j}\}_{j \geq 1}$ be the eigenvalues of $\RO$, ordered from the largest to the smallest. 

\begin{assumption}
\label{eqcondition}
If $\KK > \nd$ then $\tau_j \neq 0$ for $j=1, \ldots , \KK-\nd$ and $\tau_{\KK-\nd} \neq \tau_{\KK-\nd+1}$.
\end{assumption}

Assumption~\ref{eqcondition} tells us that the allowable values of~$\KK$, which is chosen by the practitioner, depend on the number of nonzero eigenvalues of $\RO$ defined in~\eqref{eqlongrun}. Specifically, it requires that the first $\KK-\nd$ eigenvalues of $\RO$ are nonzero and that the $(\KK-\nd)$-th eigenvalue is different from the next eigenvalue,~$\tau_{\KK-\nd+1}$. The role of the requirement $\tau_{\KK-\nd} \neq \tau_{\KK-\nd+1}$ is subtle (and may be relaxed under some additional assumptions). It is employed to make the slack extractor $\widehat{P}_{\KK}$ have a certain desirable property in our proof and can be checked using the eigenvalues of the sample counterpart of~$\RO$; see Assumption~\ref{assumvr1ll} and Theorem~\ref{thmslack2}.

In a high-dimensional Hilbert space, the number of nonzero eigenvalues of $\RO$ is generally very large (or possibly infinite). Particularly, in a typical functional time series setting with $\dim(\mathcal H) = \infty$ (e.g.\ the case considered by \citealp{NSS}), $\RO$ generally has infinitely many nonzero eigenvalues, and Assumption~\ref{eqcondition} allows $\KK$ to be any arbitrary finite integer as long as $\tau_{\KK-\nd} \neq \tau_{\KK-\nd+1}$. In a finite-dimensional space, a reasonable upper bound on $\nd$ will be relevant to choose $\KK$ to satisfy Assumption~\ref{eqcondition}.

\begin{remark}
\label{rem:Kchoice}
Consider the simple choice $\KK=\smalls_0+m$ for $m\geq 0$. Then Assumption~\ref{eqcondition} becomes more stringent as either $\smalls_0$ or $m$ increases. It is thus recommended that practitioners select a small $m$ and a $\smalls_0$ that is not much greater than~$\nd$. Since we can construct a reasonable upper bound on~$\nd$ (Remark~\ref{remslack3} and Section~\ref{sec_determine}), Assumption~\ref{eqcondition} does not appear to impose significant restrictions on the use of the VR($d_L,d_R$)-based tests in practice. 
\end{remark}

\begin{remark}
\label{rem:Kchoice2}
Even if the VR($d_L,d_R$)-based tests allow choosing $\KK = \smalls_0$, this is not recommended in practice. If $\KK = \smalls_0$, Assumption~\ref{assumvr1} requires $\widehat{P}_{\KK}$ to be an accurate estimator of $P_{\NNN}$ in the sense that $\widehat{P}_{\KK}$ is required to extract precisely the stochastic trends. If this fails, it tends to result in severe over-rejection. On the other hand, $\widehat{P}_{\KK}$ for $\KK > \smalls_0$ is less likely to miss relevant stochastic trends and will thus contribute to having correct size in finite samples. This issue is discussed in detail in \citet[e.g., Remark~13]{NSS}.   
\end{remark}
We now describe the construction of the estimator of the slack extractor. Let 
\begin{equation}	\label{eqlongrun2}
\widehat{\Lambda}_{1,P} = \widehat{\Lambda}_{1,P} (a_R) = \widehat{\Gamma}_{1,0}+ 1_{\{a_R>0\}}\sum_{s=1}^{T-1} \mathrm{k}_P (s/h_P)(\widehat{\Gamma}_{1,s} + \widehat{\Gamma}_{1,-s})
\end{equation}
denote the sample variance (resp.\ long-run variance) estimator of $X_{1,t}$ if $a_R=0$ (resp.~$a_R>0$), but allowing a new choice of bandwidth, $h_P$, and kernel,~$\mathrm{k}_P(\cdot)$. The slack extractor needs to satisfy a rank condition in Assumption~\ref{assumvr1} that depends on~$\Lambda_{\SSS,R}$, which is a variance when $a_R=0$ and a long-run variance when~$a_R>0$. For this reason, we base our estimator of the slack extractor on $\widehat{\Lambda}_{1,P}$, which depends on $a_R$ in the same way; see also Remark~\ref{rem:newremark}. The advantage is that the estimator $\widehat{P}_{\KK}$ described in Theorem~\ref{thmslack1} is universal in the sense that it can be applied for any of the VR($d_L,d_R$) tests that we consider.

\begin{theorem}
\label{thmslack1}
Suppose that Assumption~\ref{assum1} holds. If $a_R>0$ then suppose also that $\mathrm{k}_P(\cdot)$ and $h_P$ in \eqref{eqlongrun2} satisfy Assumption~\ref{assumkernel}. If $\KK > \nd$ then suppose also that Assumption~\ref{eqcondition} holds. Let $\{ \hat{f}_j\}_{j=1}^{\KK}$ be the eigenvectors corresponding to the $\KK$ largest eigenvalues of~$\widehat{\Lambda}_{1,P}$. Then $\widehat{P}_{\KK} = \sum_{j=1}^{\KK} \hat{f}_j \otimes \hat{f}_j$ satisfies Assumption~\ref{assumvr1} when $\KK \geq \nd$ and satisfies Assumption~\ref{assumvr1add} when $\KK < \nd$.
\end{theorem} 	

Note that practitioners can employ the usual functional principal component analysis (FPCA) in $\mathcal H$ to obtain the eigenvectors in Theorem~\ref{thmslack1}. Thus, Theorem~\ref{thmslack1} provides an easy-to-implement way to obtain an estimated projection operator $\widehat{P}_{\KK}$ that satisfies the high-level conditions in Assumptions~\ref{assumvr1} and~\ref{assumvr1add}.

\begin{remark}
\label{remslack}
Of course, the $\widehat{P}_{\KK}$ given in Theorem~\ref{thmslack1} is not the only choice that satisfies Assumptions~\ref{assumvr1} and~\ref{assumvr1add}. In fact, alternative estimators from the literature satisfy Assumptions~\ref{assumvr1} and~\ref{assumvr1add} under different (but stronger) low-level conditions. For example, regardless of~$a_R$, \citet{Chang2016} proposed using $\widetilde{P}_{\KK} = \sum_{j=1}^{\KK} \hat\phi_j \otimes \hat\phi_j$, where $\{\hat\phi_j\}_{j\geq 1}$ are the eigenvectors of the sample variance of $\{X_t\}_{t\geq1}$ (i.e., of~$\widehat\Lambda_{1,P}(0)$), but that requires additional conditions. If (i)~$\KK \geq \nd$, (ii)~the variance $\RO$ satisfies that $\langle v, \RO v\rangle \neq 0$ for every nonzero $v \in \ran P_{\SSS}$, and (iii)~the $(\KK-\nd)$-th largest eigenvalue of $\E [ P_{\SSS}X_{t} \otimes  P_{\SSS} X_{t}] (=\RO(0))$ is distinct from the next one, then we may deduce from \citet[Theorem~3.3]{Chang2016} and our proof of Theorem~\ref{thmslack1} that $\sum_{j=1}^{\nd} \hat\phi_j \otimes \hat\phi_j \plowto P_{\NNN}$ and that $\sum_{j=\nd+1}^{\KK} \hat\phi_j \otimes \hat\phi_j$ converges to a nonrandom projection, $P_{\SSS}^{\KK}$, satisfying $\ran P_{\SSS}^{\KK} \subseteq \ran P_{\SSS}$ and $\rank(P_{\SSS}^{\KK}\RO P_{\SSS}^{\KK}) = \KK-\nd$. That is, Assumptions~\ref{assumvr1} and~\ref{assumvr1add} hold for~$\widetilde{P}_{\KK}$. Condition~(ii), in particular, may be restrictive in a functional time series setup, but it is necessary. Without it, $\rank(P_{\SSS}^{\KK}\RO P_{\SSS}^{\KK})$ could be smaller than $\KK-\nd$, such that the I(0) components have reduced rank (long-run) variance. This is a concern when its inverse is used to construct the test statistic as in, e.g., \citet[Section~3.3]{Chang2016}; see also the discussion of Assumption~2 and (B.4) in \citet{NSS}.
\end{remark}

\begin{remark}
\label{rem:newremark}
Our proposed $\widehat{P}_{\KK}$ in Theorem~\ref{thmslack1} does not require the additional conditions in Remark~\ref{remslack}, and it therefore seems preferable for practitioners. Specifically, because of the way $\widehat{\Lambda}_{1,P}$ depends on~$a_R$, the projection $\sum_{j=1}^{\nd}\hat{f}_j\otimes \hat{f}_j$ is asymptotically equivalent to $\widehat{P}_{\KK}P_{\NNN}$ and converges to~$P_{\NNN}$, while $\sum_{j=\nd+1}^{\KK}\hat{f}_j\otimes \hat{f}_j$ is asymptotically equivalent to $\widehat{P}_{\KK}P_{\SSS}$ and converges to a projection, $P_{\SSS}^{\KK}$, onto the $(\KK-\nd)$-dimensional eigensubspace of $\RO$ that is spanned by the eigenvectors corresponding to the leading nonzero eigenvalues. It then follows by definition of $\RO$ that the rank condition, $\rank (P_{\SSS}^{\KK}\RO P_{\SSS}^{\KK}) = \KK-\nd$, is satisfied. In other words, to avoid the additional conditions in Remark~\ref{remslack}, the projection operator needs to be based on estimators of the eigenelements of~$\RO$, which depends on~$a_R$, and that is why we base our slack extractor on $\widehat{\Lambda}_{1,P}$, which similarly depends on~$a_R$.   
\end{remark}

The condition $\tau_{\KK-\nd} \neq \tau_{\KK-\nd+1}$ in Assumption~\ref{eqcondition} is generally required for the VR($d_L,d_R$)-based tests and also required for the inverse VR test when $\smalls_0$ is close to~$\nd$. Practitioners might be concerned with potential violation of this condition for a chosen~$\KK$. However, as shown in our proof of Theorem~\ref{thmslack1}, the ordered eigenvalues $\{\widehat{\tau}_{j}\}_{j \geq \nd+1}$ of $\widehat{\Lambda}_{1,P}$ converge in probability to~$\{ \tau_j \}_{j\geq 1}$. That is, the first $\nd$ eigenvalues of $\widehat{\Lambda}_{1,P}$ are associated (at least asymptotically) with the nonstationary subspace, and therefore it is the eigenvalues $\widehat\tau_j$ for $j=\nd+1,\ldots$ that correspond to to eigenvalues $\tau_j$ for $j=1,\ldots$ of~$\RO$.

We may thus avoid violation of the condition $\tau_{\KK-\nd} \neq \tau_{\KK-\nd+1}$ by choosing $\KK$ such that $\widehat{\tau}_{\KK}$ is sufficiently greater than~$\widehat{\tau}_{\KK+1}$. This approach is pragmatic and also theoretically supported under the following assumption (which is required only when~$a_R>0$).

\begin{assumption}
\label{assumvr1ll}
(i)~$\E [ \|P_{\SSS}X_t\|^8 ]<\infty$, (ii)~$\|\sum_{j=m}^\infty \tilde{\Phi}_j \|_{\op} = \smallo (m^{-\beta})$ for some $\beta > 4$, (iii)~there exists $\varphi>0$ such that $\lim_{x\to 0} (1-\mathrm{k}_P (x))/|x|^{\varphi}<\infty$, and $\tilde\varphi > \varphi$ such that $\sum_{j=-\infty}^\infty |j|^{\tilde\varphi} \| \E [ P_{\SSS}X_t \otimes P_{\SSS}X_{t-j}] \|_{\op} < \infty,$ and (iv)~$T/h_P^{2\varphi +1} =\smallo (1)$.
\end{assumption}

The conditions in Assumption~\ref{assumvr1ll} are stronger than those in the previous sections. In particular, $\{P_{\SSS}X_t\}_{t\geq 1}$ needs to satisfy a stronger moment condition than required for $\{X_t\}_{t\geq 1}$ in Section~\ref{sec_cointeg}, and the coefficients $\{\Phi_j\}_{j\geq 1}$ are required to decay at a faster rate than in Section~\ref{sec_cointeg}. The latter can be seen from the fact that $\|\sum_{j=m}^\infty \tilde{\Phi}_j \|_{\op} = \smallo (m^{-\gamma})$ for some $\gamma>1$ implies the summability condition in~\eqref{eqlinear2}. Assumption~\ref{assumvr1ll}(iii) imposes a regularity condition on $\mathrm{k}_P(\cdot)$ and a summability condition on the autovariance operators of~$\{P_{\SSS}X_t\}_{t\geq 1}$, and Assumption~\ref{assumvr1ll}(iv) requires that $h_P$ is not too small relative to~$T$. Nonetheless, even if Assumption~\ref{assumvr1ll} includes some stronger conditions, we note that they are not essential for our proposed ADI tests but are only needed if practitioners want to check if $\tau_{\KK-\nd} \neq \tau_{\KK-\nd+1}$ based on the sample eigenvalues~$\widehat\tau_j$. The following theorem establishes the desired result.

\begin{theorem}
\label{thmslack2}
Suppose that Assumption~\ref{assum1} holds. If $a_R>0$ then assume also that $\mathrm{k}_P(\cdot)$ and $h_P$ in \eqref{eqlongrun2} satisfy Assumption~\ref{assumkernel} and that Assumption~\ref{assumvr1ll} holds. If $\KK > \nd$ then
\begin{equation*}
\label{eqadd1}
(\widehat{\tau}_{\KK} - \widehat{\tau}_{\KK+1}) - ( \tau_{\KK-\nd} - \tau_{\KK-\nd+1} ) = \bigOp ( M_T^{-1/2}),
\end{equation*} 
where $M_T = T/h_P$ if $a_R>0$ and $M_T = T$ if~$a_R = 0$.
\end{theorem}

From Theorem~\ref{thmslack2} we deduce that, for any $\eta_T$ satisfying $\eta_T M_T^{-1/2} \to 0$ and $\eta_T \to \infty$, 
\begin{align*}
\eta_T (\widehat{\tau}_{\KK}-\widehat{\tau}_{\KK+1})  &\pto 0 \phantom{\infty} \quad \text{if} \quad \tau_{\KK-\nd} = \tau_{\KK-\nd+1}, \\
\eta_T (\widehat{\tau}_{\KK}-\widehat{\tau}_{\KK+1}) &\pto \infty \phantom{0} \quad \text{if} \quad \tau_{\KK-\nd} \neq \tau_{\KK-\nd+1}.
\end{align*}
This result can be applied in practice as an informal way to check whether the condition $\tau_{\KK-\nd} \neq \tau_{\KK-\nd+1}$ in Assumption~\ref{eqcondition} is likely to hold. Since we do not need to impose Assumption~\ref{eqcondition} in Theorem~\ref{thmslack1} when $\KK \leq \nd$, we do not need to consider $\KK \leq \nd$ in Theorem~\ref{thmslack2}.

\begin{remark}
\label{remslack2}
In our proof of Theorem~\ref{thmslack2}, Assumption~\ref{assumvr1ll} is used to show that, when $a_R>0$, 
\begin{equation}
\label{eqrem01}
\| T^{-1}P_{\SSS} \widehat{\Lambda}_{1,P} P_{\SSS} - \RO \|_{\op} = \bigOp  ((T/h_P)^{-1/2});
\end{equation}
see \citet[Theorems~2.2 and~2.3]{BERKES2016150}. The desired result then follows from~\eqref{eqrem01}. Thus, any other conditions that imply \eqref{eqrem01} can replace Assumption~\ref{assumvr1ll} in Theorem~\ref{thmslack2}; see also Lemma~4.1 of \citet{Rice_Shang}.
\end{remark}

\begin{remark}
\label{remslack3}
A potentially useful result that may be deduced from our proofs of Theorems~\ref{thmslack1} and~\ref{thmslack2} is that the eigenvalues $\{\widehat{\tau}_j\}_{j\geq1}$ of $\widehat{\Lambda}_{1,P}$ satisfy
\begin{equation}
\label{eqrem1}
\frac{\widehat{\tau}_j}{\widehat{\tau}_{j+1}} \pto \infty \text{ if } j=\nd \quad\text{and}\quad
\frac{\widehat{\tau}_j}{\widehat{\tau}_{j+1}} =\bigOp (1) \text{ if } j\neq\nd .
\end{equation}
This result is analogous to that in Theorem~3.3 of \citet{Chang2016} for the sample variance operator of $\{X_t\}_{t\geq 1}$ (see also Theorem~3.2 of \citealp{LRS2020nonst}). Based on \eqref{eqrem1}, we may in practice easily obtain a reasonable upper bound on $\nd$, say $\smalls_{\max}$, by computing the ratio given in \eqref{eqrem1} and set $\smalls_{\max} = \max_{1\leq j \leq \bar{s}} \{{\widehat{\tau}_j}/{\widehat{\tau}_{j+1}}\} + k$ for some large integer $\bar{s}$ and a small nonnegative integer~$k$. Of course, in view of the asymptotic properties of~$\widehat{\tau}_j$, $\max_{1\leq j \leq \bar{s}} \{{\widehat{\tau}_j}/{\widehat{\tau}_{j+1}}\}$ is itself a consistent estimator of~$\nd$. The latter is a slight modification of the eigenvalue ratio estimator of \citet{LRS2020nonst}. However, we found that the finite-sample properties of this estimator tend to be worse than those of a different eigenvalue ratio-based estimator to be discussed in Section~\ref{sec:ratio}. Nonetheless, either eigenvalue ratio-based estimator could be useful as a simple way to construct a reasonable upper bound on $\nd$ in practice.
\end{remark}	

\begin{remark}\label{remadd3}\revlab{r1_2nd_major1}
One practical advantage of our methodology is its applicability to both vector-valued and functional time series. In practice, functional data $X_t(r)$ are often observed only at discrete, regular intervals, $\{X_t(r_1), \dots, X_t(r_M)\}$, as illustrated in \citet{NSS} and Section~\ref{sec_empirical_1}. When the number of sampling points $M$ is small, applying our methodology directly to the vector-valued sequence is often more natural than smoothing the data to reconstruct functional variables. In more complicated cases with irregular or time-varying sampling grids, the process can be modeled as an error-contaminated cointegrated functional time series \citep{nam2025functional}. Notably, \citet[Section~C.2]{nam2025functional} demonstrates that a version of our VR($2,1$) approach is robust to I(0) measurement errors. A formal extension to accommodate such irregularly sampled and noisy data is left for future research.
\end{remark}

\subsection{Inclusion of deterministic components}
\label{sec_deterministic}

So far we have assumed that $\{X_t\}_{t \geq 1}$ has no deterministic component. However, in practice it is quite common that observed time series contain a deterministic component. In this section, we focus on the case with a nonzero intercept and a linear trend since those seem to be most relevant in practice.

Suppose that the observed time series $\{Y_t\}_{t\geq 1}$ is
\begin{equation}
\label{equnobs}
Y_t = \zeta_1 + \zeta_2 t + X_t, \quad t \geq1,
\end{equation}
where $\{X_t\}_{t\geq 1}$ satisfies Assumption~\ref{assum1}. We let $U_t^{(1)}$ and $U_t^{(2)}$ be defined by
\begin{equation}
\label{fresid}
U_t^{(1)} = Y_t - \bar{Y}, \quad 
U_t^{(2)} = Y_t - \bar{Y} - ( t-\bar{t} ) \frac{\sum_{t=1}^T ( t-\bar{t} ) Y_t}{\sum_{t=1}^T ( t-\bar{t} )^2}, 
\end{equation}
where $ \bar{Y} = T^{-1}\sum_{t=1}^T Y_t$ and $\bar{t} = T^{-1}\sum_{t=1}^T t = (T+1)/2$. Thus, $U_t^{(1)}$ and $U_t^{(2)}$ are, respectively, the mean-adjusted and the trend-adjusted residuals of~$Y_t$ \citep{kokoszka2016kpss}.

Our ADI variance ratio tests can be adjusted to accommodate deterministic terms by replacing the sample operators computed from $\{X_t\}_{t\geq1}$ with the corresponding operators computed from $\{U_t^{(1)}\}_{t\geq1}$ or~$\{U_t^{(2)}\}_{t\geq1}$. When the model for $Y_t$ is given by \eqref{equnobs} with $\zeta_2 =0$ then we use $U_t^{(1)}$, and otherwise we use~$U_t^{(2)}$.

This adjustment to accommodate deterministic terms produces some obvious changes in the asymptotic results given in Theorems~\ref{thmv2} and~\ref{thmvd0} as in the existing literature (e.g., \citealp{NSS}, Section~3.4; \citealp{seo2024functional}, Section~S3). For example, the extension for the VR$(d_L,1)$ test can easily be done as in Theorem~3 of \citet{NSS}. More specifically, in Theorems~\ref{thmv2} and~\ref{thmvd0}, $W_{1,\nd}$ needs to be replaced by $\nd$-dimensional demeaned standard Brownian motion (if $\zeta_2=0$) or detrended standard Brownian motion (if $\zeta_2\neq 0$), and the definition of $W_{d_L,\nd}$ needs to change accordingly. Similar changes have to be made in Theorem~\ref{thmvr2ab}, but where, as may be deduced from our proof of Theorem~\ref{thmvr2ab}, $B_{\KK - \nd}$ needs to be replaced with the standard Brownian bridge (if $\zeta_2=0$) or the second-level standard Brownian bridge (if $\zeta_2\neq 0$), while $W_{d_L,\nd}$ is the $(d_L-1)$-fold integrated demeaned (if $\zeta_2=0$) or detrended (if $\zeta_2\neq 0$) Brownian motion. The asymptotic results given in Theorems~\ref{thmslack1} and~\ref{thmslack2} do not require any changes once $\widehat{\Lambda}_{1,P}$ is constructed from the relevant residuals (our proofs of Theorems~\ref{thmslack1} and~\ref{thmslack2} contain related discussions). Following these changes to the theorems, it is quite obvious to make relevant changes to the corollaries given in the previous sections, and hence the details are omitted.

\section{Determination of dimension of nonstationary subspace}
\label{sec_determine} 

Practitioners may mostly be interested in determining the dimension of $\mathcal H_{\NNN}$ rather than testing a specific hypothesis on the dimension. In this section we first propose sequential testing procedures for doing so, and then we consider an eigenvalue ratio estimation approach.

\subsection{Estimation via sequential testing}
\label{sec:sequential}

First, we propose a ``top-down'' (TD) sequential procedure to estimate~$\nd$. Here we test~\eqref{eqhypo} sequentially (using any of the tests given in Corollaries~\ref{corthmv2} and~\ref{corthmv12}) for $\smalls_0 = \smalls_{\max}, \smalls_{\max}-1, \ldots$, where $\smalls_{\max}$ is an upper bound on the dimension of $\mathcal H_{\NNN}$ that can reasonably be chosen to be (slightly) greater than $\nd$ in practice; e.g., Remark~\ref{remslack3}. We let $\widehat\smalls_{\rm TD}$ denote the value of $\smalls_0$ under the first non-rejected null hypothesis; if the null hypothesis is rejected for every $\smalls_0 = \smalls_{\max}, \smalls_{\max}-1, \ldots , 1$, then $\widehat\smalls_{\rm TD} =0$.

Second, we consider a ``bottom-up'' (BU) procedure, where we test \eqref{eqhypo2} sequentially for $\smalls_0 = 0 , 1 , \ldots$ using any of the inverse VR tests in Corollary~\ref{corthmv2a}. We then let $\widehat\smalls_{\rm BU}$ denote the first non-rejected null hypothesis. This procedure is attractive because no $\smalls_{\max}$ needs to be specified, and the condition $\smalls_{\max} \geq \nd$ is not required; see also Remark~\ref{remadd}.

The sequential procedures can be applied with a different $\KK$ and $\widehat{P}_{\KK}$ for each value of~$\smalls_0$, i.e.\ in each step of the algorithm. Then $\KK$ should be understood as a function of~$\smalls_0$, denoted $\KK = \KK(\smalls_0)$, although we mostly suppress the dependence on $\smalls_0$ to simplify the notation. In practice, $\KK$ can simply be set to $\KK = \smalls_{\max} + m$ or $\KK = \KK(\smalls_0) = \smalls_0 + m$ for some integer $m \geq 1$ as in \citet{NSS}; see Remark~\ref{remadd} for details.

The consistency results for both procedures are given in the following theorem. 

\begin{theorem}
\label{thmseq}
Suppose that Assumptions~\ref{assum1} and~\ref{assumkernel} hold and that the nominal level of each individual test is~$\alpha$.
\begin{enumerate}[label=(\roman*)]
	\item If $\smalls_{\max} \geq \nd$ and, in each step, $\KK \geq \smalls_0$ and $\widehat{P}_{\KK}$ satisfies Assumption~\ref{assumvr1}, then 
	\begin{equation*}
		\mathrm{P}(\widehat\smalls_{\rm TD} = \nd) \to 1-\alpha 
		\quad \text{and} \quad
		\mathrm{P}(\widehat\smalls_{\rm TD} > \nd) \to 0. 
	\end{equation*} 
	\item If, in each step, $\KK > \smalls_0$ and $\widehat{P}_{\KK}$ satisfies Assumption~\ref{assumvr1add}, then 
	\begin{equation*}
		\mathrm{P}(\widehat\smalls_{\rm BU} = \nd) \to 1-\alpha
		\quad \text{and} \quad
		\mathrm{P}(\widehat\smalls_{\rm BU} < \nd) \to 0. 
	\end{equation*} 
\end{enumerate}
\end{theorem} 

The proof of Theorem~\ref{thmseq} follows directly from earlier results and is therefore omitted. For both procedures we note that if $\alpha \to 0$ as $T \to \infty$ then the procedures are consistent; that is $\mathrm{P}(\widehat\smalls_{\rm TD} = \nd)~\to~1$ and $\mathrm{P}(\widehat\smalls_{\rm BU} = \nd)~\to~1$.

\begin{remark} 
\label{remadd}
The conditions stated in Theorem~\ref{thmseq} for each of the sequential procedures simplify nicely if simple choices of $\KK$ are applied. For example, if either $\KK = \smalls_{\max} + m$ or $\KK = \KK(\smalls_0) = \smalls_0 + m$ for some integer $m \geq 1$, then the TD procedure in Theorem~\ref{thmseq}(i) only requires Assumption~\ref{assumvr1} to hold for $\KK = \smalls_{\max}+m$. If $\KK = \KK(\smalls_0) = \smalls_0 + m$, then the BU procedure in Theorem~\ref{thmseq} only requires Assumption~\ref{assumvr1add} to hold for $\KK = \nd + m$, i.e.\ Assumption~\ref{assumvr1} with $\KK = \nd + m$. Because $\smalls_{\max} \geq \nd$, the condition required for the TD procedure is stronger than that for the BU procedure.
\end{remark}

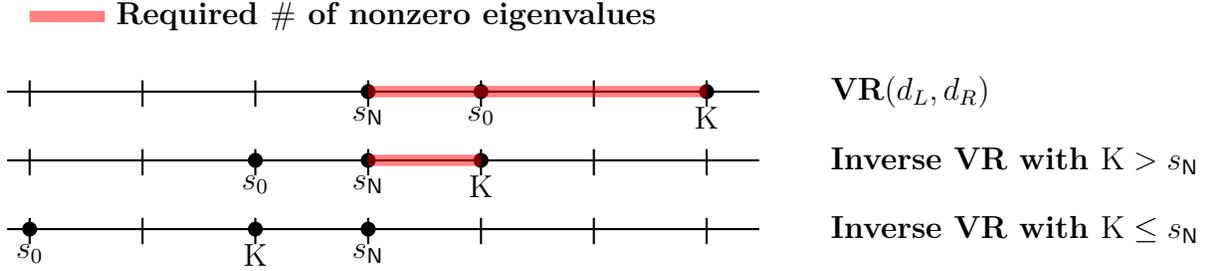
\begin{figure}
\caption{Eigenvalue requirements for the proposed tests, $\KK=\KK(\smalls_0)=\smalls_0+2$}
\label{fig:eigenvalues}
\vspace*{-\baselineskip}
\begin{flushleft}
	\begin{tikzpicture}[
		dot/.style = {circle, fill=black,inner sep=2pt, minimum size=5pt},
		every label/.append style = {inner sep=2pt},
		thick      
		]
		\draw node[right] at (1,1) {\textbf{Required $\#$ of nonzero eigenvalues}};
		\draw (0,0.15) -- + (0,-0.35) node[below] {};
		\foreach \p in {0.15, 0.3, 0.45,0.6,0.75,0.9}
		{
			\draw (10*\p,0.15) -- + (0,-0.35) node[below] {};
		}
		\draw[thick] (-0.3,0) -- + (10,0) node[right=8mm] {\textbf{VR$(d_L,d_R)$}};
		\node[dot,label=below:{$\KK$}] at (10*0.9,0) {};
		\node[dot,label=below:{$\smalls_0$}] at (10*0.6,0) {};
		\node[dot,label=below:{$\nd$}] at (10*0.45,0) {};
		(10*0.45,0) -- (10*0.90,0) node[midway,yshift=-2em]{};
		\draw[thick, red, fill=red!30,line width=1.5mm,opacity=0.5](10*0.45,0) -- (10*0.90,0) ; 
		
		\draw[thick, red, fill=red!30,line width=1.5mm,opacity=0.5](0,1) -- (1,1);

	\end{tikzpicture} \\ 
	\begin{tikzpicture}[
		dot/.style = {circle, fill=black,inner sep=2pt, minimum size=5pt},
		every label/.append style = {inner sep=2pt},
		thick      
		]
		
		\draw (0,0.15) -- + (0,-0.35) node[below] {};
		\foreach \p in {0.15, 0.3, 0.45,0.6,0.75,0.9}
		{
			\draw (10*\p,0.15) -- + (0,-0.35) node[below] {};
		}
		\draw[thick] (-0.3,0) --  + (10,0) node[right=8mm] {\textbf{Inverse VR with $\KK>\nd$}};
		\node[dot,label=below:{$\KK$}] at (10*0.60,0) {};
		\node[dot,label=below:{$\smalls_0$}] at (10*0.30,0) {};
		\node[dot,label=below:{$\nd$}] at (10*0.45,0) {};
		\draw[thick, red, fill=red!30,line width=1.5mm,opacity=0.5](10*0.45,0) -- (10*0.60,0);
	\end{tikzpicture}\\
	\begin{tikzpicture}[
		dot/.style = {circle, fill=black,inner sep=2pt, minimum size=5pt},
		every label/.append style = {inner sep=2pt},
		thick      
		]
		
		\draw (0,0.15) -- + (0,-0.35) node[below] {};
		\foreach \p in {0.15, 0.3, 0.45,0.6,0.75,0.9}
		{
			\draw (10*\p,0.15) -- + (0,-0.35) node[below] {};
		}
		\draw[thick] (-0.3,0) --  + (10,0) node[right=8mm] {\textbf{Inverse VR with $\KK\leq\nd$}};
		\node[dot,label=below:{$\KK$}] at (10*0.3,0) {};
		\node[dot,label=below:{$\smalls_0$}] at (10*0.0,0) {};
		\node[dot,label=below:{$\nd$}] at (10*0.45,0) {};
	\end{tikzpicture}
\end{flushleft}
\end{figure}

\begin{remark} 
\label{remadd2}
To implement the proposed sequential procedures under the low-level conditions discussed in Section~\ref{secprojec}, in each step we require Assumption~\ref{eqcondition} with $\KK = \KK(\smalls_0) \geq \smalls_0$ for the TD procedure or $\KK = \KK(\smalls_0) > \smalls_0$ for the BU procedure. In either case, for each~$\smalls_0$, our tests do not require knowledge of the dimension of the (possibly strict sub)space in which $X_t$ takes values, but do require $\RO$ to have $\KK - \nd$ nonzero eigenvalues. It may be deduced from Theorem~\ref{thmslack1} and Assumption~\ref{eqcondition} that the TD procedures based on VR($d_L,d_R$) tests require a stronger condition on $\RO$ than the BU procedure based on inverse VR tests. To see this, suppose that a common choice of $\KK$ (as a function of~$\smalls_0$) is used for both of the procedures, such as $\KK=\smalls_0+m$ for some $m \geq 1$. Because the BU procedure starts from $\smalls_0 \leq \nd$ and the TD procedure from $\smalls_0 \geq \nd$, the number of nonzero eigenvalues that we require for the TD procedure is always larger than that required for the BU procedure; this is illustrated for $m=2$ in Figure~\ref{fig:eigenvalues}. Moreover, the number of nonzero eigenvalues of $\RO$ required by the TD procedure is always positive when $\smalls_0 > \nd$, and this number increases as $\smalls_0$ gets larger. This suggests that it is important to have a reasonable starting point $\smalls_{\max} (\geq \nd )$ for the TD procedure. Similarly, it should also be noted that, all else equal, $\RO$ is required to have more nonzero eigenvalues as $\KK$ increases for both procedures, so it is preferable to choose $\KK$ not too much larger than~$\smalls_0$. A similar suggestion for the choice of $\KK$ can be found in \citet{NSS}, but their argument was based on computational advantages in a more restrictive setting than ours. Our finding thus gives a theoretical justification for their choice in our more general setting.      
\end{remark}

As discussed in Remarks~\ref{remadd} and~\ref{remadd2}, the conditions required for the TD procedure are stronger than those required for the BU procedure. The BU procedure has the additional advantage that it can be implemented without specifying an upper bound $\smalls_{\max}$ on~$\nd$ and, specifically, the choices of $\KK$ for the BU procedure to consistently estimate~$\nd$ (with shrinking significance levels as discussed above) depend only on~$\smalls_0$ (not~$\nd$). However, Monte Carlo simulations have shown that the TD procedure has better finite-sample properties. Thus, we combine the above two sequential procedures into an ``up-down'' (UD) hybrid procedure that enjoys the advantages of both the TD and BU procedures. Specifically, we first apply the BU procedure to obtain~$\widehat\smalls_{\rm BU}$ and use this to specify a data-driven value of the upper bound as $\smalls_{\max} = \widehat\smalls_{\rm BU} +m_\smalls$ for an integer $m_\smalls \geq 0$. We then apply the TD procedure initiated at this upper bound to obtain the UD hybrid estimate,~$\widehat\smalls_{\rm UD}$. Note that the upper bound obtained in the first (BU) part of the UD procedure can also be used to choose the same $\KK = \smalls_{\max} + m$ for an integer $m \geq 0$ in all steps in the second (TD) part of the UD procedure. The finite-sample properties of this UD hybrid procedure, along with those of the other procedures, will be investigated by simulations in Section~\ref{sec:sim}.

\subsection{Eigenvalue ratio estimator}
\label{sec:ratio}

Recent work concerning vector-valued time series provides ways to consistently estimate the number of stochastic trends using the fact that the eigenvalues of the sample autocovariance matrices have different stochastic orders if the time series contains components of different integration orders \citep[e.g.,][]{Zhang2018,Zhang_et_al2019,Chang2021}. In a functional time series setting, the eigenvalue ratio estimator proposed by \citet{LRS2020,LRS2020nonst} can be used to estimate the dimension of the component of the highest integration order. A slight modification of their estimator was discussed in Remark~\ref{remslack3}.

Alternatively, based on the asymptotic results in Section~\ref{sec_dfvrtest}, we may use the generalized eigenvalue problem \eqref{eqgev} to construct an eigenvalue ratio estimator for~$\nd$. Let $\mu_j$ be an eigenvalue of~\eqref{eqgev}. As we prove in Theorem~\ref{thmest} below, we can use Theorems~\ref{thmv2}--\ref{thmvr2ab} and arguments from their proofs to show that $\mu_j^{-1}\mu_{j+1} = \bigOp (1)$ if $j < \nd$ (resp.\ $j > \nd$) since both $n_T\mu_j$ and $n_T\mu_{j+1}$ converge (resp.\ diverge at the same rate), while $\mu_j^{-1}\mu_{j+1} \plowto \infty$ if $j=\nd$. Hence, the ratio $\mu_j^{-1}\mu_{j+1}$ will be maximized at $j=\nd$. This motivates the estimator
\begin{equation}
\label{eqeigratio}
\ddot\smalls = \underset{1\leq j\leq \smalls_{\max}}{\arg\max} \left\{\frac{\mu_{j+1}}{\mu_{j}}\right\},
\end{equation}
where $\smalls_{\max}$ is a pre-specified integer indicating an upper bound on~$\nd$. We note that these eigenvalue ratio-type estimators are only applicable when $\nd \geq 1$; hence this needs to be examined first in practical implementation of these estimators. A convenient way to do this is to test \eqref{eqhypo2} with $\smalls_0=0$ using the tests in Corollary~\ref{corthmv2a}.

\begin{theorem}
\label{thmest}
Suppose that (i)~$\smalls_{\max} \geq \nd \geq 1$, (ii)~Assumptions \ref{assum1}--\ref{assumkernel} hold with $\KK \geq \smalls_{\max}+1$, and (iii)~$a_R \geq 0$ (resp.\ $a_R > 0$) if $\mu_j$ is obtained from \eqref{eqgev} with $d_R = 1$ (resp.\ $d_R = 0$) and $d_L \geq d_R+1$. Then $\mathrm{P}(\ddot\smalls =\nd) \to 1.$
\end{theorem} 

The estimator in \eqref{eqeigratio} is very similar to the eigenvalue ratio estimator proposed by \citet{LRS2020,LRS2020nonst}, see also Remark~\ref{remslack3}, but there are some important differences. First, their estimator is constructed directly from the eigenvalues of a sample variance operator, while we use the generalized eigenvalue problem~\eqref{eqgev}. Second, their estimator requires a tuning parameter, say $\kappa >0$, to deal with estimation error associated with small eigenvalues; specifically, they set $\widehat\smalls_{\text{\tiny{LRS}}} = \arg\min_{1\leq j\leq \smalls_{\max}} \{ \hat\kappa_{j+1}/\hat\kappa_j \}$, where $\hat\kappa_j$ is the $j$-th largest eigenvalue of the sample variance operator and $\hat\kappa_j$ is regarded as zero if $\hat\kappa_j / \hat\kappa_1$ is smaller than~$\kappa$, and they define $0/0=1$. However, our estimator in \eqref{eqeigratio} does not require such a tuning parameter.

\section{Inference on subspaces}
\label{sec:tests}

In practical applications it may be of interest to test hypotheses about $\mathcal H_{\NNN}$ or~$\mathcal H_{\SSS}$. For a specified subspace~$\mathcal H_0$, consider the following hypotheses:
\begin{align} 
\label{hptest1}
&H_0 : \mathcal H_0 \subseteq \mathcal H_{\NNN} \text{ (or equivalently, $\mathcal H_0^\perp \supseteq \mathcal H_{\SSS}$)}
\quad\text{vs}\quad H_1 : H_0 \text{ is not true}, \\
\label{hptest2}
&H_0 : \mathcal H_0 \supseteq \mathcal H_{\NNN} \text{ (or equivalently, $\mathcal H_0^\perp \subseteq \mathcal H_{\SSS}$)}
\quad\text{vs}\quad H_1 : H_0 \text{ is not true}, \\
\label{hptest3}
&H_0 : \mathcal H_0 \subseteq \mathcal H_{\SSS} \text{ (or equivalently, $\mathcal H_0^\perp \supseteq \mathcal H_{\NNN}$)}
\quad\text{vs}\quad H_1 : H_0 \text{ is not true}.
\end{align}
For example, we may be interested in testing if a specific element $v \in \mathcal H$ is included in~$\mathcal H_{\NNN}$, and hence can be interpreted as one of the common stochastic trends. In that case we can use \eqref{hptest1} with $\mathcal H_0 = \spn\{v\}$. We can also test if a specified subspace, e.g.\ $\mathcal H_0 = \spn\{v_1,v_2\}$, contains the entire nonstationary subspace by using~\eqref{hptest2}. Finally, it may be of interest to examine if a specified element or subspace is in~$\mathcal H_{\SSS}$, and then we can use~\eqref{hptest3}. In a finite-dimensional setting, \eqref{hptest3} corresponds to testing if a vector or matrix is cointegrating.

Let $P_{\mathcal H_0}$ denote the projection onto~$\mathcal H_0$, let $p_0 = \dim(\mathcal H_{0})$, and let $\nd$ be known (or replaced by a consistent estimator from Section~\ref{sec_determine}). \revlab{r2_main1}The hypotheses \eqref{hptest1} and \eqref{hptest2} can then be tested by investigating the dimension of the nonstationary subspace associated with the residual series~$\{ (I-P_{\mathcal H_0}) X_t\}_{t\geq 1}$, and the hypothesis \eqref{hptest3} can be tested by investigating the dimension of the nonstationary subspace associated with the projected series~$\{ P_{\mathcal H_0} X_t\}_{t\geq 1}$. Specifically, if $H_0$ of \eqref{hptest1} is true, then $\{(I-P_{\mathcal H_0}) X_t\}_{t\geq 1}$ contains $\nd-p_0$ stochastic trends, if $H_0$ of \eqref{hptest2} is true then $\{ (I-P_{\mathcal H_0}) X_t\}_{t\geq 1}$ contains zero stochastic trends, and if $H_0$ of \eqref{hptest3} is true then $\{ P_{\mathcal H_0} X_t\}_{t\geq 1}$ contains zero stochastic trends. Under $H_1$ in \eqref{hptest1}--\eqref{hptest3}, there are more stochastic trends in the relevant time series than under the null. Thus, by analyzing $\{ (I-P_{\mathcal H_0}) X_t\}_{t\geq 1}$ or $\{P_{\mathcal H_0} X_t\}_{t\geq 1}$, \eqref{hptest1}--\eqref{hptest3} may be readily tested by our inverse VR test, assuming the projected time series satisfy the conditions for the inverse VR test.

\begin{theorem}
\label{thm:inference}
Suppose that $\{P_{\mathcal H_0}X_t\}_{t\geq 1}$ and $\{(I-P_{\mathcal H_0})X_t\}_{t\geq 1}$ satisfy the conditions imposed in Theorem~\ref{thmvr2ab}.
\begin{enumerate}[label=(\roman*)]
\item Consider the hypotheses~\eqref{hptest1}. The test statistics in Corollary~\ref{corthmv2a} computed from $\{ (I-P_{\mathcal H_0}) X_t\}_{t\geq 1}$ with $\smalls_0 = \nd-p_0 \geq 0$ satisfy \eqref{thmeqaa3} under $H_0$ and \eqref{eqcorthmv2a} under~$H_1$.
\item Consider the hypotheses~\eqref{hptest2}. The test statistics in Corollary~\ref{corthmv2a} computed from $\{ (I-P_{\mathcal H_0}) X_t\}_{t\geq 1}$ with $\smalls_0 = 0$ and $p_0\geq \nd$ satisfy \eqref{thmeqaa3} under $H_0$ and \eqref{eqcorthmv2a} under~$H_1$.
\item Consider the hypotheses~\eqref{hptest3}. The test statistics in Corollary~\ref{corthmv2a} computed from $\{ P_{\mathcal H_0} X_t\}_{t\geq 1}$ with $\smalls_0 = 0$ and $\KK = p_0$ satisfy \eqref{thmeqaa3} under $H_0$ and \eqref{eqcorthmv2a} under~$H_1$.
\end{enumerate}
\end{theorem}

Of course, the high-level conditions in Assumption~\ref{assumvr1add}, which are applied in Theorem~\ref{thm:inference}, may be replaced by appropriate low-level conditions as discussed in Theorems~\ref{thmslack1} and~\ref{thmslack2}.

The approach to testing the hypotheses \eqref{hptest1} and \eqref{hptest2} based on $\{(I-P_{\mathcal H_0})X_t\}_{t\geq 1}$ is parallel to that in \citet{seo2024functional} who relied on an FPCA-based test for examining the number of stochastic trends embedded in a cointegrated functional time series. However, \citet[Assumption~W]{seo2024functional} requires the condition that the long-run variance of $\{P_{\SSS} X_t\}_{t\geq 1}$ is positive definite on~$\mathcal H_{\SSS}$, and this may be restrictive in cases where $X_t$ only takes values on an unknown (and possibly finite-dimensional) subspace. This condition is not required under our setup.

It is not possible, in general, to investigate the hypothesis \eqref{hptest1} (resp.\ \eqref{hptest2}) by testing if there are $p_0$ (resp.\ $\nd$) stochastic trends in the projected time series~$\{ P_{\mathcal H_0} X_t\}_{t\geq 1}$. For example, let $\{f_j\}_{j\geq 1}$ be an orthonormal basis of $\mathcal H$ and suppose $\mathcal H_{\NNN} = \spn\{f_1\}$, i.e., $\nd=1$. Consider $\mathcal H_0= \spn\{g\}$ for $g = 0.5f_1+0.5f_2$, which can be a null hypothesis in either \eqref{hptest1} or~\eqref{hptest2}. Both null hypotheses are false since neither $\spn\{g\} \subseteq \spn\{f_1\}$ nor $\spn\{g\} \supseteq \spn\{f_1\}$ is true. However, the projected time series $P_{\mathcal H_0} X_t=\langle X_t,g \rangle = \langle X_t,f_1 \rangle + \langle X_t,f_2 \rangle$ is nonstationary and has one stochastic trend. \revlab{r2_main1_2}Similarly, $H_0$ in \eqref{hptest3} cannot be investigated by testing if there are $\nd$ stochastic trends in $\{ (I-P_{\mathcal H_0}) X_t\}_{t\geq 1}$. These impossibilities are discussed in Theorem~4.2 in \citet{FranchiParuolo2025}, which characterizes the structure of the parameter space under the null and alternative of hypotheses of type \eqref{hptest1} and~\eqref{hptest2}.

\begin{remark}
\label{rem:addtest}
In addition to testing the hypotheses \eqref{hptest1}--\eqref{hptest3}, it may be of independent interest to examine the number of stochastic trends in the residual series $\{(I-P_{\mathcal H_0})X_t\}_{t\geq 1}$ for a specified subspace~$\mathcal H_0$. This is particularly useful when $\mathcal H_0$ may be understood as a finite-dimensional ``model'' of~$X_t$, as in our empirical example in Section~\ref{sec:NSM}. In such cases, we may be interested in determining how well the model can capture the dominant (nonstationary) variation in the data. The number of stochastic trends in $\{ (I-P_{\mathcal H_0}) X_t \}_{t\geq 1}$, denoted~$\smalls_{\mathcal H_0}$, can be interpreted as the dimension of the nonstationary subspace that remains unexplained by the model. We can determine $\smalls_{\mathcal H_0}$ by sequential testing as in Section~\ref{sec:sequential}. Specifically, we can apply inverse VR tests on
\begin{equation}
	\label{sequential1}
	H_0:\smalls_{\mathcal H_0} = \smalls_0 \quad \text{vs} \quad H_1: \smalls_{\mathcal H_0} > \smalls_0
\end{equation}
for $\smalls_0=\max\{\nd-p_0,0\},\max\{\nd-p_0,0\}+1,\ldots,\nd$, or we can apply VR tests on
\begin{equation}
	\label{sequential2}
	H_0:\smalls_{\mathcal H_0} = \smalls_0 \quad \text{vs} \quad H_1: \smalls_{\mathcal H_0} < \smalls_0
\end{equation}
for $\smalls_0=\nd,\nd-1,\ldots, \max\{\nd-p_0,1\}$. 
\end{remark}

\section{Monte Carlo study}
\label{sec:sim}

We study the finite-sample performance of the proposed ADI tests using a Monte Carlo simulation setup similar to the functional AR(1) model in \citet{LRS2020nonst}, \citet{NSS}, and \citet{seo2024functional}. Let $\{ g_j \}_{j \geq 1}$ be the Fourier basis functions on~$[0,1]$. For each $\nd$, we let $P_{\NNN} \Delta X_t$ and $P_{\SSS} X_t$ be generated by the following stationary functional AR(1) models,
\begin{equation*} 
P_{\NNN} \Delta X_t =  \sum_{j=1}^{\nd} \alpha_j \langle g^N_j, P_{\NNN} \Delta X_{t-1} \rangle g^N_j  + P_{\NNN} \varepsilon_t,  \quad   
P_{\SSS} X_t =  \sum_{j=1}^{12} (0.9)^{j-1} \beta_j \langle g^S_j,  P_{\SSS} X_{t-1} \rangle g^S_j + P_{\SSS} \varepsilon_t,  
\end{equation*}
where $\{ g_j^N \}_{j=1}^{\nd}$ are randomly drawn from $\{ g_1 , \ldots , g_{\nd+3} \}$ ($\nd \leq 8$ in our simulation experiments), $\{ g_j^S \}_{j=1}^{12}$ are randomly drawn from $\{ g_{15}, \ldots , g_{30} \}$, and $\varepsilon_t$ is given by $\sum_{j=1}^{40} (0.9)^{j-1} \theta_{j,t} g_j$ for standard normal random variables $\{ \theta_{j,t} \}_{j \geq 1}$ that are independent across $j$ and~$t$. We let $\alpha_j$ and $\beta_j$ be independent uniform random variables on $[-0.8,0.8]$ for each realization of the data generating process~(DGP). In practice, it is common to have a nonzero intercept or a linear time trend as in Section~\ref{sec_deterministic}. We here consider the former case and add an intercept~$\zeta_1$, which is also randomly chosen, to each realization of the DGP. Specifically, $\zeta_1 = \sum_{j=1}^{30} (0.9)^{j-1} \tilde{\theta}_j g_j$, where $\{ \tilde{\theta}_j \}_{j=1}^{30}$ are independent standard normal random variables. Viewed as a time series taking values in the $L^2[0,1]$ Hilbert space, we may obtain the eigenelements of the sample (long-run) variance of $X_t$ using FPCA (see e.g., \citealp{Chang2016}; \citealp{NSS}; \citealp{LRS2020nonst}). To this end, we represent $X_t$, which is assumed to be observed on 200 regularly spaced points of~$[0,1]$, using the first 40 Legendre basis functions.

To choose bandwidths, we follow Remarks~\ref{remvr21}, \ref{remLRV}, \ref{remInvVR}, and~\ref{rem:newremark}. For VR tests with $d_R = 1$ we let $h_P=h_L=h_R=0$ (Remarks~\ref{remvr21} and~\ref{rem:newremark}). For the remaining tests, we let $h_P=[T^{1/3}]$ because a large bandwidth is supported by Assumption~\ref{assumvr1ll}. For VR tests with $d_R = 0$ we let $h_R = h_{\rm opt}$, which is the \citet{andrews1991} optimal bandwidth (Remark~\ref{remLRV}), but for the inverse tests the series are nonstationary, so we let $h_R=[T^{1/5}]$, which is the same rate as~$h_{\rm opt}$ (Remark~\ref{remInvVR}). We consider both $h_L=0$ and $h_L=[ T^{1/3}]$ (Remark~\ref{remLRV}). Throughout, we let $\km(\cdot)$ be the Tukey-Hanning kernel for $m \in \{ L,R,P \}$, all test statistics apply the trace test functional~$\Ftr$, and the slack extraction uses $\KK = \smalls_0 +2$ following Remark~\ref{remadd}.

\revlab{ce_2nd_major3}We present simulation results for the VR(2,1), VR(2,0), VR(1,0), and inverse VR tests (the latter based on the (2,1) problem). The VR choices (2,1), (1,0), and the inverse VR are motivated by well-known tests from finite-dimensional Euclidean space, see Remark~\ref{revremadd1} and Appendix~\ref{secexisting}, and the VR(2,0) is in-between. We also performed simulations for VR tests based on larger values of~$d_L$, but found that their performances were inferior to the tests reported here (results available upon request).
 
\begin{table}[tbp]
\caption{Rejection frequencies of VR(2,1), VR(2,0), VR(1,0), and inverse VR tests}
\label{tab:mc1}
\vskip -8pt
\small
\begin{tabular*}{\linewidth}{@{\extracolsep{\fill}}lcccccccccccc}
\toprule
&&\multicolumn{5}{c}{$T = 250$} &&\multicolumn{5}{c}{$T = 500$}\\
\cmidrule{3-7}\cmidrule{9-13}
& $\smalls_0$ & $\nd =0$ & $\nd =1$ & $\nd =3$ & $\nd =5$ & $\nd =7$ &  & $\nd =0$ & $\nd =1$ & $\nd =3$ & $\nd =5$ & $\nd =7$ \\
\midrule 
VR(2,1) & $ \nd $ &  & 0.049 & 0.053 & 0.064 & 0.105 &  &  & 0.051 & 0.050 & 0.059 & 0.061 \\ 
& $ \nd +1$ & 1.000 & 0.998 & 0.999 & 0.999 & 0.999 &  & 1.000 & 1.000 & 1.000 & 1.000 & 1.000 \\ 
& $ \nd +2$ & 1.000 & 1.000 & 1.000 & 1.000 & 1.000 &  & 1.000 & 1.000 & 1.000 & 1.000 & 1.000 \\ 
\midrule 
\multicolumn{13}{c}{$h_L=0$}\\  
VR(2,0) & $ \nd $ &  & 0.043 & 0.022 & 0.008 & 0.004 &  &  & 0.046 & 0.031 & 0.018 & 0.009 \\ 
& $ \nd +1$ & 1.000 & 0.899 & 0.599 & 0.374 & 0.213 &  & 1.000 & 0.982 & 0.857 & 0.678 & 0.498 \\ 
& $ \nd +2$ & 1.000 & 0.915 & 0.667 & 0.451 & 0.286 &  & 1.000 & 0.992 & 0.919 & 0.770 & 0.608 \\ 
VR(1,0) & $ \nd $ &  & 0.027 & 0.005 & 0.000 & 0.000 &  &  & 0.036 & 0.010 & 0.003 & 0.000 \\ 
& $ \nd +1$ & 1.000 & 0.636 & 0.144 & 0.024 & 0.002 &  & 1.000 & 0.709 & 0.374 & 0.155 & 0.036 \\ 
& $ \nd +2$ & 1.000 & 0.671 & 0.223 & 0.093 & 0.010 &  & 1.000 & 0.729 & 0.417 & 0.246 & 0.122 \\ 
Inv.VR& $ \nd $ & 0.092 & 0.070 & 0.070 & 0.065 & 0.080 &  & 0.084 & 0.073 & 0.075 & 0.082 & 0.097 \\
& $\nd-1$ &    & 0.982 & 0.940 & 0.923 & 0.890 &  &    & 0.998 & 0.998 & 0.999 & 0.999 \\
& $\nd-2$ &    &    & 1.000 & 1.000 & 1.000 &  &    &    & 1.000 & 1.000 & 1.000 \\ 
\midrule 
\multicolumn{13}{c}{$h_L=[T^{1/3}]$}\\  
VR(2,0) & $ \nd $ &  & 0.043 & 0.024 & 0.010 & 0.005 &  &  & 0.046 & 0.032 & 0.021 & 0.011 \\ 
& $ \nd +1$ & 1.000 & 0.902 & 0.613 & 0.402 & 0.252 &  & 1.000 & 0.982 & 0.865 & 0.698 & 0.534 \\ 
& $ \nd +2$ & 1.000 & 0.920 & 0.689 & 0.491 & 0.330 &  & 1.000 & 0.992 & 0.928 & 0.799 & 0.646 \\ 
VR(1,0) & $ \nd $ &  & 0.052 & 0.046 & 0.047 & 0.050 &  &  & 0.058 & 0.046 & 0.052 & 0.049 \\ 
& $ \nd +1$ & 1.000 & 0.879 & 0.638 & 0.480 & 0.362 &  & 1.000 & 0.954 & 0.795 & 0.668 & 0.547 \\ 
& $ \nd +2$ & 1.000 & 0.899 & 0.698 & 0.552 & 0.433 &  & 1.000 & 0.967 & 0.850 & 0.743 & 0.637 \\  
Inv.VR & $\nd$ & 0.089 & 0.066 & 0.054 & 0.032 & 0.025 &  & 0.082 & 0.070 & 0.064 & 0.057 & 0.051 \\
& $\nd-1$ &    & 0.982 & 0.930 & 0.882 & 0.773 &  &    & 0.998 & 0.997 & 0.998 & 0.996 \\
& $\nd-2$ &    &    & 1.000 & 1.000 & 0.998 &  &    &    & 1.000 & 1.000 & 1.000 \\ 
\bottomrule 
\end{tabular*}
\vskip 4pt
{\footnotesize Notes: Based on 10,000 Monte Carlo replications. The DGP true dimension is~$\nd$, the null hypothesized value is~$\smalls_0$, and the nominal size is~5\%.} 
\end{table} 

Table~\ref{tab:mc1} reports rejection frequencies. We consider true dimension $\nd \in \{ 0,1,3,5,7 \}$. For the VR tests we consider the null $\smalls_0 = \nd$ and alternatives $\smalls_0 = \nd +1$ and $\smalls_0 = \nd +2$, and for the inverse VR test we consider the null $\smalls_0 = \nd$ and alternatives $\smalls_0 = \nd -1$ and $\smalls_0 = \nd -2$. The results show that the VR(1,0) and VR(2,0) tests are seriously under-sized when $a_L=0$, as expected based on \citet{shintani2001simple}, especially for the larger values of $\nd$ considered, and both tests suffer from lower power in those cases as a consequence. Size control is much better for $h_L=[T^{1/3}]$ for the VR(1,0) test, but not for the VR(2,0) test. The VR(2,1) test has very good size properties in all cases considered except the largest value of $\nd$ with the smallest sample size, and it has excellent power. Finally, we see that the inverse VR test is slightly over-sized in some cases. Its power is nearly as impressive as that of the VR(2,1) test, but of course the alternatives are different for the inverse VR test, and it does not have the natural upper bound $\smalls_{\max}$ that the VR tests have.

\begin{table}[tbp]
\caption{Frequencies of correctly estimating $\nd$}
\label{tab1a} 
\vskip -8pt
\small
\begin{tabular*}{\linewidth}{@{\extracolsep{\fill}}lcccccccccccc}
\toprule
&& \multicolumn{5}{c}{$T=250$}&&\multicolumn{5}{c}{$T=500$}\\
\cmidrule{3-7}\cmidrule{9-13} 
Method & & $\nd =0$ & $\nd =1$ & $\nd =3$ & $\nd = 5$ & $\nd =7$&  &$\nd =0$ & $\nd =1$ & $\nd =3$ & $\nd = 5$ & $\nd =7$   \\ 
\midrule 
$\ddot{\smalls}$ &       &       &  0.916  &  0.622  &  0.543  &  0.356  &       &       &  0.958  &  0.832  &  0.862  &  0.830  \\ 
$\widehat{\smalls}_{\text{\tiny{LRS}}}$ &       &       &  0.998  &  0.275  &  0.062  &  0.016  &       &       &  0.996  &  0.478  &  0.190  &  0.071  \\ 
$\widehat{\smalls}_{\rm TD}$: VR(2,1)  &       &  1.000  &  0.949  &  0.946  &  0.935  &  0.894  &       &  1.000  &  0.949  &  0.950  &  0.941  &  0.939  \\ 
$\widehat{\smalls}_{\rm UD}$: VR(2,1)  &       &  1.000  &  0.949  &  0.946  &  0.935  &  0.894  &       &  1.000  &  0.949  &  0.950  &  0.941  &  0.939  \\ 
\midrule
\multicolumn{13}{c}{$h_L=0$}\\  
$\widehat{\smalls}_{\rm TD}$: VR(2,0)  &       &  1.000  &  0.810  &  0.552  &  0.357  &  0.207  &       &  1.000  &  0.925  &  0.805  &  0.650  &  0.487  \\ 
$\widehat{\smalls}_{\rm TD}$: VR(1,0)  &       &  1.000  &  0.602  &  0.134  &  0.023  &  0.002  &       &  1.000  &  0.668  &  0.363  &  0.152  &  0.036  \\ 
$\widehat{\smalls}_{\rm BU}$: Inv.VR &       &  0.908  &  0.915  &  0.872  &  0.860  &  0.812  &       &  0.916  &  0.925  &  0.923  &  0.917  &  0.901  \\ 
$\widehat{\smalls}_{\rm UD}$: VR(2,0)  &       &  1.000  &  0.810  &  0.552  &  0.358  &  0.207  &       &  1.000  &  0.925  &  0.804  &  0.650  &  0.487  \\ 
$\widehat{\smalls}_{\rm UD}$: VR(1,0)  &       &  1.000  &  0.602  &  0.133  &  0.023  &  0.002  &       &  1.000  &  0.668  &  0.362  &  0.152  &  0.036  \\ 
\midrule 
\multicolumn{13}{c}{$h_L=[T^{1/3}]$}\\  
$\widehat{\smalls}_{\rm TD}$: VR(2,0)  &       &  1.000  &  0.817  &  0.568  &  0.384  &  0.243  &       &  1.000  &  0.928  &  0.815  &  0.668  &  0.519  \\ 
$\widehat{\smalls}_{\rm TD}$: VR(1,0)  &       &  1.000  &  0.823  &  0.591  &  0.432  &  0.312  &       &  1.000  &  0.893  &  0.748  &  0.615  &  0.498  \\ 
$\widehat{\smalls}_{\rm BU}$: Inv.VR  &       &  0.911  &  0.918  &  0.878  &  0.851  &  0.749  &       &  0.918  &  0.928  &  0.933  &  0.941  &  0.945  \\ 
$\widehat{\smalls}_{\rm UD}$: VR(2,0)  &       &  1.000  &  0.817  &  0.569  &  0.384  &  0.243  &       &  1.000  &  0.928  &  0.815  &  0.668  &  0.519  \\ 
$\widehat{\smalls}_{\rm UD}$: VR(1,0)  &       &  1.000  &  0.822  &  0.591  &  0.432  &  0.312  &       &  1.000  &  0.893  &  0.748  &  0.615  &  0.498  \\  
\bottomrule
\end{tabular*}
\vskip 4pt
{\footnotesize Notes: Based on 10,000 Monte Carlo replications. The DGP true dimension is~$\nd$, the nominal size is~5\%, and $\smalls_{\max} = \nd+5$ for the TD methods.} 
\end{table}

In Table~\ref{tab1a} we investigate the finite-sample performance of the tests when they are applied to estimate $\nd$ using one of the sequential procedures described in Section~\ref{sec:sequential}. For comparison, we also consider the eigenvalue ratio estimator $\ddot\smalls$ computed from the VR(2,1) problem (see Theorem~\ref{thmest}), and \citepos{LRS2020nonst} estimator, which is denoted~$\widehat\smalls_{\text{\tiny{LRS}}}$. For each method and each value of~$\nd$, the table reports the frequencies of correctly estimating the true dimension~$\nd$. Overall, the direct estimators $\ddot\smalls$ and $\widehat\smalls_{\text{\tiny{LRS}}}$ perform worse than the VR(2,1) and inverse VR sequential testing procedures, and between the two eigenvalue ratio estimators, $\ddot\smalls$ tends to perform better than~$\widehat\smalls_{\text{\tiny{LRS}}}$. The sequential TD procedure based on the VR(2,1) tests performs very well, and the BU procedure (based on the inverse VR test) performs nearly as well. On the other hand, the TD procedures based on the VR(2,0) and VR(1,0) tests perform very poorly for $\nd \geq 3$, which was expected based on their poor power observed in Table~\ref{tab:mc1}. The choice of $h_L$ seems to significantly affect the performance of the VR(1,0) tests, especially when $\nd$ is large. Of course, this is not the case for the TD procedure based on the VR(2,1) test, and indeed, this method appears to perform the best overall.

Results for the UD hybrid procedure are also reported in Table~\ref{tab1a}. This enjoys the advantage of a data-driven selection of $\smalls_{\max} = \widehat\smalls_{\rm BU}+5$. This $\smalls_{\max}$ is random, but the good performance of the BU procedure, the addition of 5 from~$\widehat\smalls_{\rm BU}$, and the strong robustness of the TD procedures to the choice of~$\smalls_{\max}$, imply that in fact the UD procedures enjoy nearly identical frequencies of correct estimation of $\nd$ as the corresponding TD procedures. Thus, the advantage of the data-driven selection of $\smalls_{\max}$ comes without cost, at least for these sample sizes, and consequently the UD procedure based on VR(2,1) is the preferred method in our Monte Carlo simulations.

Finally, recall that the functional observations are constructed from 200 regularly spaced points on $[0,1]$ by smoothing using the first 40 Legendre basis functions. However, as discussed earlier, our methodology can also be applied to the time series of 200-dimensional vectors of discrete realizations of $X_t$ with no additional theoretical modification. We thus repeated the analysis in Table~\ref{tab1a} letting $\{X_t\}_{t\geq 1}$ be viewed as a 200-dimensional vector-valued time series, but the results were nearly identical and are hence omitted.

\begin{remark} \label{remarkimplement}
Based on our simulation evidence, the hybrid procedure, combining the inverse VR (with $h_L=0$ and $h_R>0$) and the VR($2,1$) sequential tests, appears to perform well with $\KK = \smalls_0+m$ (see Remark~\ref{remadd}). We provide a detailed practical implementation guide in Section~\ref{sec:guide} in the supplement. In that section, we also discuss how to implement the procedure when the VR($2,1$) test is replaced by other variants.
\end{remark}

\section{Empirical applications}
\label{sec_empirical}
In this section we illustrate our methdology with two real-world data examples.

\subsection{Corporate bond yield curves}
\label{sec_empirical_1}

We apply the proposed tests to examine the number of common stochastic trends in the monthly time series of high quality market corporate bond yield curves (a.k.a.\ the HQM yield curve). The data is from the US Department of Treasury available at \url{https://home.treasury.gov/} and spans Jan.\ 1984 to Dec.\ 2018. At each time~$t$, we observe a corporate bond with 200 different maturities ranging from 6~months to 100~years. This data was studied by \citet{BT2022}, who assumed that this high-dimensional time series is driven by a few factors of deterministic trends and I(1) and I(0) processes, and estimated the number of those factors using their own testing procedure. However, it may be more natural to view this time series as realizations of yield curves that are either finite-dimensional (of high and unknown dimension) or curve-valued. With this point of view, we apply our methodology to determine the number of stochastic trends. To implement our methodology, we represent the HQM yield data by the first 40 Legendre basis functions as in Section~\ref{sec:sim}.

\begin{figure}[tbp]
\caption{HQM yield curves}
\begin{subfigure}{.5\textwidth}
	\subcaption{Yields at 200 maturities}
	\label{fig_yield1}
	\includegraphics[width =  \textwidth, height = .2\textheight, trim = {5cm 5cm 5cm 10cm},clip]{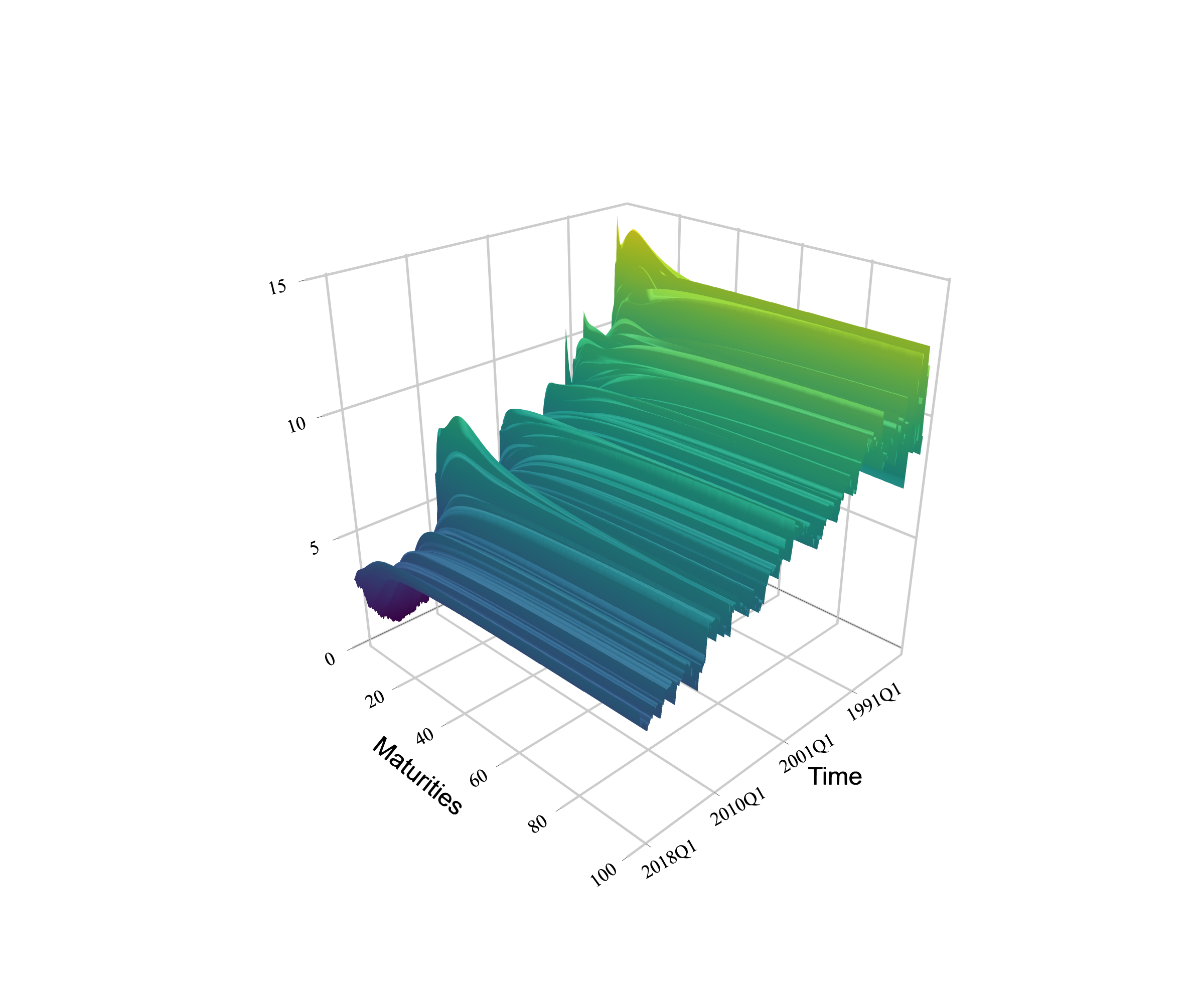}
\end{subfigure}
\begin{subfigure}{.4\textwidth}
	\subcaption{Yields at a few different maturities}
	\label{fig_yield2}
	\includegraphics[width =  \textwidth, height = .2\textheight]{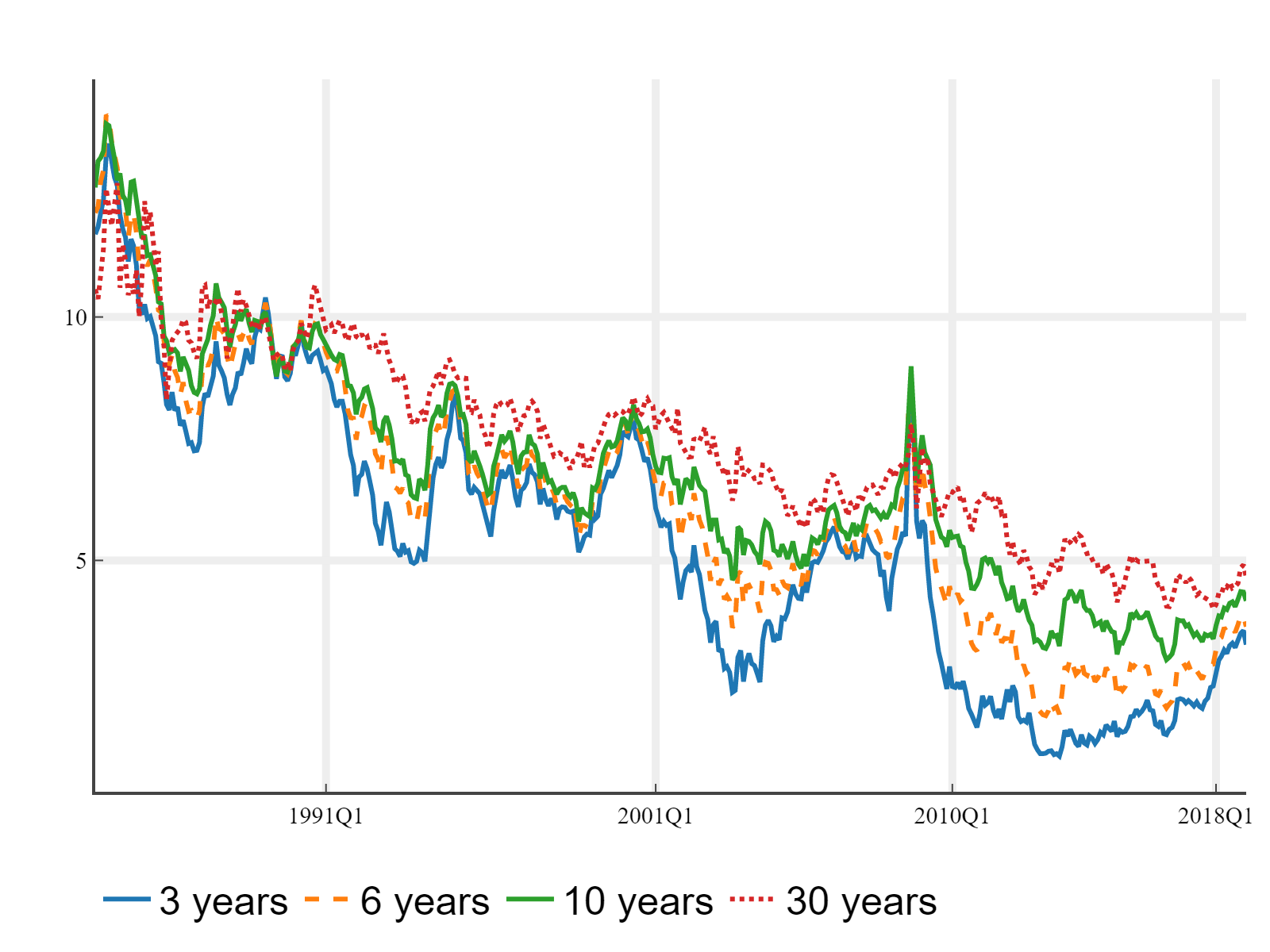}
\end{subfigure}
\end{figure}

Figure~\ref{fig_yield1} shows the time series of HQM yield curves (observed at 200 different maturities) and Figure~\ref{fig_yield2} shows the time series of yields at fixed maturities of 3, 6, 10, and 30 years. The figures suggest that the HQM yield data includes a linear time trend. The existence of a linear trend is tested and confirmed by \citet{BT2022} for the data span of Jan.\ 1985 to Sep.\ 2017. Thus, we apply the trend-adjusted testing procedures in Section~\ref{sec_deterministic}.

\begin{table}[tbp] 
\caption{Empirical results for HQM yield curve data}
\label{tab:bonds} 
\vskip -8pt
\small
\begin{tabular*}{\linewidth}{@{\extracolsep{\fill}}lrrrrrrrr}
\toprule
\multicolumn{9}{c}{Eigenvalue ratio estimates}\\
Statistic &&$j=1$\phantom{$^{**}$} & $j=2$\phantom{$^{**}$} & $j=3$\phantom{$^{**}$} & $j=4$\phantom{$^{**}$} & $j=5$\phantom{$^{**}$} &$j=6$\phantom{$^{**}$} &$j=7$\phantom{$^{**}$} \\
\midrule
\multicolumn{2}{l}{$\mu_{j+1}/\mu_j$} & 2.89\phantom{$^{**}$} & 3.00\phantom{$^{**}$} & 2.02\phantom{$^{**}$} & 1.91\phantom{$^{**}$} & 1.15\phantom{$^{**}$} & 3.25\phantom{$^{**}$} & 2.15\phantom{$^{**}$} \\
\multicolumn{2}{l}{$\hat{\kappa}_{j+1}/\hat{\kappa}_j$} &0.42\phantom{$^{**}$} & 0.14\phantom{$^{**}$} & 0.08\phantom{$^{**}$} & 0.29\phantom{$^{**}$} & {0.00}\phantom{$^{**}$} & 1\phantom{$.00^{**}$} & 1\phantom{.00$^{**}$} \\
\midrule
\multicolumn{9}{c}{Variance ratio test statistics}\\
Statistic & $\smalls_0 = 0$ & $\smalls_0 = 1$\phantom{$^{**}$} & $\smalls_0 = 2$\phantom{$^{*}$} & $\smalls_0 = 3$\phantom{$^{**}$} & $\smalls_0 = 4$\phantom{$^{**}$} & $\smalls_0 = 5$\phantom{$^{**}$} & $\smalls_0 = 6$\phantom{$^{**}$} & $\smalls_0 = 7$\phantom{$^{**}$} \\
\midrule
Inv.VR  & 0.94$^{**}$ & 0.60$^{**}$ & 0.21$^{**}$ & 0.12$^{*}$\phantom{$^{*}$}  & 0.10$^{*}$\phantom{$^{*}$}  & 0.07$^{*}$\phantom{$^{*}$}  & 0.05$^{\dagger}$\phantom{$^{*}$} & 0.03\phantom{$^{**}$} \\ 
VR(2,1) &  & 151.61\phantom{$^{**}$} & 409.16\phantom{$^{**}$} & 1355.09$^{*}$\phantom{$^{*}$}  & 3484.67$^{**}$ & 7175.13$^{**}$ & 10623.79$^{**}$ & 23154.51$^{**}$ \\ 
VR(2,0) &  & 7007.65\phantom{$^{**}$} & 13261.54\phantom{$^{**}$} & 117284.83$^{*}$\phantom{$^{*}$}  & 522860.44$^{**}$ & 1208358.03$^{**}$ & 2215359.22$^{**}$ & 6936298.27$^{**}$ \\ 
VR(1,0) &  & 24.25\phantom{$^{**}$} & 77.66$^{\dagger}$\phantom{$^{*}$} & 259.58$^{**}$ & 599.41$^{**}$ & 1050.22$^{**}$ & 1836.25$^{**}$ & 3825.57$^{**}$ \\
\bottomrule
\end{tabular*}
\vskip 4pt
{\footnotesize Notes: $^{\dagger}$, $^{*}$, and $^{**}$ indicate significance at 10\%, 5\%, and 1\% level, respectively.}
\end{table}

The empirical results are reported in Table~\ref{tab:bonds}. We implement all procedures as in Section~\ref{sec:sim} using $h_L=[T^{1/3}]$ for the VR(2,0), VR(1,0), and inverse VR tests, and using $\KK = \smalls_0 +2$ for slack extraction. The eigenvalue ratio estimates in the first part of the table are $\ddot\smalls = 6$ and $\widehat\smalls_{\text{\tiny{LRS}}} = 5$ (using~$\kappa=10^{-4}$), although the ratios are close to suggesting $\ddot\smalls = 2$ and $\widehat\smalls_{\text{\tiny{LRS}}} = 3$.

In the second part of Table~\ref{tab:bonds} we report the VR and inverse VR test statistics for a range of~$\smalls_0$. The inverse VR tests imply that $\widehat\smalls_{\rm BU} = 3$ (at 1\% level) or $\widehat\smalls_{\rm BU} = 6$ (at 5\% level). Given this result, the choice $\smalls_{\max} = 7$ seems like a reasonable input to the UD procedures. The latter give either $\widehat\smalls_{\rm UD} = 2$ or $\widehat\smalls_{\rm UD} = 3$ depending on the choice of ($d_L,d_R)$ and level. However, the BU and TD procedures (with (2,1) and (2,0)) only agree at 1\% level, in which case all three find $\widehat\smalls = 3$. Overall, the evidence thus supports $\nd = 3$ stochastic trends.

\subsection{The Nelson-Siegel term structure model}
\label{sec:NSM}

Given an estimated dimension, it is possible to examine various hypotheses on the nonstationary subspace as in Section~\ref{sec:tests}. A popular model for the term structure of interest rates was proposed by \citet{NelsonSiegel87}; henceforth the N-S model. They represent the term structure of yield curves by a linear combination of three parametric functions,
\begin{equation*}
\gamma_0 = 1, \quad 
\gamma_1 = \gamma_1 (\varsigma, \tau) = \varsigma (1-e^{-\tau/\varsigma})/\tau , \quad 
\gamma_2 = \gamma_2 (\varsigma, \tau) = \varsigma (1-e^{-\tau/\varsigma})/\tau - e^{-\tau/\varsigma} ,
\end{equation*}
where $\tau$ is time to maturity, $\varsigma$ is a shape parameter, and $\gamma_0$, $\gamma_1$, and $\gamma_2$ denote level, slope, and curvature factors. Because $\gamma_1$ and $\gamma_2$ are nearly collinear for large $\tau$, we subsequently consider only data for $\tau \in [0,30]$ years. This has almost no impact on the results reported in Table~\ref{tab:bonds}. We now investigate to what extent the N-S model can explain the dominant (nonstationary) variation in the data. To this end, we need to specify the shape parameter,~$\varsigma$. Since our data is annualized, we set $\varsigma = 1.37$ following \citet{DieboldLi2006}.

We first test the hypothesis $H_0$ in \eqref{hptest3} with $\mathcal H_0 = \spn \{ \gamma_0 , \gamma_1, \gamma_2 \}$ using Theorem~\ref{thm:inference}(iii). That is, we test whether ``fitted values,'' $\{ P_{\mathcal H_0}X_t \}_{t\geq 1}$, from the N-S model are stationary. Not surprisingly, this hypothesis is strongly rejected. The same conclusion holds for tests of \eqref{hptest3} when $\mathcal H_0$ is specified as the span of any one or two of the factors $\gamma_i$,~$i=0,1,2$. This confirms that none of the factors are stationary.

Next, we test if the level, slope, and curvature factors span the nonstationary subspace as in~\eqref{hptest2}. From our earlier results in Table~\ref{tab:bonds}, we set the dimension of the latter to $\nd = 3$ (testing at 1\% level). Thus, there is no possibility that the N-S model with only one or two factors can yield stationary residuals. With three factors, it is of interest to test if the N-S model can explain all the nonstationary variation in the yield curve data and, as a result, can yield stationary residuals. In this sense, the hypothesis $H_0 : \mathcal H_0 = \spn \{ \gamma_0 , \gamma_1, \gamma_2 \} \supseteq \mathcal H_{\NNN}$ in \eqref{hptest2} is very interesting, and its non-rejection would be the strongest conclusion that we could obtain. However, implementing the test as in Theorem~\ref{thm:inference}(ii) with $\smalls_0 = 0$ and $p_0 = 3$, the inverse VR statistic is $1.47$ (see Table~\ref{tab:NSM}) and rejects the null at 1\% level.

We proceed to test $H_0$ in \eqref{hptest1}, where $\mathcal H_0$ is specified as various choices of factor(s). That is, we test whether any one factor can reduce the dimension of the nonstationary subspace by one, and also whether any two factors can reduce the dimension by two. Of course, the latter is less likely to be accepted than the former. For one factor, implementing the test as in Theorem~\ref{thm:inference}(i), the inverse VR statistics with $\smalls_0 = 2$ are $0.24$, $0.21$, and $0.22$ for $\mathcal H_0 = \spn \{ \gamma_0 \}$, $\mathcal H_0 = \spn \{ \gamma_1 \}$, $\mathcal H_0 = \spn \{ \gamma_2 \}$, respectively, in each case rejecting the null at 1\% level. For two factors, the inverse VR statistics with $\smalls_0 = 1$ are $0.40$ and $0.38$ for $\mathcal H_0 = \spn \{ \gamma_0 , \gamma_1 \}$ and $\mathcal H_0 = \spn \{ \gamma_1 , \gamma_2 \}$, respectively, in both cases rejecting the null at 1\% level. Thus, there is no evidence to suggest that any one factor is in the nonstationary subspace, i.e., that any one factor can be interpreted as one of the common stochastic trends.

\begin{table}[tbp] 
\caption{Test results for $( I - P_{\mathcal H_0})X_t$ with $\mathcal H_0 = \spn\{\gamma_0,\gamma_1 ,\gamma_2 \}$ and $\varsigma=1.37$} 
\label{tab:NSM}
\vskip -8pt
\small
\begin{tabular*}{\linewidth}{@{\extracolsep{\fill}}lrrrrr}
\toprule
Statistic & $\smalls_0=0$\phantom{$^{**}$} & $\smalls_0=1$\phantom{$^{**}$} & $\smalls_0=2$\phantom{$^{**}$} & $\smalls_0=3$\phantom{$^{**}$} \\ 
\midrule 
Inv.VR  & 1.47$^{**}$ & 0.27$^{**}$ & 0.24$^{**}$ & 0.12$^{*}$\phantom{$^{*}$}    \\ 
VR(2,1) &  & 113.48\phantom{$^{**}$} & 629.06$^{\dagger}$\phantom{$^{*}$} & 1577.24$^{*}$\phantom{$^{*}$}   \\ 
VR(2,0) &  & 6842.81\phantom{$^{**}$} & 39832.64$^{*}$\phantom{$^{*}$} & 123243.78$^{**}$  \\ 
VR(1,0) &  & 72.34$^{**}$ & 167.27$^{**}$ & 346.79$^{**}$   \\ 
\bottomrule
\end{tabular*}
\vskip 4pt
{\footnotesize Notes: $^{\dagger}$, $^{*}$, and $^{**}$ indicate significance at 10\%, 5\%, and 1\% level, respectively.}
\end{table}

Finally, following Remark~\ref{rem:addtest}, we investigate if the nonstationary subspace of the residual series, $\{ (I-P_{\mathcal H_0})X_t \}_{t\geq 1}$ with $\mathcal H_0 = \spn \{ \gamma_0 , \gamma_1, \gamma_2 \}$, is of smaller dimension than that of $X_t$ itself. Letting the former be denoted~$\smalls_{\mathcal H_0}$, we apply the sequential procedures in \eqref{sequential1} and~\eqref{sequential2}. The results are reported in Table~\ref{tab:NSM}. The BU procedure yields $\widehat\smalls_{\mathcal H_0 , {\rm BU}} = 3$ at 1\% level and rejects $\smalls_0 = 3$ at 5\% level. Applying the TD sequential procedure starting from $\smalls_{\max} = 3$, we find $\widehat\smalls_{\mathcal H_0 , {\rm TD}} = 2$ with the preferred VR(2,1) test at 5\% level (but $\widehat\smalls_{\mathcal H_0 , {\rm TD}} = 1$ at 10\% level with the VR(2,1) test, $\widehat\smalls_{\mathcal H_0 , {\rm TD}} = 1$ at 5\% level with the VR(2,0) test, and $\widehat\smalls_{\mathcal H_0 , {\rm TD}} = 0$ at 1\% level with the VR(1,0) test). Thus, we do find some evidence that three factors in the N-S model explain some of the nonstationary variation in the data.

The empirical results for the N-S model are specific to the value of $\varsigma$, at least for the slope and curvature factors. As a robustness check, we repeated the analysis for $\varsigma = 1,2,3,4,5$, and the results were qualitatively the same. In conclusion, the three parametric factors of the N-S model cannot fully explain the nonstationarity in the corporate yield curve data, so the residuals from the model will still exhibit nonstationarity. However, the model can capture some of the nonstationarity, in the sense that the dimension of the nonstationary subspace of the residuals is smaller than that of the original yield curve data.

\subsection{Labor market indices}
\label{sec_empirical_2} 

In empirical (macro) analysis, high-dimensional observations are often given by a collection of potentially nonstationary variables that are closely related to each other but do not have a natural ordering. In such cases, not only is each observation naturally understood as a realization of a high-dimensional random vector that cannot generally be smoothed to a curve (unlike in our previous example of yield curve data), but also the number of linearly independent I(1) stochastic trends may be substantially smaller than the total number of variables. As discussed, in this case as well, our proposed methodology can be applied to study the number of stochastic trends and/or stationarity without any modifications. To illustrate this versatility of our methodology, we now consider another empirical example analyzing 29 monthly labor market indices provided in the \texttt{FRED-MD} data set \citep{McCracken2016}. The indices include the civilian labor force, employment/unemployment rate, hourly wage, and other labor-related economic variables; a detailed list can be found in the table labeled ``Group 2'' in the appendix of \citet{McCracken2016}. We apply log-transformation to the data.\footnote{The log transformation was suggested by \citet{McCracken2016} for some variables. We found that the remaining variables are often very large in scale (e.g., ``HWI'') or have only positive values. Thus, those variables are also log-transformed in our testing procedure.} The time span used in this analysis  is Jan.\ 1990 to Dec.\ 2019.

\begin{figure}[tbp]
\caption{Time series of $\langle X_t,\widehat{\nu}_j \rangle$}
\label{fig:levels}
\begin{subfigure}{.32\linewidth}
	\subcaption{$\langle X_t,\widehat{\nu}_1 \rangle$}
	\includegraphics[width = \linewidth]{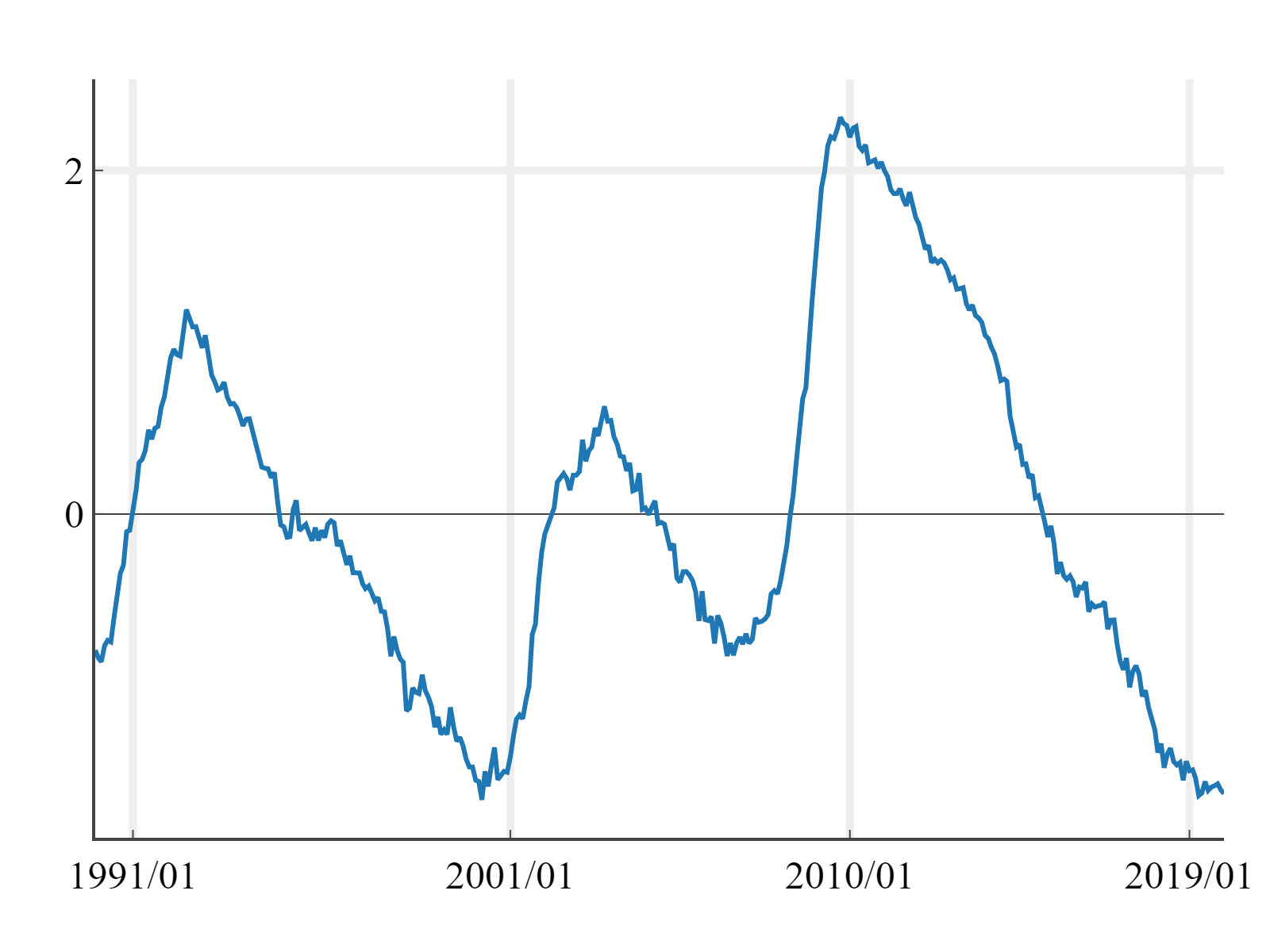}
\end{subfigure}
\begin{subfigure}{.32\linewidth}
	\subcaption{$\langle X_t,\widehat{\nu}_5 \rangle$}
	\includegraphics[width = \linewidth]{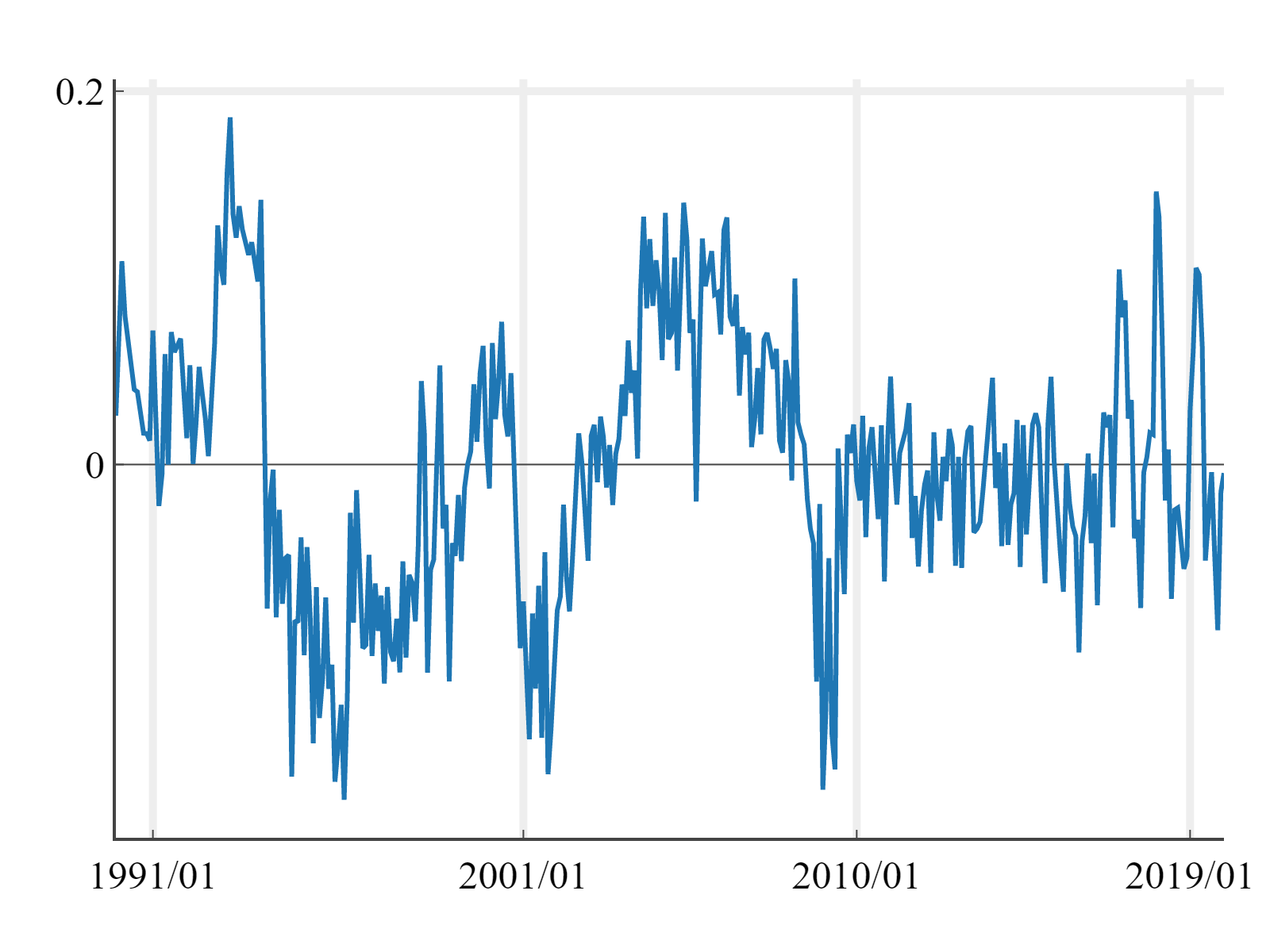}
\end{subfigure}
\begin{subfigure}{.32\linewidth}
	\subcaption{$\langle X_t,\widehat{\nu}_{10} \rangle$}
	\includegraphics[width = \linewidth]{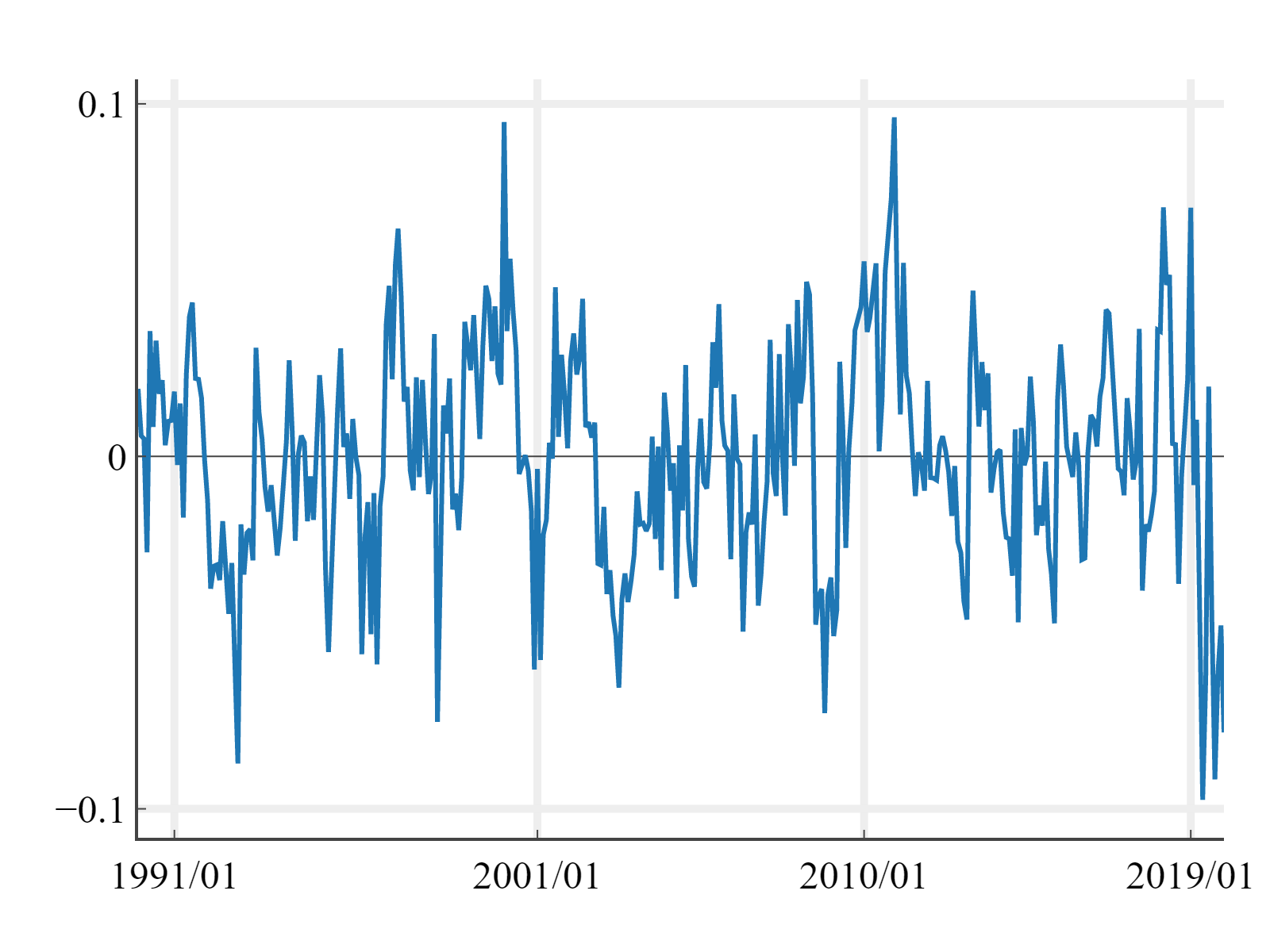}
\end{subfigure}\\
{\footnotesize Notes: Time series of $\langle X_t, \widehat{\nu}_j \rangle $ for some selected eigenvectors $\widehat{\nu}_j$ of $\widehat\Lambda_{0,R}$.}
\end{figure}

\begin{figure}[tbp]
\caption{Time series of $\langle \Delta X_t,\widehat{\nu}_j \rangle$}
\label{figscore}
\begin{subfigure}{.32\linewidth}
	\subcaption{$\langle \Delta X_t,\widehat{\nu}_1 \rangle$}
	\includegraphics[width = \linewidth]{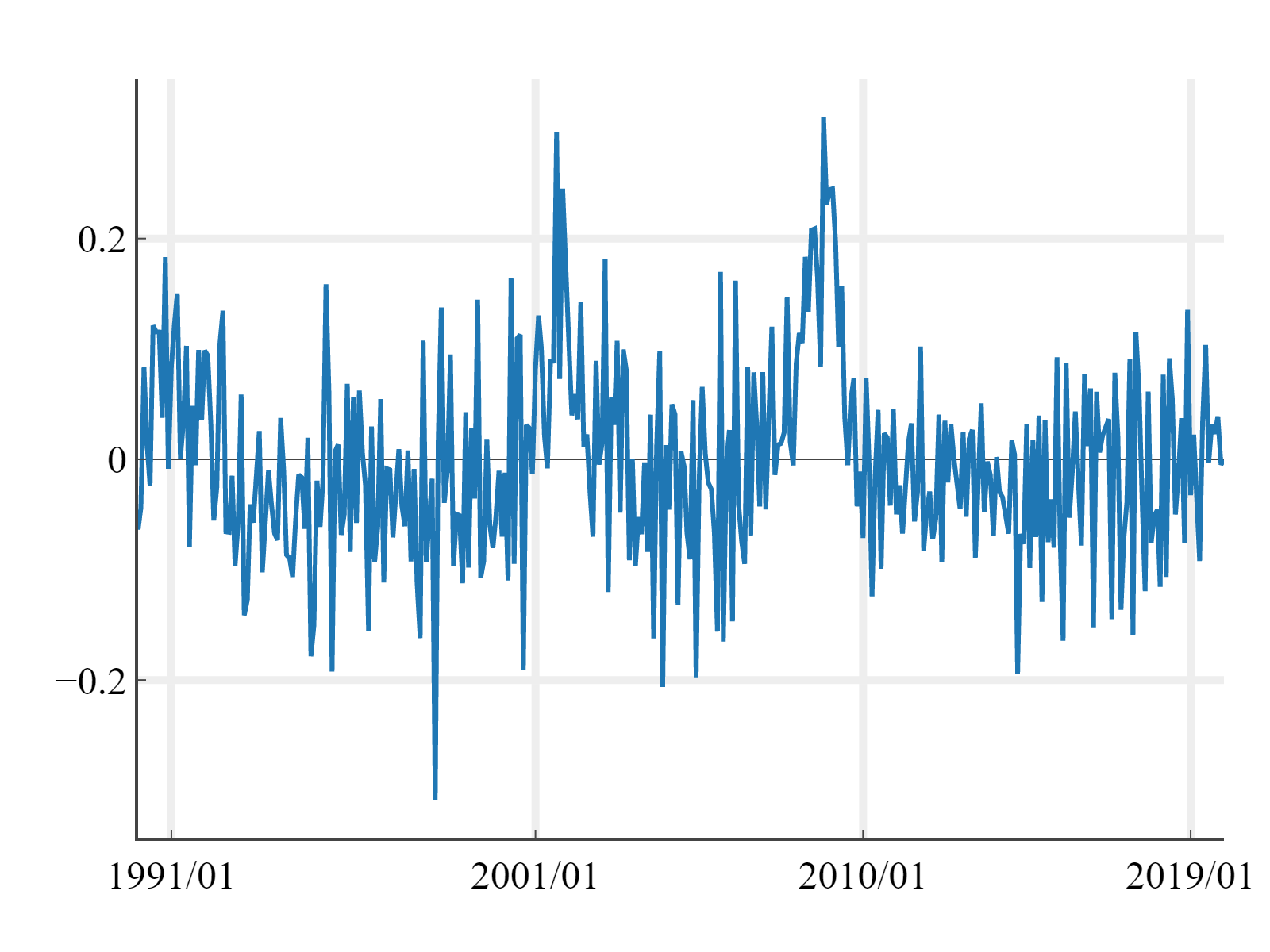}
\end{subfigure}
\begin{subfigure}{.32\linewidth}
	\subcaption{$\langle \Delta X_t,\widehat{\nu}_2 \rangle$}
	\includegraphics[width = \linewidth]{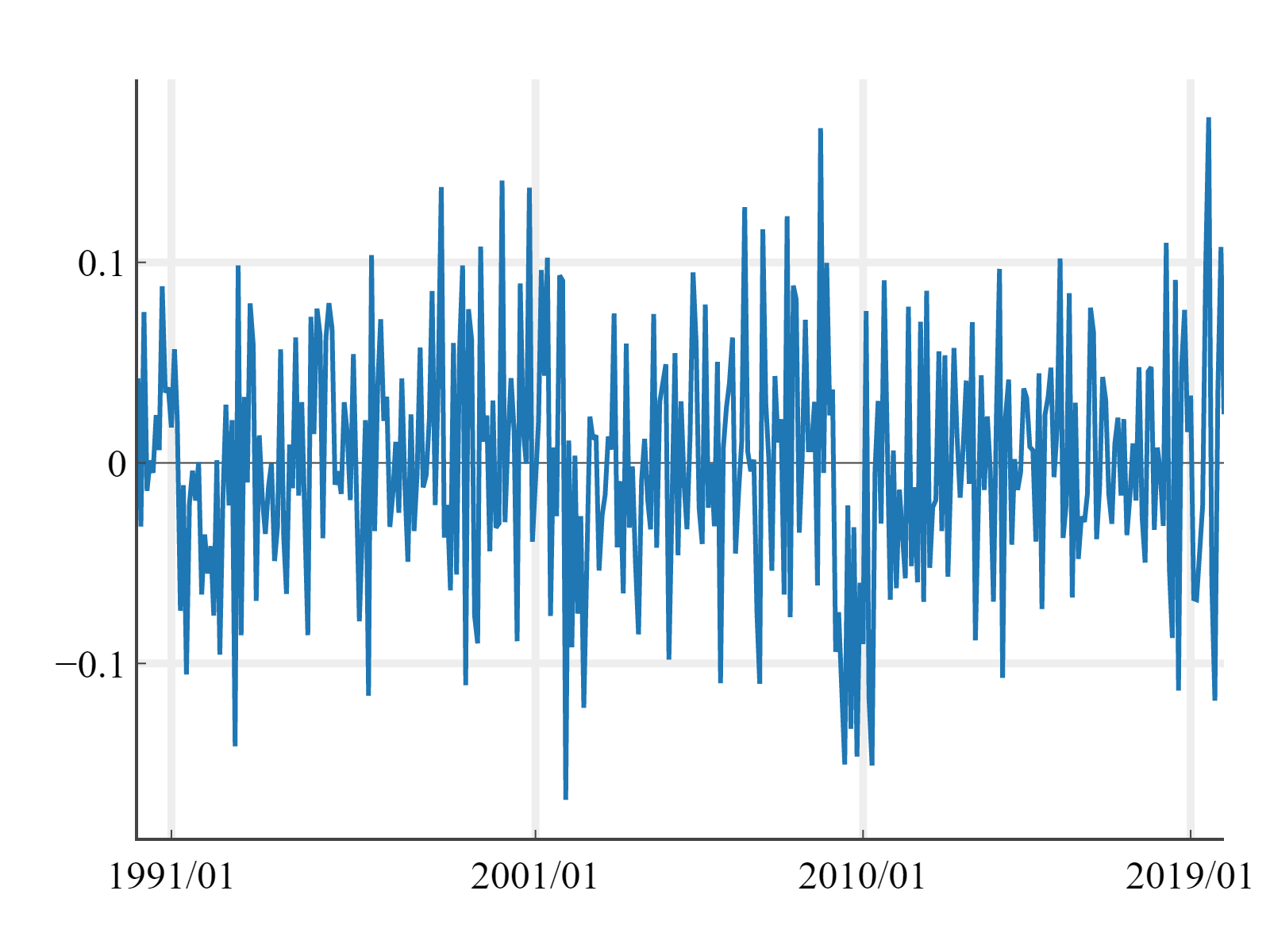}
\end{subfigure}
\begin{subfigure}{.32\linewidth}
	\subcaption{$\langle \Delta X_t,\widehat{\nu}_3 \rangle$}
	\includegraphics[width = \linewidth]{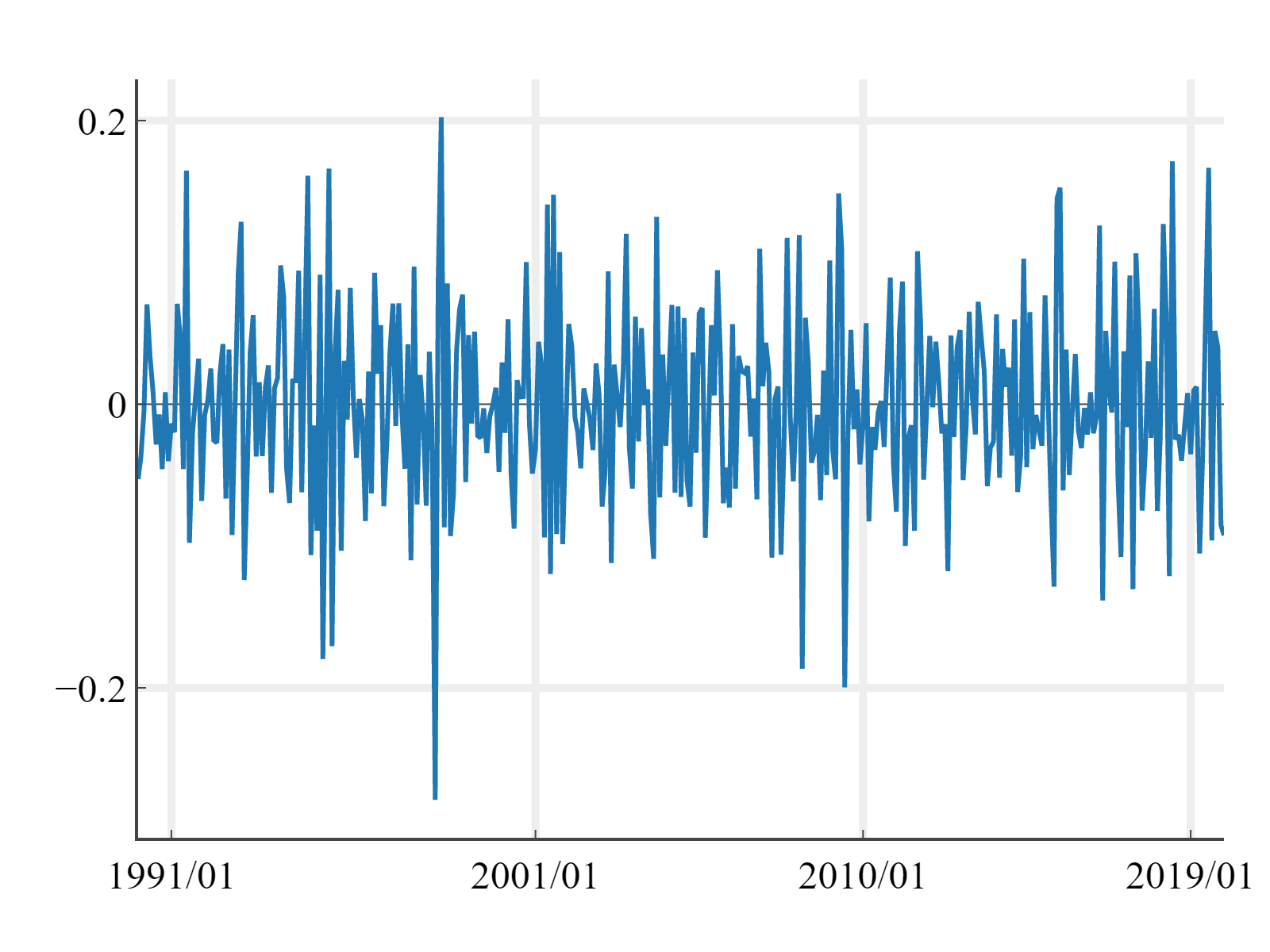}
\end{subfigure}\\
{\footnotesize Notes: Time series of the first three score functions of~$\Delta X_t$. The estimated eigenvectors $\widehat{\nu}_1,\widehat{\nu}_2$, and $\widehat{\nu}_3$ are computed from~$\widehat\Lambda_{1,R}$.}
\end{figure}

\begin{table}
\caption{Empirical results for labor indices in first-differences}
\label{tab:labor-diff} 
\vskip -8pt
\small
\begin{tabular*}{\linewidth}{@{\extracolsep{\fill}}lrrrrrr}
\toprule
\multicolumn{7}{c}{Eigenvalue ratio estimates}\\
Statistic & & $j=1$\phantom{$^{**}$} & $j=2$\phantom{$^{**}$} & $j=3$\phantom{$^{**}$} & $j=4$\phantom{$^{**}$} & $j=5$\phantom{$^{**}$} \\
\midrule
$\mu_{j+1}/\mu_j$ & & 2.06\phantom{$^{**}$} & 2.77\phantom{$^{**}$} & 2.67\phantom{$^{**}$} & 2.32\phantom{$^{**}$} & 1.88\phantom{$^{**}$} \\ 
$\hat{\kappa}_{j+1}/\hat{\kappa}_j$ &  & 0.57\phantom{$^{**}$} & 0.82\phantom{$^{**}$} & 0.96\phantom{$^{**}$} & 0.62\phantom{$^{**}$} & 0.70\phantom{$^{**}$} \\ 
\midrule
\multicolumn{7}{c}{Variance ratio test statistics}\\
Statistic & $\smalls_0=0$\phantom{$^{**}$} & $\smalls_0=1$\phantom{$^{**}$} & $\smalls_0=2$\phantom{$^{**}$} & $\smalls_0=3$\phantom{$^{**}$} & $\smalls_0=4$\phantom{$^{**}$} & $\smalls_0=5$\phantom{$^{**}$} \\  
\midrule  Inv.VR  & 0.72$^{\dagger}$\phantom{$^{*}$} & 0.07\phantom{$^{**}$} & 0.04\phantom{$^{**}$} & 0.01\phantom{$^{**}$} & 0.01\phantom{$^{**}$} & 0.01\phantom{$^{**}$} \\ 
VR(2,1) &  & 423.84$^{**}$ & 10347.35$^{**}$ & 39933.38$^{**}$ & 101141.36$^{**}$ & 139311.88$^{**}$ \\ 
VR(2,0) &  & 17922.13$^{**}$ & 481545.37$^{**}$ & 812958.94$^{**}$ & 2993476.52$^{**}$ & 5194233.83$^{**}$ \\ 
VR(1,0) &  & 117.55$^{**}$ & 558.39$^{**}$ & 1291.18$^{**}$ & 1695.46$^{**}$ & 3269.15$^{**}$ \\ 
\bottomrule 	
\end{tabular*} 
\vskip 4pt
{\footnotesize Notes: $^{\dagger}$, $^{*}$, and $^{**}$ indicate significance at 10\%, 5\%, and 1\% level, respectively.}
\end{table}

In Figure~\ref{fig:levels} we display time series of $\langle X_t,\widehat{\nu}_j \rangle$ for some eigenvectors $\widehat{\nu}_j$ that are computed from~$\widehat\Lambda_{0,R}$. These are the $j^{\rm th}$ score processes (see \citealp{Chang2016} and Section~5 of \citealp{NSS}). It is clear from Figure~\ref{fig:levels} that there is some nonstationary behavior (e.g.,~$j=1$), but this seems to eventually disappear leaving stationary processes (e.g.,~$j=10$).

\citet{McCracken2016} suggested that some of these labor market indices may be I(2) variables.\footnote{Specifically, FRED ID variables CES0600000008, CES2000000008, and CES3000000008.} To examine the existence of this type of higher-order stochastic trends, we first display in Figure~\ref{figscore} the first three score processes, $\{\langle \Delta X_t,\widehat{\nu}_j \rangle\}_{t=1}^T$. Some of these must behave as a unit root process if the original time series is~I(2), and that does not appear to be the case. Next, we apply our testing procedures to the first differenced time series, $\{\Delta X_t\}_{t=1}^T$. The bandwidth parameters are set to those applied in Section~\ref{sec_empirical_1}, and in this part of the analysis, we use mean-adjusted statistics (Section~\ref{sec_deterministic}). The results are reported in Table~\ref{tab:labor-diff}, where we see that $\ddot\smalls= 2$, $\widehat\smalls_{\text{\tiny{LRS}}} =1$ (but recall that they cannot be zero), whereas $\widehat\smalls_{\rm BU} = \widehat\smalls_{\rm TD} = \widehat\smalls_{\rm UD} = 0$ regardless of the choice of test statistic and level (except inv.VR at 10\% level). From Figure~\ref{figscore} and Table~\ref{tab:labor-diff}, it seems unlikely that there are any I(2) stochastic trends.

We therefore proceed to apply our methodology to the level time series, $\{X_t\}_{t=1}^T$, to determine the number of I(1) stochastic trends. We now correct for a mean and linear time trend. The results are presented in Table~\ref{tab:labor-levels}. For the eigenvalue ratio estimators we find $\ddot\smalls = \widehat\smalls_{\text{\tiny{LRS}}} = 1$. However, the sequential testing results are quite different. Although the inverse VR test results are reported only for $\smalls_0=6,\ldots, 11$, the null hypotheses $\smalls_0=0,1,\ldots , 5$ are all rejected at 1\% level. Thus, we find that $\widehat\smalls_{\rm BU} = 10$ at 5\% level and $\widehat\smalls_{\rm BU} = 9$ at 1\% level, and we use this to set $\smalls_{\max} = 15$. We report results for the VR tests for $\smalls_0=6,7,\ldots,11$. The larger values of $\smalls_0$ are rejected at 1\% level for VR(2,1) and VR(1,0), but not for VR(2,0). This is consistent with the simulation findings in Table~\ref{tab:mc1}. We then find $\widehat\smalls_{\rm UD} = 8$ based on VR(1,0) at 5\% level, $\widehat\smalls_{\rm UD}=9$ based on VR(2,1) at 5\% level or VR(1,0) at 1\% level, and $\widehat\smalls_{\rm UD}=10$ based on VR(2,1) at 1\% level. On balance, we conclude that there are likely 9 stochastic trends. This means that the number of cointegrating relationships is about 2/3 of the dimension of the original data. Given this result, one should be careful in the application of cointegration rank tests that require a small number of cointegrating relationships (Remark~\ref{rem3}).

\begin{table}[tbp] 
\caption{Empirical results for labor indices in levels}
\label{tab:labor-levels}
\vskip -8pt
\small
\begin{tabular*}{\linewidth}{@{\extracolsep{\fill}}lrrrrrr}
\toprule
\multicolumn{7}{c}{Eigenvalue ratio estimates}\\ 
Statistic & $j=1$\phantom{$^{**}$} & $j=2$\phantom{$^{**}$} & $j=3$\phantom{$^{**}$} & $j=4$\phantom{$^{**}$} & $j=5$\phantom{$^{**}$} & $j=6$\phantom{$^{**}$}  \\
\midrule
$\mu_{j+1}/\mu_j$ & 2.06\phantom{$^{**}$} & 2.02\phantom{$^{**}$} & 1.74\phantom{$^{**}$} & 1.37\phantom{$^{**}$} & 1.34\phantom{$^{**}$} & 1.42\phantom{$^{**}$}  \\ 
$\hat{\kappa}_{j+1}/\hat{\kappa}_j$   & 0.07\phantom{$^{**}$} & 0.27\phantom{$^{**}$} & 0.34\phantom{$^{**}$} & 0.82\phantom{$^{**}$} & 0.80\phantom{$^{**}$} & 0.50\phantom{$^{**}$}   \\ 
\midrule
\multicolumn{7}{c}{Variance ratio test statistics}\\ 
Statistic & $\smalls_0=6$\phantom{$^{**}$} & $\smalls_0=7$\phantom{$^{**}$} & $\smalls_0=8$\phantom{$^{**}$} & $\smalls_0=9$\phantom{$^{**}$} & $\smalls_0=10$\phantom{$^{**}$} & $\smalls_0=11$\phantom{$^{**}$}  \\
\midrule
Inv.VR  & 0.15$^{**}$ & 0.11$^{**}$ & 0.06$^{**}$ & 0.03$^{*}$\phantom{$^{*}$} & 0.03\phantom{$^{**}$} & 0.02\phantom{$^{**}$}  \\ 
VR(2,1) & 2557.11\phantom{$^{**}$} & 4531.11\phantom{$^{**}$} & 6119.56\phantom{$^{**}$} & 10730.08\phantom{$^{**}$} & 16691.44$^{*}$\phantom{$^{*}$} & 22714.94$^{**}$  \\ 
VR(2,0) & 74227.09\phantom{$^{**}$} & 157057.93\phantom{$^{**}$} & 235784.03\phantom{$^{**}$} & 738700.29\phantom{$^{**}$} & 1581152.06\phantom{$^{**}$} & 2481609.80$^{*}$\phantom{$^{*}$}  \\ 
VR(1,0) & 169.35\phantom{$^{**}$} & 229.45\phantom{$^{**}$} & 300.84\phantom{$^{**}$} & 564.11$^{*}$\phantom{$^{*}$} & 762.54$^{**}$ & 898.93$^{**}$   \\ 
\bottomrule 	
\end{tabular*}
\vskip 4pt
{\footnotesize Notes: $^{\dagger}$, $^{*}$, and $^{**}$ indicate significance at 10\%, 5\%, and 1\% level, respectively.}
\end{table}

Finally, for both empirical applications we investigate the robustness of $\widehat\smalls_{\rm TD}$ to the initial hypothesis,~$\smalls_{\max}$, by computing $\widehat\smalls_{\rm TD}$ using the VR(2,1), VR(2,0), and VR(1,0) tests at 5\% level for a range of~$\smalls_{\max}$. In the yield example, $\widehat\smalls_{\rm TD} = 2$ for $\smalls_{\max}=2,\ldots ,20$ using either of the tests. In the labor index example in levels, $\widehat\smalls_{\rm TD}$ is unchanged for $\smalls_{\max} = 9,\ldots ,20$ using VR(2,1) and VR(1,0). The VR(2,0) test is sensitive to $\smalls_{\max}$ as expected given the simulation results in Table~\ref{tab:mc1}. The former two tests are remarkably robust to the choice of~$\smalls_{\max}$.

\section{Conclusion}
\label{sec:conc}

We have considered statistical inference on unit roots and cointegration for time series taking values in a Hilbert space of an arbitrarily large dimension or a subspace of possibly unknown dimension. This has wide applicability in practice; for example, in cointegrated vector time series of finite dimension, in a high-dimensional factor model that includes a finite number of nonstationary factors, in cointegrated curve-valued (or function-valued) time series, and in nonstationary dynamic functional factor models. We considered mainly determination of the dimension of the nonstationary subspace (number of common stochastic trends), but we also considered hypothesis testing on the stationary and nonstationary subspaces themselves.

We provided limit theory for general variance ratio-type statistics based on partial summation and/or differencing, and we demonstrated how to apply these variance ratio statistics to sequentially test for the dimension of the nonstationary subspace or to test hypotheses on the subspaces. Our theory covers many possible tests, and choosing among them is not trivial. Our preferred test is the VR(2,1) test due to several theoretical and practical advantages. Most importantly, it does not require the choice of any bandwidth parameters as it can be implemented only with estimated sample variance operators and not long-run variance operators. This is not only a practical advantage, but also implies a consistency property not shared by tests that need to estimate a long-run variance \citep{muller2007,muller2008}. Also, critical values are readily available \citep{Breitung2002} by the isomorphism with~$\mathbb{R}^\KK$. To illustrate our methods, we included a small Monte Carlo simulation study, wherein the VR(2,1) test also showed superior performance, as well as two empirical illustrations to the term structure of interest rates and labor market indices, respectively.

\appendix 
\numberwithin{equation}{section} 
\numberwithin{figure}{section} 
\numberwithin{table}{section} 

\makeatletter 
\def\@seccntformat#1{\@ifundefined{#1@cntformat}
	{\csname the#1\endcsname\quad}
	{\csname #1@cntformat\endcsname}}
\newcommand\section@cntformat{}
\makeatother

\section{Appendix A: Notation, linear operators, and random elements in $\mathcal H$}
\label{sec:app-notation}

For a real, separable Hilbert space $\mathcal H$ with inner product $\langle\cdot,\cdot\rangle$ and norm $\Vert\cdot\Vert$, let $\mathcal L_{\mathcal H}$ denote the collection of bounded linear operators acting on~$\mathcal H$ equipped with the uniform operator norm, $\|A\|_{\op} = \sup_{\|v\|\leq1}\|Av\|$. For any $A \in \mathcal L_{\mathcal H}$, let $\ker A =\{v\in \mathcal H: Av=0\}$ denote the kernel of $A$ and let $\ran A=\{Av : v\in \mathcal H\}$ denote the range of~$A$. The dimensions of $\ker A$ and $\ran A$ are, respectively, called the nullity and rank of~$A$. We let $A^\ast$ denote the adjoint operator of~$A$, which is uniquely defined by the property $\langle A v_1 , v_2 \rangle = \langle v_1 , A^* v_2 \rangle$ for all $v_1, v_2 \in \mathcal H$. An operator $A \in \mathcal L_{\mathcal H}$ is compact if it is the limit of a sequence of finite-rank operators in~$\mathcal L_{\mathcal H}$. Moreover, whenever it is convenient, we will let $[A]$ denote the matrix representation for any operator $A$ with respect to some orthonormal set of vectors $\{ g_j\}_{j=1}^m$ for some $m>0$ that will be specified depending on the context; that is, $[A]_{ij} = \langle g_i, A g_j \rangle$ as discussed in Remark~\ref{remisomorphism}.

An $\mathcal H$-valued random variable $X$ is defined as a measurable map from the underlying probability space, say~$\mathcal S$, to~$\mathcal H$, where $\mathcal H$ is equipped with the usual Borel $\sigma$-field. The random element $X$ is square-integrable if $\E \|X\|^2 < \infty$. For any square-integrable $X$, its expectation $\E [X]$ is defined as a unique element in $\mathcal H$ satisfying $\E \langle X,v\rangle = \langle \E [X],v\rangle$ for any $v \in \mathcal H$. The variance operator of $X$ is defined as $C_X = \E [(X-\E [X])\otimes (X - \E [X])]$, where $\otimes$ denotes the tensor product on~$\mathcal H$. The operator $C_X$ is self-adjoint (i.e., $C_X = C_X^\ast$) and allows only nonnegative eigenvalues. We also note the well-known property $\| X_{t-s} \otimes X_t \|_{\op} \leq \| X_{t-s} \| \|X_t \|$.

We will sometimes consider convergence of a sequence of random bounded linear operators. For such a sequence $\{A_j\}_{j \geq1}$, we write $A_j \plowto A$ if $\|A_j-A\|_{\op} \plowto 0$. We also write $A_j = \bigOp (a_n)$ (resp.\ $A_j= \smallop (a_n)$) if $\|A_j\|_{\op} = \bigOp (a_n)$ (resp.\ $\|A_j\|_{\op} = \smallop (a_n)$). For example, under Assumption~\ref{assumvr1}, $\widehat{P}_{\KK}$ converges in probability and in operator norm to $P_{\NNN} + P_{\SSS}^{\KK} = P_{\KK}$.  

When Assumption~\ref{assum1} is satisfied, \citet[Prop.~3.1 and~3.2]{Beare2017} showed that the long-run variance operator $\Lambda_{\Delta X} (= \Phi(1) C_\epsilon \Phi^\ast (1))$ of $\Delta X_t$ is a finite-rank operator of rank~$\nd$. Because $\Lambda_{\Delta X}$ is also compact and self-adjoint, it allows the spectral decomposition $\Lambda_{\Delta X}=\sum_{j=1}^{\nd} r_j \varpi_j \otimes \varpi_j$, where $\{r_j, \varpi_j\}_{j = 1}^{\nd}$ denote the pairs of eigenvalues and eigenvectors of~$\Lambda_{\Delta X}$ (see \citealp{Bosq2000}, pp.\ 34--35). Obviously, $\{\varpi_j\}_{j = 1}^{\nd}$ is an orthonormal basis of~$\mathcal H_{\NNN}$.  
We then let $\{\mathcal W_{1,\nd}(r)\}_{r\in[0,1]} $ denote the $\mathcal H$-valued random element given by $\mathcal W_{1,\nd}(r) = \sum_{j=1}^\nd W_j(r) \varpi_j$ (e.g., \citealp{Berkes2013}, p.\ 386), where $\{W_j(r)\}_{r\in[0,1]}$ are independent standard Brownian motions. Noting that $\Lambda_{\Delta X}^{1/2}$ is well-defined in terms of its eigenvalues and eigenvectors, $\{r_j^{1/2}, \varpi_j\}_{j = 1}^{\nd}$, we find that $\Lambda_{\Delta X}^{1/2}\mathcal W_{1,\nd}(r)=\sum_{j=1}^\nd r_j^{1/2} W_j(r) \varpi_j$ is $\mathcal H$-valued Brownian motion with variance~$\Lambda_{\Delta X}$ (e.g., \citealp{Chen1998}, p.\ 268). 
We also recursively define the $(d-1)$-fold integrated Brownian motion $\mathcal W_{d,\nd}(r)  = \int_0^r \mathcal W_{d-1,\nd} (u)du$ for $d \geq 2$.

Under Assumption~\ref{assum1}, it is well known that
\begin{equation}
\label{eqa1}
\sup_{0\leq r\leq 1}\| T^{-1/2} \sum_{t=1}^{\lfloor Tr \rfloor} \Delta X_t - \Lambda_{\Delta X}^{1/2}\mathcal W_{1,\nd}(r) \| \pto 0;
\end{equation}
see \citet[Theorem~1]{Berkes2013}. We will repeatedly, and sometimes implicitly, use this result in the subsequent sections. Moreover, in the case where the unobserved components model \eqref{equnobs} is considered, and thus $U_t^{(1)}$ or $U_t^{(2)}$ is used in the analysis, we note that 
\begin{align}
&\sup_{0\leq r\leq 1}\| T^{-1/2} \sum_{t=1}^{\lfloor Tr \rfloor} \Delta U_{t}^{(1)} -  \Lambda_{\Delta X}^{1/2}\mathcal W_{1,\nd}^{(1)}(r) \| \pto 0, \label{eqaa1} \\
&\sup_{0\leq r\leq 1}\| T^{-1/2} \sum_{t=1}^{\lfloor Tr \rfloor} \Delta U_{t}^{(2)} -  \Lambda_{\Delta X}^{1/2}\mathcal W_{1,\nd}^{(2)}(r) \| \pto 0, \label{eqaa2} 
\end{align}
where $\mathcal W_{1,\nd}^{(1)}$ (resp.\ $\mathcal W_{1,\nd}^{(2)}$) is the demeaned (resp.\ detrended) standard Brownian motion taking values in~$\mathcal H_{\NNN}$; see \citet[Lemma~3]{NSS}. Results \eqref{eqaa1} and \eqref{eqaa2} are essential to extend our theoretical results to the unobserved components model~\eqref{equnobs}. 

In the sequel, we will often consider operators that can be written as $\widehat{P}_{\KK}A \widehat{P}_{\KK}$ for some $A \in \mathcal L_{\mathcal H}$. Let $\mathcal H_{\KK} = \ran \widehat{P}_{\KK}$. Given that $\widehat{P}_{\KK}v = 0$ and $\langle \widehat{P}_{\KK}A \widehat{P}_{\KK}v,w\rangle=0$ for any $v,w\in  \mathcal H_{\KK}^\perp$, $\widehat{P}_{\KK}A\widehat{P}_{\KK}$ can be understood as a map on~$\mathcal H_{\KK}$. Let $Q_1$ be any orthogonal projection on $\mathcal H$ and let $Q_2=I-Q_1$. Then it holds that $\mathcal H_{\KK} = \mathcal H_{\KK,1} \oplus \mathcal H_{\KK,2}$, where $\mathcal H_{\KK,1} = [\ker Q_1\widehat{P}_{\KK}]^\perp$ and $\mathcal H_{\KK,2}=\ker Q_1\widehat{P}_{\KK} \cap \mathcal H_{\KK}$.
Note that $\ker Q_1\widehat{P}_{\KK}$ includes $\mathcal H_{\KK}^\perp = [\ran \widehat{P}_{\KK}]^\perp$, and hence it needs to be restricted to be a subspace of~$\mathcal H_{\KK}$, whereas $[\ker Q_1\widehat{P}_{\KK}]^\perp = \ran \widehat{P}_{\KK} Q_1$ \citep[Theorem~2.19]{Conway1990} and thus is already a subspace of~$\mathcal H_{\KK}$.
It follows that any $v \in \mathcal H_{\KK}$ can be written as $v=v_1+v_2$, where $v_1  \in \mathcal H_{\KK,1}$ and  $v_2 \in \mathcal H_{\KK,2}$, and we may understand $v$ as the tuple~$(\begin{smallmatrix} v_1 \\ v_2 \end{smallmatrix})$. Let $\widehat{\Pi}$ be the orthogonal projection onto $\mathcal H_{\KK,2}$ (i.e., $[\ran \widehat{P}_{\KK}Q_1]^\perp =\ker Q_1 \widehat{P}_{\KK}$ in~$\mathcal H_{\KK}$) such that $(I-\widehat{\Pi}) \widehat{P}_{\KK}Q_1= \widehat{P}_{\KK}Q_1 = \mathcal H_{\KK,1}$. Then any $w \in \mathcal H_{\KK}$ can be written as $(I-\widehat{\Pi})w + \widehat{\Pi}w$, where $(I-\widehat{\Pi})w \in \mathcal H_{\KK,1}$ and $\widehat{\Pi}w \in \mathcal H_{\KK,2}$, and thus $w$ may be understood as the tuple~$\big(\begin{smallmatrix}(I-\widehat{\Pi})w\\ \widehat{\Pi}w \end{smallmatrix}\big)$. 
From this result and properties of~$\widehat{\Pi}$, we may understand $w=\widehat{P}_{\KK}A \widehat{P}_{\KK}v =  \widehat{P}_{\KK}Q_1A Q_1\widehat{P}_{\KK}v_{1}+\widehat{P}_{\KK}Q_1A Q_2\widehat{P}_{\KK}v_{2}  $+$\widehat{P}_{\KK}Q_2A Q_1\widehat{P}_{\KK}v_{1}+\widehat{P}_{\KK}Q_2A Q_2\widehat{P}_{\KK}v_{2}$ as the tuple~$(\begin{smallmatrix}w_1\\w_2 \end{smallmatrix})$, where 
$w_1=(I-\widehat{\Pi}) w = (\widehat{P}_{\KK}Q_1A Q_1\widehat{P}_{\KK} +  \widehat{\Upsilon}_{1})v_{1} +(\widehat{P}_{\KK}Q_1A Q_2\widehat{P}_{\KK} + \widehat{\Upsilon}_{2})v_{2}$ and $w_2=\widehat{\Pi}w=\widehat{\Pi} \widehat{P}_{\KK} Q_2AQ_1\widehat{P}_{\KK}v_1 +  \widehat{\Pi} \widehat{P}_{\KK}Q_2AQ_2\widehat{P}_{\KK}v_2$ with $\widehat{\Upsilon}_1 = (I-\widehat{\Pi})\widehat{P}_{\KK}Q_2A Q_1\widehat{P}_{\KK}$ and $\widehat{\Upsilon}_2 = (I-\widehat{\Pi})\widehat{P}_{\KK}Q_2A Q_2\widehat{P}_{\KK}$. 
To sum up, the operator $\widehat{P}_{\KK}A \widehat{P}_{\KK}$ on ${\mathcal H}_{\KK}$ may be understood as a transformation via the (operator) matrix $\mathcal A$, viewing the aforementioned tuples as vectors:
\begin{equation}
\label{eqopmatrix}
\left(\begin{matrix}w_1\\ w_2 \end{matrix}\right) = \mathcal A
\left(\begin{matrix}v_{1}\\ v_{2} \end{matrix}\right), \quad \text{where} \quad
\mathcal A= \left(\begin{matrix} 
\widehat{P}_{\KK}Q_1AQ_1 \widehat{P}_{\KK} + \widehat{\Upsilon}_1&\widehat{P}_{\KK} Q_1AQ_2\widehat{P}_{\KK}+ \widehat{\Upsilon}_2\\
\widehat{\Pi} \widehat{P}_{\KK}	Q_2AQ_1\widehat{P}_{\KK} & \widehat{\Pi} \widehat{P}_{\KK}Q_2AQ_2\widehat{P}_{\KK}
\end{matrix}\right).
\end{equation}
Note that $\mathcal A$ is a standard characterization of the map $\widehat{P}_{\KK}A\widehat{P}_{\KK}$, viewed as a map from ${\mathcal H}_{\KK}$ to itself when ${\mathcal H}_{\KK}$ allows the bipartite decomposition ${\mathcal H}_{\KK} = \mathcal H_{\KK,1} \oplus \mathcal H_{\KK,2}$. Obviously, the \mbox{$(i,j)$-th} entry of $\mathcal A$ may be understood as a map from ${\mathcal H}_{\KK,j}$ to ${\mathcal H}_{\KK,i}$ for $i,j=1,2$. 

For any operator $\widehat{P}_{\KK}A \widehat{P}_{\KK}$ whose operator matrix form is $\big(\begin{smallmatrix}A_{11}&A_{12}\\A_{21} &A_{22}
\end{smallmatrix}\big)$, such as $\mathcal A$ in~\eqref{eqopmatrix}, we let $D_m$ be the diagonal operator corresponding to the operator matrix $\big(\begin{smallmatrix}m^{-1/2}I_{\dim(\mathcal H_{\KK,1})}&0\\0& I_{\dim(\mathcal H_{\KK,2})}\end{smallmatrix}\big)$ for appropriately defined identity operators $I_{\dim(\mathcal H_{\KK,1})}$ on~$\mathcal H_{\KK,1}$ and $I_{\dim(\mathcal H_{\KK,2})}$ on~$\mathcal H_{\KK,2}$ so that $D_{m}\widehat{P}_{\KK}A \widehat{P}_{\KK}$ and  $ \widehat{P}_{\KK}A \widehat{P}_{\KK} D_{m}$, respectively, result in the operator matrices
\begin{equation}
\label{eqdm}
\left(\begin{matrix}
	m^{-1/2}A_{11} & m^{-1/2}A_{12}\\
	A_{21} & A_{22}
\end{matrix}\right) \quad \text{and} \quad 
\left(\begin{matrix}
	m^{-1/2}A_{11} & A_{12}\\
	m^{-1/2} A_{21} & A_{22}
\end{matrix}\right).
\end{equation}
Subsequent proofs with such a $D_m$ will often have $\dim(\mathcal{H}_{\KK,1}) = \nd$ and $\dim(\mathcal{H}_{\KK,2}) = \KK - \nd$.

\section{Appendix B: Proofs of theorems}
\label{sec:app-proofs}

In this appendix we give the proofs of all theorems. The corollaries follow by the continuous mapping theorem. We first state a lemma, which is proven in the supplementary appendix.

\begin{lemma}
	\label{lem1}
	Under Assumptions~\ref{assum1}--\ref{assumkernel}, the following hold:
	\begin{enumerate}[label=(\roman*)]
		\item \label{lem1a}
		$P_{\NNN} \widehat{\Lambda}_{d,m} P_{\NNN} = \bigOp (h_m T^{2d})$ for $d \geq 1$
		and $P_{\NNN} \widehat{\Lambda}_{0,m} P_{\NNN} = \bigOp (T)$.
		\item
		$P_{\NNN}\widehat{\Lambda}_{d,m} P_{\SSS}^{\KK} = \bigOp (h_m T^{2d-1})$ for $d \geq 1$
		and $P_{\NNN} \widehat{\Lambda}_{0,m} P_{\SSS}^{\KK} = \bigOp (T/h_m)$.
		\item
		\mbox{$P_{\SSS}^{\KK} \widehat{\Lambda}_{d,m} P_{\SSS}^{\KK} = \bigOp (h_m T^{2d-2})$ for $d \geq 2$,
		$P_{\SSS}^{\KK} \widehat{\Lambda}_{1,m} P_{\SSS}^{\KK} = \bigOp (T)$,
		and $P_{\SSS}^{\KK} \widehat{\Lambda}_{0,m} P_{\SSS}^{\KK} = \bigOp (T/h_m^2)$.}
	\end{enumerate} 
	Finally, of course, $P_{\SSS}^{\KK}\widehat{\Lambda}_{d,m} P_{\NNN}$ is the same order as $P_{\NNN}\widehat{\Lambda}_{d,m} P_{\SSS}^{\KK}$ for $d\geq 0$.
\end{lemma}

\subsection{Proof of Theorem~\ref{thmv2}}

We only give the proof for the case with $a_m > 0$ for $m \in \{ L,R \}$. If $a_m = 0$, such that $\widehat\Lambda_{d_m, m}$ is a sample variance operator, then we set $h_m =1$ and $\km(u)=1_{\{u=0\}}$ and use the same proof.

First note that the eigenvalues $\mu_j$ from the problem \eqref{eqgev} with $d_L \geq 2$ and $d_R = 1$ are identical to those in
\begin{equation}
\label{eqpf:gev1}
(T^{2d_L-2} h_Lh_R^{-1} \mu_j ) \left( \frac{h_R}{h_L T^{2d_L-1}} D_{h_RT} \widehat{P}_{\KK} \widehat{\Lambda}_{2,L}\widehat{P}_{\KK} D_{h_RT} \right) \nu_j = \left( \frac{1}{T} D_{h_RT}\widehat{P}_{\KK}  \widehat{\Lambda}_{1,R}\widehat{P}_{\KK}D_{h_RT} \right) \nu_j,
\end{equation}
where $D_{h_RT}$ is defined as in~\eqref{eqdm}. Next, consider the decomposition $\widehat{P}_{\KK} = \widehat{P}_{\KK}  P_{\NNN}+  \widehat{P}_{\KK} P_{\SSS} = \widehat{P}^{\KK} _{\NNN}  + \widehat{P}_{\SSS} ^{\KK}$. Then, for $d \geq 1$ and $m \in \{L,R\}$,
\begin{equation}
\widehat{P}_{\KK} \widehat{\Lambda}_{d,m}\widehat{P}_{\KK} = \widehat{P}_{\NNN} ^{\KK} \widehat{\Lambda}_{d,m} (\widehat{P}_{\NNN} ^{\KK})^\ast + \widehat{P}_{\SSS} ^{\KK} \widehat{\Lambda}_{d,m} (\widehat{P}_{\NNN} ^{\KK})^\ast +\widehat{P}_{\NNN} ^{\KK} \widehat{\Lambda}_{d,m} ( \widehat{P}_{\SSS} ^{\KK} )^\ast+\widehat{P}_{\SSS} ^{\KK} \widehat{\Lambda}_{d,m} (\widehat{P}_{\SSS} ^{\KK} )^\ast . 
\end{equation}
As in Appendix~\ref{sec:app-notation}, $\widehat{P}_{\KK} \widehat{\Lambda}_{d,m} \widehat{P}_{\KK}$ may be understood as the operator matrix given by
\begin{equation}
\label{eqmatrixrep} 
\left( \begin{matrix}
	\widehat{P}^{\KK} _{\NNN}\widehat{\Lambda}_{d,m}(\widehat{P}^{\KK} _{\NNN})^\ast + \widehat{\Upsilon}_{d,m}^{\NNN} & \widehat{P}^{\KK} _{\NNN}  \widehat{\Lambda}_{d,m}(\widehat{P}^{\KK} _{\SSS} )^\ast  +\widehat{\Upsilon}_{d,m}^{\SSS}\\ \widehat{\Pi}\widehat{P}^{\KK} _{\SSS} \widehat{\Lambda}_{d,m} (\widehat{P}^{\KK} _{\NNN})^\ast  &\widehat{\Pi} \widehat{P}^{\KK} _{\SSS} \widehat{\Lambda}_{d,m}(\widehat{P}^{\KK} _{\SSS} )^\ast 
\end{matrix} \right),
\end{equation} 
where $\widehat{\Pi}$ is defined so that $I-\widehat{\Pi}$ becomes the orthogonal projection onto~$\ran \widehat{P}^{\KK}_{\NNN}$. Hence,
\begin{equation}
\label{equpsilon}
\widehat{\Upsilon}_{d,m}^{\NNN} = (I-\widehat{\Pi})\widehat{P}_{\SSS}^{\KK}\widehat{\Lambda}_{d,m}(\widehat{P}^{\KK}_{\NNN})^\ast \quad\text{and}\quad
\widehat{\Upsilon}_{d,m}^{\SSS} = (I-\widehat{\Pi})\widehat{P}_{\SSS}^{\KK}\widehat{\Lambda}_{d,m} (\widehat{P}_{\SSS}^{\KK})^\ast ,
\end{equation} 
see \eqref{eqopmatrix}, where we replaced $A$, $Q_1$, and $Q_2$ with $\widehat{\Lambda}_{d,m}$, $P_{\NNN}$, and $P_{\SSS}$, respectively. Noting that $\|\widehat{P}^{\KK}_{\NNN}-P_{\NNN}\|_{\op} = \|\widehat{P}_{\KK}P_{\NNN}-P_{\NNN}\|_{\op} \plowto 0$ and $\|A\|_{\op} = \|A^\ast\|_{\op}$ for any $A \in \mathcal L_{\mathcal H}$, it follows that $\|(I-\widehat{\Pi}) P_{\SSS}\|_{\op}=\|P_{\SSS}(I-\widehat{\Pi})\|_{\op} = \smallop(1)$, and hence $\|\widehat{\Pi}P_{\SSS}-P_{\SSS}\|_{\op} = \smallop(1)$ and also 
\begin{equation}
\label{eqaddadd01}
\|\widehat{\Pi}P_{\SSS}^{\KK}-P_{\SSS}^{\KK}\|_{\op} = \smallop(1),
\end{equation}
where $P_{\SSS}^{\KK}$ is defined in Assumption~\ref{assumvr1}(iii). We next obtain the limits of the operators in~\eqref{eqpf:gev1}.

First, $h_L^{-1}T^{1-2d_L}h_R D_{h_RT} \widehat{P}_{\KK} \widehat{\Lambda}_{2,L} \widehat{P}_{\KK} D_{h_RT} $ is understood as the operator matrix (Appendix~\ref{sec:app-notation})
\begin{equation*}
\widehat{\mathfrak D}_L = \left(\begin{matrix}
h_L^{-1} T^{-2d_L}  (\widehat{P}_{\NNN} ^{\KK} \widehat{\Lambda}_{d_L,L} (\widehat{P}_{\NNN} ^{\KK}) ^\ast +   \widehat{\Upsilon}_{d_L,L}^{\NNN}) &h_L^{-1}h_R^{ 1/2} T^{1/2-2d_L}(\widehat{P}_{\NNN} ^{\KK}\widehat{\Lambda}_{d_L,L} (\widehat{P}_{\SSS} ^{\KK}) ^\ast + \widehat{\Upsilon}_{d_L,L}^{\SSS})\\
h_L ^{-1}h_R^{1/2}  T^{1/2-2d_L} \widehat{\Pi}\widehat{P}_{\SSS} ^{\KK} \widehat{\Lambda}_{d_L,L}(\widehat{P}_{\NNN} ^{\KK}) ^\ast&h_L ^{-1} h_R T^{1-2d_L}  \widehat{\Pi} \widehat{P}_{\SSS} ^{\KK}\widehat{\Lambda}_{d_L,L} (\widehat{P}_{\SSS} ^{\KK}) ^\ast
\end{matrix}\right).
\end{equation*}
Let $\widehat{\delta}_{\KK} = \widehat{P}_{\KK}- P_{\KK}$, where $P_{\KK}=P_{\NNN}+P_{\SSS}^{\KK}$. Using the fact that $P_{\KK}P_{\NNN}= P_{\NNN}$ and the definition $\widehat{P}_{\NNN}^{\KK}=\widehat{P}_{\KK} P_\NNN$, we find $\widehat{P}_{\NNN} ^{\KK} \widehat{\Lambda}_{d_L,L} (\widehat{P}_{\NNN} ^{\KK})^\ast = P_{\NNN} \widehat{\Lambda}_{d_L,L} P_{\NNN} + \widehat{\delta}_{\KK} P_{\NNN}\widehat{\Lambda}_{d_L,L} P_{\NNN}\widehat{P}_{\KK} + P_{\NNN}\widehat{\Lambda}_{d_L,L} P_{\NNN} \widehat{\delta}_{\KK}^\ast$, where Assumption~\ref{assumvr1} implies $\widehat{\delta}_{\KK}= \smallop (1)$ and $\widehat{P}_{\KK}=\bigOp (1)$. Combining these results with Lemma~\ref{lem1}, we find that 
\begin{align}
\nonumber
\frac{1}{h_LT^{2d_L}} \widehat{P}_{\NNN}^{\KK} \widehat{\Lambda}_{d_L,L} (\widehat{P}_{\NNN} ^{\KK})^\ast
&= \frac{1}{h_LT^{2d_L}} P_{\NNN} \widehat{\Lambda}_{d_L,L} P_{\NNN} + \smallop (1) 
= \frac{1}{h_LT^{2d_L}} \sum_{s=1-T} ^{T-1} \mathrm{k}_L \left (\frac{s}{h_L}\right) \widehat{\Gamma}_{d_L,s}^{\NNN} + \smallop (1) \\
&\dto c_L \Lambda_{\Delta X}^{1/2} \left( \int \mathcal{W}_{d_L,\nd} \otimes \mathcal{W}_{d_L,\nd} \right)\Lambda_{\Delta X}^{1/2},
\label{eqpf01a}
\end{align}
where $\widehat{\Gamma}_{d,s}^{\NNN} = \sum_{t=s+1} ^T P_{\NNN}	X_{d,t-s} \otimes P_{\NNN}X_{d,t}$ for $s \geq 0$ and $\widehat{\Gamma}_{d,s}^{\NNN}=\sum_{t=s+1} ^TP_{\NNN} X_{d,t} \otimes P_{\NNN} X_{d,t-s}$ for $s < 0$. The first equality in \eqref{eqpf01a} tells us that the limiting behavior of $h_L^{-1}T^{-2d_L} \widehat{P}_{\NNN}^{\KK} \widehat{\Lambda}_{d_L,L} (\widehat{P}_{\NNN} ^{\KK})^\ast$ is the same as that of the finite-rank operator $h_L^{-1}T^{-2d_L} P_{\NNN} \widehat{\Lambda}_{d_L,L} P_{\NNN}$. The second equality follows by definition of~$P_{\NNN}\widehat{\Lambda}_{d_L,L} P_{\NNN}$. Due to the isomorphism between $\mathbb{R}^{\nd}$ and the $\nd$-dimensional subspace $\mathcal H_{\NNN}=\ran P_{\NNN}$, $P_{\NNN}X_t$ can be understood as the $\nd$-dimensional vector $x_{\nd,t} = (\langle X_t, f_1 \rangle,\ldots \langle X_t, f_{\nd} \rangle)'$, where $\{ f_j\}_{j=1}^{\nd}$ is any orthonormal basis of~$\mathcal H_{\NNN}$; see Remark~\ref{remisomorphism}. Because $\nd$ is a fixed constant, we then deduce the convergence in distribution given in \eqref{eqpf01a} from \eqref{eqa1}, \citet[unnumbered equation between (A.10) and~(A.11)]{phillips1988spectral} and the continuous mapping theorem (applied to the integral functional on~$[0,1]$).

By similar arguments, using Assumption~\ref{assumvr1}, $\widehat{\delta}_{\KK}= \smallop (1)$, \eqref{eqaddadd01}, Lemma~\ref{lem1}, and the fact that $h_R/T \to 0$ (Assumption~\ref{assumkernel}), we can further show that
\begin{align}
\label{eqpf05}
h_L^{-1}  h_R^{1/2}T^{1/2-2d_L}\widehat{P}_{\NNN} ^{\KK} \widehat{\Lambda}_{d_L,L} (\widehat{P}_{\SSS} ^{\KK})^\ast &=  h_L^{-1}  h_R^{1/2}T^{1/2-2d_L}{P}_{\NNN}  \widehat{\Lambda}_{d_L,L} {P}_{\SSS} ^{\KK} + \smallop (1) = \smallop (1),\\ 
h_L^{-1}  h_R^{1/2}T^{1/2-2d_L} \widehat{\Pi} \widehat{P}_{\SSS}^{\KK} \widehat{\Lambda}_{d_L,L} (\widehat{P}_{\NNN}^{\KK})^\ast &= h_L^{-1}  h_R^{1/2}T^{1/2-2d_L} P_{\SSS}^{\KK} \widehat{\Lambda}_{d_L,L} P_{\NNN} + \smallop (1) = \smallop(1),\\
h_L ^{-1} h_R T^{1-2d_L} \widehat{\Pi}\widehat{P}_{\SSS} ^{\KK} \widehat{\Lambda}_{d_L,L} (\widehat{P}_{\SSS} ^{\KK}) ^\ast  &= h_L ^{-1} h_R T^{1-2d_L} {P}_{\SSS} ^{\KK} \widehat{\Lambda}_{d_L,L} {P}_{\SSS} ^{\KK} 
+ \smallop (1) = \smallop (1).
\label{eqpf06}
\end{align}
From the latter two equations together with $\widehat\delta_{\KK} = \smallop (1)$, we also find that
\begin{align}
h_L^{-1} T^{-2d_L} \widehat{\Upsilon}_{d_L,L}^{\NNN} &= h_L^{-1} T^{-2d_L}  (I-\widehat{\Pi})\widehat{P}_{\SSS}^{\KK}\widehat{\Lambda}_{d_L,L}(\widehat{P}^{\KK}_{\NNN})^\ast = \smallop (h_R^{-1/2}T^{-1/2}) , \label{eqpf06a}\\
h_L^{-1}h_R^{ 1/2} T^{1/2-2d_L}\widehat{\Upsilon}_{d_L,L}^{\SSS} &=h_L^{-1}h_R^{ 1/2} T^{1/2-2d_L} (I-\widehat{\Pi})\widehat{P}_{\SSS}^{\KK}\widehat{\Lambda}_{d_L,L} (\widehat{P}_{\SSS}^{\KK})^\ast = \smallop (h_R^{-1/2}T^{-1/2}) .\label{eqpf06b}
\end{align}

Second, $T^{-1} D_{h_RT} \widehat{P}_{\KK}\widehat{\Lambda}_{1,R}\widehat{P}_{\KK} D_{h_RT}$ can be understood as (see Appendix~\ref{sec:app-notation})
\begin{equation*}
\widehat{\mathfrak D}_R = \left( \begin{matrix}
	h_R^{-1} T^{-2}  ( \widehat{P}_{\NNN} ^{\KK} \widehat{\Lambda}_{1,R} (\widehat{P}_{\NNN} ^{\KK}) ^\ast +  \widehat\Upsilon_{1,R}^{\NNN}) &h_R^{- 1/2} T^{-3/2}(\widehat{P}_{\NNN} ^{\KK} \widehat{\Lambda}_{1,R} (\widehat{P}_{\SSS} ^{\KK}) ^\ast + \widehat\Upsilon_{1,R}^{\SSS})\\
	h_R^{-1/2}  T^{-3/2} \widehat{\Pi}\widehat{P}_{\SSS} ^{\KK} \widehat{\Lambda}_{1,R} (\widehat{P}_{\NNN} ^{\KK}) ^\ast& T^{-1} \widehat{\Pi} \widehat{P}_{\SSS} ^{\KK} \widehat{\Lambda}_{1,R} (\widehat{P}_{\SSS} ^{\KK}) ^\ast
\end{matrix} \right) .
\end{equation*}
As in \eqref{eqpf01a}--\eqref{eqpf06b}, it holds that
\begin{align}
\label{eqpf07}
h_R^{-1} T^{-2} (\widehat{P}_{\NNN}^{\KK} \widehat{\Lambda}_{1,R} (\widehat{P}_{\NNN} ^{\KK}) ^\ast + \widehat\Upsilon_{1,R}^{\NNN}) &\dto c_R \Lambda_{\Delta X}^{1/2} \left( \int \mathcal{W}_{1,\nd} \otimes \mathcal{W}_{1,\nd} \right) \Lambda_{\Delta X}^{1/2}, \\
\label{eqpf07b}
h_R^{-1/2} T^{-3/2}(\widehat{P}_{\NNN} ^{\KK} \widehat{\Lambda}_{1,R} (\widehat{P}_{\SSS} ^{\KK}) ^\ast  + \widehat\Upsilon_{1,R}^{\SSS}) &= \smallop (1), \\
h_R^{-1/2} T^{-3/2} \widehat{\Pi}\widehat{P}_{\SSS} ^{\KK} \widehat{\Lambda}_{1,R} (\widehat{P}_{\NNN} ^{\KK}) ^\ast &= \smallop (1).
\label{eqpf08}
\end{align}
We also find that
\begin{equation}
\label{eqpf08a}
T^{-1} \widehat{\Pi} \widehat{P}_{\SSS}^{\KK} \widehat{\Lambda}_{1,R} (\widehat{P}_{\SSS} ^{\KK}) ^\ast = T^{-1} P_{\SSS} ^{\KK} \widehat{\Lambda}_{1,R} P_{\SSS} ^{\KK} + \smallop (1) \pto \ROK ,  
\end{equation}
where $\ROK=P_{\KK}\RO P_{\KK}$ and the convergence follows from Theorem~2 of \citet{horvath2013estimation} (also see Theorem~5.3 of \citealp{kokoszka2016kpss}) since $\{P_{\SSS}X_{1,t}\}_{t\geq 1}$ is an $L^4$-$q$-approximable sequence \citep[Proposition~2.1]{hormann2010} with $q\sum_{j=q+1}^{\infty} \|\tilde{\Phi}_j\|_{\op} =\smallo (1)$ under the summability condition $\sum_{j=0}^\infty j\|\tilde \Phi_j\|_{\op}< \infty$ implied by~\eqref{eqlinear2}.

From \eqref{eqpf01a}--\eqref{eqpf08a}, $\widehat{\mathfrak{D}}_L$ and $\widehat{\mathfrak{D}}_R$ 
converge in distribution to random bounded linear operators acting on~$\ran P_{\KK}$. Specifically,
\begin{align}
\label{DLlimit}
\widehat{\mathfrak{D}}_L &\dto \mathfrak D_L = \left( \begin{matrix}
	c_L \Lambda_{\Delta X}^{1/2} ( \int \mathcal{W}_{d_L,\nd} \otimes \mathcal{W}_{d_L,\nd} ) \Lambda_{\Delta X}^{1/2} & 0 \\ 0 & 0
\end{matrix} \right) \text{ and} \\
\label{DRlimit}
\widehat{\mathfrak{D}}_R &\dto \mathfrak D_R = \left( \begin{matrix}
	c_R \Lambda_{\Delta X}^{1/2} ( \int \mathcal{W}_{1,\nd} \otimes \mathcal{W}_{1,\nd} ) \Lambda_{\Delta X}^{1/2} & 0 \\ 0 & \ROK 
\end{matrix} \right) .
\end{align}

Using the isomorphism between $\mathbb{R}^{\KK}$ and any $\KK$-dimensional Hilbert space, $\mathfrak D_L$ and $\mathfrak D_R$ may be understood as $\KK \times \KK$ matrices, say $[\mathfrak D_L]$ and $[\mathfrak D_R]$; see Remark~\ref{remisomorphism}. Moreover, any orthonormal basis of $\ran P_{\KK}$ can be used for these matrix representations since the eigenvalues that we are interested in are not affected by a change of basis. Hence, we may assume that $\{T^{2d_L-2} h_Lh_R^{-1} \mu_j \}_{j=1}^{\KK}$ in \eqref{eqpf:gev1} converge to the eigenvalues associated with $[\mathfrak D_L]$ and $[\mathfrak D_R]$, and these are represented with respect to some orthonormal basis $\{ f_j\}_{j=1}^{\KK}$ of~$\ran P_{\KK}$. Under Assumption~\ref{assumvr1}, $[\ROK]$ has rank $\KK-\nd$ and hence $[\mathfrak D_{R}]$ is invertible almost surely. We thus find that $\{(T^{2d_L-2} h_Lh_R^{-1} \mu_j )^{-1}\}_{j=1}^{\nd}$ converge to the  $\nd$ largest eigenvalues of  $[\mathfrak D_{R}]^{-1}[\mathfrak D_L]$ while $(T^{2d_L-2} h_Lh_R^{-1} \mu_j )^{-1} \to 0$ for $j \geq \nd+1$. This proves~\eqref{eqthm2}. From the expressions for $[\mathfrak D_L]$ and $[\mathfrak D_R]$, we further find that the $\nd$ largest eigenvalues of $[\mathfrak D_R]^{-1}[\mathfrak D_L]$ are distributionally identical to those of $(c_L/c_R) [\Lambda_{\Delta X}^{1/2} ( \int \mathcal{W}_{1,\nd} \otimes \mathcal{W}_{1,\nd} ) \Lambda_{\Delta X}^{1/2}]^{-1} [\Lambda_{\Delta X}^{1/2} ( \int \mathcal{W}_{d_L,\nd} \otimes \mathcal{W}_{d_L,\nd} ) \Lambda_{\Delta X}^{1/2}]$. Then \eqref{eqthm1} follows from the continuous mapping theorem.

\subsection{Proof of Theorem~\ref{thmvd0}}

We consider only the case with $a_L>0$. The proof can easily be modified to deal with $a_L =0$.

As in the proof of Theorem~\ref{thmv2}, the eigenvalues $\mu_j$ from \eqref{eqgev} with $d_L \geq 1$ and $d_R=0$ are identical to those in
\begin{equation}
\label{eqppgeq1}
(h_L T^{2d_L-1} \mu_j ) ( T^{-2d_L} h_L^{-1} h_R^2 D_{h_R }^2 \widehat{P}_{\KK} \widehat{\Lambda}_{d_L,L} \widehat{P}_{\KK} D_{h_R}^2 ) \nu_j = ( T^{-1} h_R^2 D_{h_R }^2 \widehat{P}_{\KK} \widehat{\Lambda}_{0,R} \widehat{P}_{\KK} D_{h_R}^2 ) \nu_j.
\end{equation}
Note that $h_R^2 h_L^{-1} T^{-2d_L} D_{h_R} ^2\widehat{P}_{\KK} \widehat{\Lambda}_{d_L,L} \widehat{P}_{\KK}D_{h_R} ^2$ and $h_R^2 T^{-1} D_{h_R} ^2 \widehat{P}_{\KK}\widehat{\Lambda}_{0,R}\widehat{P}_{\KK} D_{h_R} ^2$ may, respectively, be understood as (see Appendix~\ref{sec:app-notation}) 
\begin{align}
\label{eqppgeq2}
\widehat{\mathfrak D}_L^{(1)} &= \left( \begin{matrix}
h_L^{-1} T^{-2d_L}(\widehat{P}_{\NNN} ^{\KK} \widehat{\Lambda}_{d_L,L} (\widehat{P}_{\NNN} ^{\KK})^\ast + \widehat{\Upsilon}_{d_L,L}^{\NNN}) & h_L^{-1} h_R T^{-2d_L}(\widehat{P}_{\NNN} ^{\KK} \widehat{\Lambda}_{d_L,L} (\widehat{P}_{\SSS} ^{\KK})^{\ast}+ \widehat{\Upsilon}_{d_L,L}^{\SSS})\\
h_L^{-1} h_R T^{-2d_L}\widehat{\Pi}\widehat{P}_{\SSS} ^{\KK} \widehat{\Lambda}_{d_L,L} (\widehat{P}_{\NNN} ^{\KK})^{\ast}& h_L^{-1}h_R^2 T^{-2d_L}\widehat{\Pi}\widehat{P}_{\SSS} ^{\KK} \widehat{\Lambda}_{d_L,L} (\widehat{P}_{\SSS} ^{\KK})^{\ast}
\end{matrix} \right) , \\ 
\label{eqppgeq2a}
\widehat{\mathfrak D}_R^{(1)} &= \left( \begin{matrix}
T^{-1}(\widehat{P}_{\NNN} ^{\KK} \widehat{\Lambda}_{0,R} (\widehat{P}_{\NNN} ^{\KK})^\ast+ \widehat{\Upsilon}_{0,R}^{\NNN})&   h_RT^{-1}(\widehat{P}_{\NNN} ^{\KK} \widehat{\Lambda}_{0,R} (\widehat{P}_{\SSS} ^{\KK})^{\ast}+ \widehat{\Upsilon}_{0,R}^{\SSS})\\
h_R T^{-1}\widehat{\Pi}\widehat{P}_{\SSS} ^{\KK} \widehat{\Lambda}_{0,R} (\widehat{P}_{\NNN} ^{\KK})^{\ast}&  h_R^2T^{-1}\widehat{\Pi}\widehat{P}_{\SSS} ^{\KK} \widehat{\Lambda}_{0,R} (\widehat{P}_{\SSS} ^{\KK})^{\ast}
\end{matrix} \right),
\end{align}
where $\widehat{\Upsilon}_{d,m}^{\NNN}$ and $\widehat{\Upsilon}_{d,m}^{\SSS}$ are defined as in~\eqref{equpsilon}.

We first obtain the limiting behavior of the operator in~\eqref{eqppgeq2}. When $d_L \geq 2$ this is given in~\eqref{DLlimit}. When $d_L = 1$, we reverse the roles of $L$ and~$R$, use \eqref{eqpf07}--\eqref{eqpf08a} and the facts that $h_R T^{-1/2} \to 0$ and $h_L^{-1}h_R^2T^{-1} \to 0$ (Assumption~\ref{assumkernel}), and \eqref{DLlimit} still applies. Combining these results, $\widehat{\mathfrak D}_L^{(1)} \dlowto \mathfrak D_L$ for $d_L \geq 1$, where $\mathfrak D_L$ is given in~\eqref{DLlimit}.

We next establish the limiting behavior of the operator given in~\eqref{eqppgeq2a}. Assumption~\ref{assumvr1}, Lemma~\ref{lem1}, and the fact that $\{P_{\NNN}X_{0,t}\}_{t\geq 1}$ is $L^4$-$q$-approximable with $q\sum_{j=q+1}^{\infty} \|{\Phi}_j\|_{\op}=\smallo (1)$ imply that $T^{-1} \widehat{P}_{\NNN}^{\KK} \widehat{\Lambda}_{0,R} (\widehat{P}_{\NNN} ^{\KK}) ^\ast = T^{-1} P_{\NNN}  \widehat{\Lambda}_{0,R} P_{\NNN} + \smallop (1)$ and that $T^{-1} P_{\NNN} \widehat{\Lambda}_{0,R} P_{\NNN}$ converges to the long-run variance of $\{P_{\NNN}X_{0,t}\}_{t\geq1}$, which is $\Lambda_{\Delta X}$ \citep[Section~3.1]{Beare2017}. Thus,
\begin{equation}
\label{eqpf02a}
T^{-1} \widehat{P}_{\NNN} ^{\KK} \widehat{\Lambda}_{0,R} (\widehat{P}_{\NNN} ^{\KK}) ^\ast  \pto \Lambda_{\Delta X}.
\end{equation}
Similarly, from Assumption~\ref{assumvr1}, Lemma~\ref{lem1}, and~\eqref{eqaddadd01}, we know that $h_R^2T^{-1}\widehat{\Pi}\widehat{P}_{\SSS} ^{\KK} \widehat{\Lambda}_{0,R} (\widehat{P}_{\SSS} ^{\KK})^{\ast} = h_R^2T^{-1} P_{\SSS} ^{\KK} \widehat{\Lambda}_{0,R} P_{\SSS} ^{\KK} + \smallop (1)$, where $P_{\SSS} ^{\KK} \widehat{\Lambda}_{0,R} P_{\SSS}^{\KK}$ is the sample long-run variance computed from $\{{P}_{\SSS} ^{\KK} \Delta X_t\}_{t=1}^T$, which may be understood as a finite-dimensional vector-valued process (Remark~\ref{remisomorphism}). Then, from similar arguments used in the proof of Lemma~8.1(a) of \citet{phillips1995fully}, we may deduce that $h_R^2 T^{-1} P_{\SSS} ^{\KK} \widehat{\Lambda}_{0,R} P_{\SSS} ^{\KK} \plowto -\mathrm{k}_R^{\prime\prime} (0) P_{\SSS}^{\KK}\RO P_{\SSS}^{\KK}$. Since $P_{\SSS}^{\KK}\RO P_{\SSS}^{\KK}=\ROK$, which is defined in our proof of Theorem~\ref{thmv2},
\begin{equation}
\label{eqpf04}
h_R^2T^{-1}\widehat{\Pi}\widehat{P}_{\SSS} ^{\KK} \widehat{\Lambda}_{0,R} (\widehat{P}_{\SSS} ^{\KK})^{\ast}  \pto -\mathrm{k}_R^{\prime\prime} (0) \ROK.
\end{equation}
In the same way, from Assumption~\ref{assumvr1}, Lemma~\ref{lem1}, and \eqref{eqaddadd01}, we find that $h_R T^{-1}\widehat{P}_{\NNN} ^{\KK} \widehat{\Lambda}_{0,R} (\widehat{P}_{\SSS} ^{\KK})^{\ast} =  h_R T^{-1} P_{\NNN} \widehat{\Lambda}_{0,R} P_{\SSS} ^{\KK}+ \smallop (1)$ and $h_R T^{-1}\widehat{\Pi}\widehat{P}_{\SSS} ^{\KK} \widehat{\Lambda}_{0,R} (\widehat{P}_{\NNN} ^{\KK})^{\ast} =  h_R T^{-1}P_{\SSS} ^{\KK} \widehat{\Lambda}_{0,R}  P_{\NNN}+ \smallop (1)$, and by the same arguments used in the proof of Lemma~8.1(b) of \citet{phillips1995fully}, both $P_{\NNN} \widehat{\Lambda}_{0,R} P_{\SSS} ^{\KK}$ and $P_{\SSS} ^{\KK} \widehat{\Lambda}_{0,R} P_{\NNN}$ are $\bigOp (h_R^{-1}) + \bigOp ( (h_R/T)^{1/2})$, from which we find that
\begin{align}
\label{eqpf04a}
h_R T^{-1}\widehat{P}_{\NNN} ^{\KK} \widehat{\Lambda}_{0,R} (\widehat{P}_{\SSS} ^{\KK})^{\ast} &=  \bigOp (h_R^{-1}) + \bigOp ( (h_R/T)^{1/2}) + \smallop (1) =  \smallop (1),  \\
\label{eqpf04b}
h_R T^{-1}\widehat{\Pi}\widehat{P}_{\SSS} ^{\KK} \widehat{\Lambda}_{0,R} (\widehat{P}_{\NNN} ^{\KK})^{\ast} &= \bigOp (h_R^{-1}) + \bigOp ( (h_R/T)^{1/2}) + \smallop (1)=  \smallop (1).
\end{align}
From \eqref{eqpf04}, \eqref{eqpf04b}, and the fact that $(I-\widehat{\Pi})\widehat{P}_{\SSS}^{\KK} = \smallop(1)$ (by \eqref{eqaddadd01} and $\widehat\delta_{\KK} \plowto 0$), we have 
\begin{align}
T^{-1} \widehat{\Upsilon}_{0,R}^{\NNN} &= T^{-1} (I-\widehat{\Pi})\widehat{P}_{\SSS} ^{\KK} \widehat{\Lambda}_{0,R} (\widehat{P}_{\NNN} ^{\KK})^{\ast} = \smallop (h_R^{-1}), \label{eqpf04c}\\
h_RT^{-1}\widehat{\Upsilon}_{0,R}^{\SSS}  &=h_RT^{-1} (I-\widehat{\Pi})\widehat{P}_{\SSS} ^{\KK} \widehat{\Lambda}_{0,R} (\widehat{P}_{\SSS} ^{\KK})^{\ast}= \smallop (h_R^{-1}).\label{eqpf04d}
\end{align}
Combining \eqref{eqpf02a}--\eqref{eqpf04d}, we find that 
\begin{equation}
\label{eqpf002}
\widehat{\mathfrak D}_R^{(1)} \pto \mathfrak D_R^{(1)} = \left(\begin{matrix}
	\Lambda_{\Delta X} & 0 \\ 0 & -\mathrm{k}_R^{\prime\prime} (0) \ROK  
\end{matrix} \right).
\end{equation} 

As in our proof of Theorem~\ref{thmv2}, we may assume that $\{h_L T^{2d_L-1} {\mu_j}\}_{j=1}^{\KK}$ converge in distribution to the eigenvalues associated with $[\mathfrak D_L^{(1)}]$ and $[\mathfrak D_R^{(1)}]$, which are matrix representations of $\mathfrak D_L^{(1)}$ and $\mathfrak D_R^{(1)}$ with respect to some orthonormal basis of~$\ran P_{\KK}$. Because $[\mathfrak D_R^{(1)}]$ is invertible under Assumption~\ref{assumvr1}, the desired results follow from nearly identical arguments used to conclude our proof of Theorem~\ref{thmv2}.

\subsection{Proof of Theorem~\ref{thmvr2ab}}

As in our proof of Theorem~\ref{thmv2}, we only consider the case with $a_L >0$. We first consider the case with $\KK > \nd$ and rewrite the VR($d_L$,1) eigenvalue problem with $d_L \geq 2$ as
\begin{equation}
	\label{eq001aa}
	h_L^{-1}T^{2-2d_L}D_T^2 \widehat P_{\KK} \widehat\Lambda_{d_L,L} \widehat P_{\KK} D_T^2 \nu_j = (T^{2d_L-3} h_L{\mu_j})^{-1} T^{-1}D_T^2 \widehat P_{\KK} \widehat\Lambda_{1,R}\widehat P_{\KK} D_T^2  \nu_j ,
\end{equation}
where ${h_L^{-1}T^{2-2d_L}} D_T^2\widehat{P}_{\KK} \widehat\Lambda_{d_L,L} \widehat{P}_{\KK} D_T^2$ and $T^{-1}D_T^2 \widehat{P}_{\KK} \widehat \Lambda_{1,R} \widehat{P}_{\KK} D_T^2$ are, respectively, understood as
\begin{align}
	\label{eq001aa1}
	\widehat{\mathfrak D}_L^{(2)} &= \left( \begin{matrix}
		T^{-2d_L}h_L^{-1} (\widehat{P}^{\KK}_{\NNN} \widehat\Lambda_{d_L,L} (\widehat{P}^{\KK}_{\NNN})^\ast + \widehat\Upsilon_{d_L,L}^{\NNN})  & T^{1-2d_L}h_L^{-1} (\widehat{P}^{\KK}_{\NNN}  \widehat\Lambda_{d_L,L} (\widehat{P}^{\KK}_{\SSS})^\ast + \widehat\Upsilon_{d_L,L}^{\SSS})  \\
		T^{1-2d_L}h_L^{-1}    \widehat{\Pi}\widehat{P}^{\KK}_{\SSS} \widehat\Lambda_{d_L,L} (\widehat{P}^{\KK}_{\NNN})^\ast &  T^{2-2d_L}h_L^{-1}  \widehat{\Pi} \widehat{P}^{\KK}_{\SSS} \widehat\Lambda_{d_L,L} (\widehat{P}^{\KK}_{\SSS})^\ast 
	\end{matrix} \right) , \\
	\widehat{\mathfrak D}_R^{(2)} &= \left( \begin{matrix}
		T^{-3} (\widehat{P}^{\KK}_{\NNN} \widehat\Lambda_{1,R} (\widehat{P}^{\KK}_{\NNN})^\ast + \widehat\Upsilon_{1,R}^{\NNN}) & T^{-2} (\widehat{P}^{\KK}_{\NNN} \widehat\Lambda_{1,R} (\widehat{P}^{\KK}_{\SSS})^\ast+ \widehat\Upsilon_{1,R}^{\SSS}) \\
		T^{-2} \widehat{\Pi}\widehat{P}^{\KK}_{\SSS} \widehat\Lambda_{1,R} (\widehat{P}^{\KK}_{\NNN})^\ast & T^{-1}\widehat{\Pi} \widehat{P}^{\KK}_{\SSS} \widehat\Lambda_{1,R} (\widehat{P}^{\KK}_{\SSS})^\ast
	\end{matrix} \right) ,
\label{eq001aa2} 
\end{align}
see Appendix~\ref{sec:app-notation}. By Assumption~\ref{assumvr1}, Lemma~\ref{lem1}, \eqref{eqaddadd01}, and the conditions on $h_L,h_R$ in Assumption~\ref{assumkernel}, we may replace $\widehat{P}_{\NNN}^{\KK}$ (resp.~$\widehat{P}_{\SSS}^{\KK}$ and $\widehat\Pi \widehat{P}_{\SSS}^{\KK}$) with $P_{\NNN}$ (resp.~$P_{\SSS}^{\KK}$) in \eqref{eq001aa1} and \eqref{eq001aa2} and add a $\smallop (1)$ component. Moreover, from \eqref{eqpf06a}--\eqref{eqpf07b}, we find that $T^{-2d_L}h_L^{-1}\widehat\Upsilon_{d_L,L}^{\NNN} = \smallop(1)$,  $T^{1-2d_L}h_L^{-1}\widehat\Upsilon_{d_L,L}^{\SSS} = \smallop(1)$, $T^{-3}\widehat\Upsilon_{1,R}^{\NNN} = \smallop(1)$, and $T^{-2}\widehat\Upsilon_{1,R}^{\SSS} = \smallop(1)$. Let $\{g_j\}_{j=1}^{\KK-\nd}$ be eigenvectors of $(\ROK)^{1/2}$ (where $\ROK=P_{\KK}\RO P_{\KK}$ as in our proof of Theorem~\ref{thmv2}, and $(\ROK)^{1/2}$ is well defined), and let $\mathcal B_{\KK-\nd}(r) = \sum_{j=1}^{\KK-\nd} W_j(r)g_j$ be an $\mathcal H$-valued random element such that $(\langle \mathcal B_{\KK-\nd}(r),g_1 \rangle,\ldots,\langle \mathcal B_{\KK-\nd}(r),g_{\KK-\nd} \rangle)'$ is a $(\KK-\nd)$-dimensional vector-valued standard Brownian motion. From the functional central limit theorem and continuous mapping theorem, we find that $\sum_{s=1}^t P_{\SSS}^{\KK}X_s$ and $\sum_{s=1}^t P_{\NNN}X_s$, viewed as partial sum processes, jointly converge weakly to $(\ROK)^{1/2}\mathcal B_{\KK-\nd}$ and $\Lambda_{\Delta X}^{1/2}\mathcal W_{2,\nd}$, respectively. Then, from similar arguments used in \citet[proof of Theorem~3.1(a) and unnumbered equation between (A.10) and~(A.11)]{phillips1988spectral}, we deduce that
\begin{equation}
	\label{eqdlconv}
		\widehat{\mathfrak D}_L^{(2)} \dto \mathfrak D_L^{(2)} = c_L \left( \begin{matrix}
		\Lambda_{\Delta X}^{1/2} ( \int \mathcal W_{d_L,\nd} \otimes \mathcal W_{d_L,\nd} ) \Lambda_{\Delta X}^{1/2} & \Lambda_{\Delta X}^{1/2} ( \int \mathcal B_{\KK-\nd}\otimes \mathcal W_{d_L,\nd}) (\ROK)^{1/2} \\
		(\ROK)^{1/2} ( \int \mathcal W_{d_L,\nd} \otimes \mathcal B_{\KK-\nd} ) \Lambda_{\Delta X}^{1/2} & (\ROK)^{1/2} ( \int \mathcal B_{\KK-\nd} \otimes \mathcal B_{\KK-\nd} ) (\ROK)^{1/2}
	\end{matrix} \right) . 
\end{equation}
We also find that 
\begin{equation}
	\label{eqpf12}
	\widehat{\mathfrak D}_R^{(2)}
	\pto \mathfrak D_R^{(2)} = \left( \begin{matrix}
		0 & 0 \\ 0 & \ROK
	\end{matrix} \right).
\end{equation}

As in our proof of Theorem~\ref{thmv2}, we let $[\mathfrak D_L^{(2)}]$ and $[\mathfrak D_R^{(2)}]$ be matrix representations of $\mathfrak D_L^{(2)}$ and $\mathfrak D_R^{(2)}$ for some orthonormal basis of~$\ran P_{\KK}$. It then follows that $\{T^{2d_L-3} h_L c_L \mu_j \}_{j=\nd+1}^{\KK}$ converges in distribution to the $\KK-\nd$ largest eigenvalues of $c_L [\mathfrak D_L^{(2)}]^{-1}[\mathfrak D_R^{(2)}]$. The limit of the latter matrix simplifies to $(\int B_{\KK-\nd} B_{\KK-\nd}' - \int B_{\KK-\nd} W_{d_L,\nd}' ( \int W_{d_L,\nd} W_{d_L,\nd}' )^{-1} \int W_{d_L,\nd} B_{\KK-\nd}' )^{-1}$ without affecting the distributional properties of the eigenvalues (see the proof of Theorem~\ref{thmv2}), where $B_{\KK-\nd}$ is a $(\KK-\nd)$-dimensional vector-valued standard Brownian motion independent of~$W_{d_L,\nd}$. Then \eqref{thmeqaa1} follows immediately. Moreover, from Theorem~\ref{thmv2} we know $T^{2d_L-3} h_L\mu_j \plowto 0$ holds for $j\leq \nd$, which implies that $(T^{2d_L-3} c_Lh_L\mu_{j})^{-1} \plowto \infty$ for such~$j$.

If $\KK\leq \nd$, and thus $\|\widehat{P}_{\KK}P_{\SSS}\|_{\op} = \smallop (1)$, it holds that $\|\widehat{P}_{\KK} - P_{\NNN}^{\KK}\|_{\op} \plowto 0$ for some (possibly random) orthogonal projection $P_{\NNN}^{\KK}$ onto a subspace of~$\mathcal H_{\NNN}$. We deduce from our previous discussion that both $T^{-2d_L} h_L^{-1} \widehat{P}_{\KK} \widehat\Lambda_{d_L,L} (\widehat{P}_{\KK})^\ast$ and $h_R^{-1} T^{-2} \widehat{P}_{\NNN}^{\KK} \widehat{\Lambda}_{1,R} (\widehat{P}_{\NNN}^{\KK})^\ast$ converge to random bounded linear operators allowing $\KK$ (almost surely) positive eigenvalues. From these results and because $T/h_R \to \infty$ by Assumption~\ref{assumkernel}, $(T^{2d_L-3} c_Lh_L\mu_j)^{-1} \plowto \infty$ follows for all~$j \leq \KK$.

\subsection{Proof of Theorem~\ref{thmslack1}}

In this proof, we let $\widehat{P}_{\NNN} = \sum_{j=1}^{\nd} \hat{f}_j \otimes \hat{f}_j$ and $\widehat{P}_{\SSS} = I-\widehat{P}_{\NNN}$. We first consider the case $a_R>0$ (and hence $h_P>0$). Note that, under Assumption~\ref{assumkernel}, $\mathrm{k}_P (\cdot)$ has bounded support $[-\ell, \ell]$ for some $\ell <\infty$ and $|\mathrm{k}_P (\cdot)|\leq 1$. Without loss of generality, we assume that $\ell=1$, and then $\mathrm{k}_P (s/h_P) = 0$ holds if and only if $s \notin [-h_P, h_P]$. 

First, consider~$\KK > \nd$. We then argue that
\begin{equation}
\label{lemeq000aa1}
\|h_P^{-1}T^{-2} \widehat{\Lambda}_{1,P} -  h_P^{-1}T^{-2}P_{\NNN} \widehat{\Lambda}_{1,P} P_{\NNN} \|_{\op} = \smallop (1).
\end{equation}
This holds because
\begin{equation*}
\frac{1}{h_PT^2} \sum_{s=0}^{h_P} \sum_{t=1}^T ( \|P_{\NNN}X_{t-s}\| \|P_{\SSS}X_t\| + \|P_{\SSS}X_{t-s}\| \|P_{\NNN}X_t\| 
+ \|P_{\SSS}X_{t-s}\| \|P_{\SSS}X_{t}\| ) = \smallop (1),
\end{equation*}
which follows from the facts that $\sup_{1\leq t\leq T}\|P_{\NNN}X_t/\sqrt{T}\| = \bigOp (1)$ and $T^{-1}\sum_{t=1}^T \E  \|P_{\SSS}X_t\|^2 = \bigO (1)$, which both hold under our assumptions, along with the Cauchy-Schwarz inequality. By \eqref{lemeq000aa1}, we then find, from the same arguments used in the proof of Proposition~3.2 of \citet{Chang2016}, that $\widehat{P}_{\NNN} - P_{\NNN} = \bigOp (T^{-1})$, $\widehat{P}_{\SSS} - P_{\SSS} = \bigOp  (T^{-1})$, and $\widehat{P}_{\SSS} P_{\NNN} = \bigOp  (T^{-1})$. These results combined with Lemma~\ref{lem1} imply that $T^{-1}\widehat{P}_{\SSS}P_{\NNN}\widehat{\Lambda}_{1,P} P_{\NNN}\widehat{P}_{\SSS}$, $T^{-1}\widehat{P}_{\SSS}P_{\NNN}\widehat{\Lambda}_{1,P} P_{\SSS}\widehat{P}_{\SSS}$, and  $T^{-1}\widehat{P}_{\SSS}P_{\NNN}\widehat{\Lambda}_{1,P} P_{\SSS}\widehat{P}_{\SSS}$ are all $\smallop (1)$, and, in addition, that
\begin{equation}
\label{eqpfslack0}
T^{-1}\widehat{P}_{\SSS} P_{\SSS} \widehat{ \Lambda}_{1,P} P_{\SSS} \widehat{P}_{\SSS} \pto \RO,
\end{equation}
where the convergence follows from \citet[Theorem~2]{horvath2013estimation}; see also our proof of Theorem~\ref{thmv2}. We may then deduce from \citet[Lemma~4.2]{Bosq2000} that $\sup_{j\geq 1}|\widehat{\tau}_{j+\nd} - \tau_j| \plowto 0$. Moreover, from \citet[Lemmas~4.3 and~4.4]{Bosq2000}, we find that \eqref{eqpfslack0} implies that
\begin{equation}
\label{eqpfslack1}
\Big\|\sum_{j=\nd+1}^{\KK} \hat{f}_j \otimes \hat{f}_j - P_{\SSS}^{\KK}\Big\|_{\op} \pto 0 ,
\end{equation}
where $P_{\SSS}^{\KK}$ is the projection onto the span of $\{f_j\}_{j=\nd+1}^{\KK}$, and we know that $P_{\SSS}^{\KK}$ is a nonrandom projection since $\tau_{\KK-\nd} \neq \tau_{\KK-\nd+1}$. Therefore, $\widehat{P}_{\KK}$ satisfies Assumption~\ref{assumvr1}.

Next, if $\KK = \nd$ then Assumption~\ref{assumvr1} holds for $P_{\SSS}^{\KK}=0$ obviously.

Finally, consider~$\KK < \nd$. From \eqref{eqpf07}--\eqref{eqpf08a}, we find that $\|h_P^{-1}T^{-2}\widehat{\Lambda}_{1,P} - c_P \Lambda_{\Delta X}^{1/2} (\int \mathcal{W}_{1,\nd} \otimes \mathcal{W}_{1,\nd}) \Lambda_{\Delta X}^{1/2}\|_{\op} = \smallop  (1)$ (where $c_P= 2 \int_0^1 \mathrm{k}_P(u) du$).  Since the nonzero eigenvalues of $\Lambda_{\Delta X}^{1/2} ( \int \mathcal{W}_{1,\nd} \otimes \mathcal{W}_{1,\nd} ) \Lambda_{\Delta X}^{1/2}$ are almost surely distinct and $\ran \Lambda_{\Delta X}^{1/2} = \mathcal H_{\NNN}$, we deduce from \citet[Lemma~4.3]{Bosq2000} that 
\begin{equation}
\label{eqpfslack1add}
\sup_{1\leq j \leq \KK}\|\hat{f}_j - \mathrm{sgn} (\langle \hat{f}_j, f_j \rangle) f_j\| \pto 0,
\end{equation}
where $\{f_j\}_{j=1}^{\KK}$ is an orthonormal set in~$\mathcal H_{\NNN}$. Thus we find that $\|(\sum_{j=1}^{\KK} \hat{f}_j \otimes \hat{f}_j)P_{\SSS}\|_{\op} = \smallop (1)$.

In the case $a_R =0$ and $\KK\geq\nd$, the result \eqref{lemeq000aa1} reduces to $\|T^{-2} \widehat{\Lambda}_{1,P} -  T^{-2}P_{\NNN}\widehat{\Lambda}_{1,P}P_{\NNN}\|_{\op} = \smallop(1)$ and \eqref{eqpfslack0} still holds with $a_R= 0$ (whereby $\RO$ becomes $\mathbb{E}[{P}_{\SSS}X_t\otimes {P}_{\SSS}X_{t}]$). Using nearly identical arguments, we then find that \eqref{eqpfslack1} is established and $\widehat{P}_{\KK}$ satisfies the requirements in Assumption~\ref{assumvr1}. In the case $\KK<\nd$, we similarly deduce from \eqref{eqpf07}--\eqref{eqpf08a} that $\|T^{-2} \widehat{\Lambda}_{1,P} -  \Lambda_{\Delta X}^{1/2} (\int \mathcal{W}_{1,\nd} \otimes \mathcal{W}_{1,\nd}) \Lambda_{\Delta X}^{1/2}\|_{\op} = \smallop(1)$, from which the desired result follows.

The above proof does not require any significant changes even if we allow for a nonzero intercept or a linear trend as in Section~\ref{sec_deterministic}. Provided that $\widehat{\Lambda}_{1,P}$ is computed from the relevant residuals, i.e., $U_t^{(1)}$ or $U_t^{(2)}$ given in~\eqref{fresid}, we may show that \eqref{lemeq000aa1} holds and that $\widehat{P}_{\NNN} - P_{\NNN} = \bigOp (T^{-1})$ and $\widehat{P}_{\SSS} - P_{\SSS} = \bigOp  (T^{-1})$. Then $T^{-1}\widehat{P}_{\SSS}P_{\SSS}\widehat{\Lambda}_{1,R} P_{\SSS}\widehat{P}_{\SSS} \plowto \RO$ follows from \citet[Theorem~5.3]{kokoszka2016kpss}; c.f.~\eqref{eqpfslack0}. We thus find from \citet[Lemmas~4.3 and~4.4]{Bosq2000} that \eqref{eqpfslack1} holds if $\hat{f}_j$ is an eigenvector of $\widehat{\Lambda}_{1,P}$ computed from the relevant residuals.

\subsection{Proof of Theorem~\ref{thmslack2}}

We first consider the case $a_R>0$ and hence $h_P>0$. We note that
\begin{equation*}
T^{-1} \widehat{\Lambda}_{1,P} =  T^{-1}P_{\NNN}\widehat{\Lambda}_{1,P}P_{\NNN} + T^{-1}(P_{\NNN}\widehat{\Lambda}_{1,P}P_{\SSS}+P_{\SSS}\widehat{\Lambda}_{1,P}P_{\NNN}) + T^{-1} P_{\SSS}\widehat{\Lambda}_{1,P}P_{\SSS},
\end{equation*}
where the first term is $\bigOp (h_P T)$ by Lemma~\ref{lem1}. The second term is $\bigOp (h_P)$, and this is deduced from similar arguments used in our proof of Lemma~\ref{lem1} (see also Lemma~3.1 of \citealp{Chang2016}). The third term is $\bigOp (1)$, which results from \citet[Theorem~2]{horvath2013estimation}. Since $\widehat{P}_{\NNN} - P_{\NNN} = \bigOp  (T^{-1})$ and $\widehat{P}_{\SSS} - P_{\SSS} = \bigOp  (T^{-1})$,  $T^{-1}\widehat{P}_{\SSS}P_{\NNN}\widehat{\Lambda}_{1,P} P_{\NNN}\widehat{P}_{\SSS}$, $T^{-1}\widehat{P}_{\SSS}P_{\NNN}\widehat{\Lambda}_{1,P} P_{\SSS}\widehat{P}_{\SSS}$, and  $T^{-1}\widehat{P}_{\SSS}P_{\SSS}\widehat{\Lambda}_{1,P} P_{\NNN}\widehat{P}_{\SSS}$ are all $\bigOp  (h_P/T)$, and, in addition, \eqref{eqpfslack0} holds. Combining all these results, 
\begin{align}
\label{eqpfslack3}
T^{-1}\widehat{P}_{\SSS} \widehat{\Lambda}_{1,P} \widehat{P}_{\SSS} - \RO = T^{-1} \widehat{P}_{\SSS} (P_{\SSS} \widehat{ \Lambda}_{1,P} P_{\SSS}- \RO)\widehat{P}_{\SSS}  + \bigOp  (h_P/T) \pto 0,
\end{align} 
where the first equality follows from the facts that $\widehat{P}_{\NNN} \RO$ and $\RO \widehat{P}_{\NNN}$ are both~$\bigOp (T^{-1})$.
Under the conditions in Assumption~\ref{assumvr1ll}, we may deduce from \citet[Theorems~2.2, 2.3, and their proofs]{BERKES2016150} that $T^{-1}P_{\SSS}\widehat{\Lambda}_{1,R} P_{\SSS}-\RO = \bigOp((T/h_P)^{-1/2}) + \bigOp(T^{-1}) +\bigOp(h_P^{-\varphi})$. This result, combined with \eqref{eqpfslack3}, implies that $(T/h_P)^{1/2} (T^{-1}\widehat{P}_{\SSS} \widehat{\Lambda}_{1,P} \widehat{P}_{\SSS}-\RO) = \bigOp(1) + \bigOp((T/h_P)^{1/2} h_P^{-\varphi})$, which is $\bigOp(1)$ under Assumption~\ref{assumvr1ll}(iv). We thus find that $T^{-1}\widehat{P}_{\SSS} \widehat{\Lambda}_{1,P} \widehat{P}_{\SSS} =  \RO + \bigOp((h_P/T)^{1/2})$. We then deduce from Lemma~4.2 of \citet{Bosq2000} that the eigenvalues $\{\widehat{\tau}_{\nd +j} \}_{j\geq 1}$ of $T^{-1}\widehat{P}_{\SSS}\widehat{\Lambda}_{1,P} \widehat{P}_{\SSS}$ satisfy  
\begin{equation}
\label{eqpfslack4}
\sup_{j\geq 1} |\widehat{\tau}_{\nd+j} - \tau_j| \leq  \|T^{-1}\widehat{P}_{\SSS}\widehat{\Lambda}_{1,P} \widehat{P}_{\SSS} - \RO\|_{\op} = \bigOp  ( (h_P/T)^{1/2}).
\end{equation}
From \eqref{eqpfslack4} we find that
\begin{equation}
\label{eqpfslack6}
\widehat{\tau}_{\nd+j}-\widehat{\tau}_{\nd+j+1} = (\widehat{\tau}_{\nd+j}-\tau_j) + (\tau_j - \tau_{j+1}) + (\tau_{j+1}-\widehat{\tau}_{\nd+j+1}) = \tau_j - \tau_{j+1} + \bigOp  ( (h_P/T)^{1/2}) ,
\end{equation}
which is the desired result for $a_R > 0$. In the case $a_R = 0$, and hence $P_{\SSS}\widehat{\Lambda}_{1,P}P_{\SSS}$ becomes the sample variance operator of $P_{\SSS}X_t$, we deduce from Theorems~3.1 and~3.2 \citet{hormann2010} that $\sup_{j\geq 1}|\widehat{\tau}_{\nd+j}- \tau_j| = \bigOp  (T^{-1/2})$.

The above proof does not require any significant changes if $\widehat{\Lambda}_{1,P}$ is computed from $U_t^{(1)}$ or $U_t^{(2)}$ given in \eqref{fresid} to accommodate a nonzero intercept or linear trend. In those cases, \eqref{eqpfslack3} still holds and $(T/h_P)^{1/2} (T^{-1}P_{\SSS} \widehat{ \Lambda}_{1,P} P_{\SSS}-\RO) = \bigOp (1)$ (see Theorem~5.3 of \citealp{kokoszka2016kpss} and Theorem~2.1 of \citealp{BERKES2016150}). Then the rest of the proof is almost identical and hence omitted.

\subsection{Proof of Theorem~\ref{thmest}}

If $\{\mu_j\}_{j=1}^{\smalls_{\max}+1}$ are the eigenvalues of the VR($d_L$,1) problem with $d_L \geq 2$, then the consistency result follows immediately from Theorems~\ref{thmv2} and~\ref{thmvr2ab}.

For the VR($d_L$,0) problem with $d_L \geq 2$, we consider only $a_L >0$ and rewrite \eqref{eqgev} as 
\begin{equation}
\label{eq001a}
h_L^{-1}T^{2-2d_L}D_T^2 \widehat P_{\KK} \widehat\Lambda_{d_L,L} \widehat P_{\KK} D_T^2  \nu_j = (T^{2d_L-3} h_Lh_R^2{\mu_j})^{-1} T^{-1}h_R^2 D_T^2 \widehat P_{\KK} \widehat\Lambda_{0,R}\widehat P_{\KK} D_T^2  \nu_j. 
\end{equation}
The operator matrix corresponding to the left-hand side of \eqref{eq001a} is $\widehat{\mathfrak D}_L^{(2)}$ from our proof of Theorem~\ref{thmvr2ab}. From \eqref{eqdlconv} we know that $\widehat{\mathfrak D}_L^{(2)} \dlowto \mathfrak D_L^{(2)}$, where $\mathfrak D_L^{(2)}$ is given in~\eqref{eqdlconv}. We let $\widehat{\mathfrak D}_R^{(3)}$ be the operator matrix corresponding to $T^{-1}h_R^2 D_T^2 \widehat P_{\KK} \widehat\Lambda_{0,R}\widehat P_{\KK} D_T^2$, which can be constructed similarly to $\widehat{\mathfrak D}_R^{(2)}$ in the proof of Theorem~\ref{thmvr2ab}. From \eqref{eqpf02a}--\eqref{eqpf04d} and the fact that $h_RT^{-1} \to 0$, we deduce that
\begin{equation}
\label{eq:lambda0R}
\widehat{\mathfrak D}_R^{(3)} \pto -\mathrm{k}_R^{\prime\prime} (0) \mathfrak D_R^{(2)} ,
\end{equation}
where $\mathfrak D_R^{(2)}$ is given in~\eqref{eqpf12}. We thus find that $T^{2d_L-3} h_L h_R^2 c_L \mu_j /\mathrm{k}_R^{\prime\prime} (0)$ converges to well-defined nonzero nonrandom eigenvalues for $j=\nd+1,\ldots, \KK$ (jointly), but converges to zero for $j \leq \nd$. From this result and Theorem~\ref{thmvd0}, the desired consistency result follows. 

Next, suppose $\{\mu_j\}_{j=1}^{\smalls_{\max}+1}$ are the eigenvalues of the VR(1,0) problem. 
As before, we only consider $a_L>0$ and rewrite \eqref{eqgev} as
\begin{equation} 
	\label{eq:10problem}
	T^{-1}D_{h_LT} \widehat P_{\KK}	\widehat\Lambda_{1,L} \widehat P_{\KK} D_{h_LT} \nu_j =	(h_R^2 \mu_j )^{-1} T^{-1}h_R^2 D_{h_LT} \widehat P_{\KK} \widehat\Lambda_{0,R}\widehat P_{\KK} D_{h_LT} \nu_j,
\end{equation}
and let $\widehat{\mathfrak D}_L^{(4)}$ (resp.\ $\widehat{\mathfrak D}_R^{(4)}$) be the operator matrix corresponding to $T^{-1}D_{h_LT} \widehat P_{\KK}	\widehat\Lambda_{1,L} \widehat P_{\KK} D_{h_LT}$ (resp.\ $T^{-1}h_R^2 D_{h_LT} \widehat P_{\KK} \widehat\Lambda_{0,R}\widehat P_{\KK} D_{h_LT}$). Let $\ROL$ be defined as in \eqref{eqlongrun} but with $a_R$ replaced by~$a_L$. From similar arguments used to obtain \eqref{DLlimit} and the facts that $h_L^{-1/2}h_RT^{-1/2} \to 0$ and $T^{-1}\widehat{\Pi}\widehat{P}_{\SSS} ^{\KK} \widehat{\Lambda}_{1,L} (\widehat{P}_{\SSS} ^{\KK})^{\ast} \plowto \ROKL=P_{\KK}\ROL P_{\KK}$, see \eqref{eqpf08a}, we find that 
\begin{equation}
	\label{eq:lambda1L}
	\widehat{\mathfrak D}_L^{(4)}
	\dto \mathfrak D_L^{(4)} = \left( \begin{matrix}
		c_L	 \Lambda_{\Delta X}^{1/2} ( \int \mathcal{W}_{1,\nd} \otimes \mathcal{W}_{1,\nd} ) \Lambda_{\Delta X}^{1/2} & 0 \\
		0 & \ROKL
	\end{matrix}\right) .
\end{equation}
From \eqref{eqpf02a}--\eqref{eqpf04d} and the fact that $h_L^{-1/2}h_R T^{-1/2} \to 0$, we find, as in \eqref{eq:lambda0R}, that
\begin{equation}
	\label{eq:lambda0R1}
	\widehat{\mathfrak D}_R^{(4)} \pto -\mathrm{k}_R^{\prime\prime} (0) \mathfrak D_R^{(2)} ,
\end{equation}
where $\mathfrak D_R^{(2)}$ is given in~\eqref{eqpf12}. From \eqref{eq:10problem}--\eqref{eq:lambda0R1}, the conclusions follow by the same arguments as for the VR($d_L$,0) problem with $d_L \geq 2$.

\subsection{Proof of Theorem~\ref{thm:inference}}

Consider first the hypotheses in~\eqref{hptest1}. Under $H_0$, the dimension of the nonstationary subspace of the residual time series $\{(I-P_{\mathcal H_0})X_t\}$ is $\nd-p_0$, while it is greater than $\nd-p_0$ under~$H_1$. We thus know from Corollary~\ref{corthmv2a} that under $H_0$ the statistics $\mathcal F_{\max}(\{\tilde{\mu}_j^{-1}\}_{j=\nd-p_0}^{\KK})$ and $\mathcal F_{\tr}(\{\tilde{\mu}_j^{-1}\}_{j=\nd-p_0}^{\KK})$ converge to the maximum eigenvalue and trace, respectively, of $\mathcal A$ with $\smalls_0=\nd-p_0$, where $\mathcal A$ is defined in Corollary~\ref{corthmv2a}. On the other hand, also by Corollary~\ref{corthmv2a}, the statistics both diverge to infinity under~$H_1$.

Next, consider the hypotheses in~\eqref{hptest2}. Under $H_0$, the residual time series $\{(I-P_{\mathcal H_0})X_t\}$ is stationary, while it is nonstationary under~$H_1$. In this case, we know from Corollary~\ref{corthmv2a} that under $H_0$ the statistics $\mathcal F_{\max}(\{\tilde{\mu}_j^{-1}\}_{j=1}^{\KK})$ and $\mathcal F_{\tr}(\{\tilde{\mu}_j^{-1}\}_{j=1}^{\KK})$ converge to the maximum eigenvalue and trace, respectively, of $\int B_{\KK} B_{\KK}^\prime$, while they diverge to infinity under~$H_1$.

The arguments for the hypotheses in \eqref{hptest3} are similar and omitted for brevity.

\newpage

\phantomsection
\addcontentsline{toc}{section}{Supplementary Appendix}

\begin{center}
{\LARGE \vspace*{1in}}

{\LARGE Supplementary Appendix to}
{\LARGE \vspace*{0.25in}}

{\LARGE ``Inference on common trends in functional time series''}

\bigskip
\bigskip

{\Large by}

\bigskip
\bigskip

{\Large M.\ \O .\ Nielsen, W.-K.\ Seo, and D. Seong}

\end{center}

\bigskip

{\LARGE \vspace*{1in}}
This supplementary appendix includes four sections. Section~\ref{sec:guide} provides a detailed guide on implemention of our proposed methods for determining the number of stochastic trends and computing the test statistics. Section~\ref{secexisting} discusses the relationships between our proposed tests and existing tests, mostly in finite-dimensional spaces. Section~\ref{appsim} presents a Monte Carlo simulation study in a finite-dimensional VAR setup. Finally, Section~\ref{sec:prooflemma} contains the proof of Lemma~\ref{lem1} stated in Appendix~\ref{sec:app-proofs}. 

Equation references (S.$n$) for $n \geq 1$ refer to equations in this supplementary appendix and other equation references are to the main paper.

\setcounter{page}{1}
\newpage

\setcounter{section}{0}
\def\thesection{S.\arabic{section}}
\def\thesubsection{S.\arabic{section}.\arabic{subsection}}
\renewcommand{\theequation}{S.\arabic{equation}}
\setcounter{equation}{0}
\renewcommand{\thetable}{S.\arabic{table}}
\setcounter{table}{0}
\renewcommand{\thefigure}{S.\arabic{figure}}
\setcounter{figure}{0}

\section{S.1 \quad Guide to practical implementation}
\label{sec:guide} 

In Section~\ref{sec:implementation}, we first discuss choice of tuning parameters and implementation of the sequential test procedure. A replication package for \textsf{R} is also available at 
\begin{center}
{\small
\url{https://github.com/sdkseong/Inference-on-common-trends-in-functional-time-series}}
\end{center}
Next, in Section~\ref{revremadd2}, we present details about computational aspects.

\subsection{Implementation of the up-down hybrid procedure}
\label{sec:implementation}

In this section, we illustrate the practical implementation of our proposed method for determining the number of stochastic trends. There are several possible choices of test statistics, i.e.\ of $(d_L,d_R)$, as discussed in Remark~\ref{revremadd1}. Our simulation experiments suggest that the up-down hybrid procedure, based on the inverse VR and VR($2,1$) tests, appears to offer a good compromise between ease of implementation and finite sample performance. We therefore focus on this procedure in the following discussion, while also briefly discussing possible variants. 

The proposed hybrid procedure is described by the following four steps.

\subsubsection*{Step 1: Eigenvector estimation for the BU procedure}
The first step is to estimate the eigenvectors of $\widehat{\Lambda}_{1,P}$ defined in~\eqref{eqlongrun2}, which serve as a crucial input for the BU procedure. Since the inverse VR test is implemented with $h_L=0$ and $h_R>0$ in the BU procedure (Remark~\ref{remInvVR}), the key tuning parameter to be determined in this step is the bandwidth~$h_P$. Ideally, a larger $h_P$ is preferable if the underlying stationary component, $P_{\SSS}X_t$, exhibits high persistence. In practice, we recommend using $h_P=[T^{1/3}]$ because a large bandwidth is supported by Assumption~\ref{assumvr1ll}. However, the choice of $h_P$ is not critical for the overall procedure because (i)~the BU procedure is solely used to conjecture an upper bound for~$\nd$, and (ii)~we will incorporate a buffer integer $m_{\smalls}$ to alleviate cases where the BU procedure might underestimate the number of stochastic trends. Once $\widehat{\Lambda}_{1,P}$ is constructed, its eigenvectors, $\{\hat{f}_j\}_{j\geq1}$, can be computed using standard~FPCA.

\subsubsection*{Step 2: Determining an upper bound $\smalls_{\max}$ via the BU procedure}
Next, we implement the BU procedure to obtain a reasonable upper bound $\smalls_{\max}$ for~$\nd$. Following Remark~\ref{remInvVR}, the inverse VR tests are implemented with $h_L=0$ and $h_R>0$, so the parameters to be determined by practitioners are $h_R$, $m$ (to determine~$\KK$), and~$m_{\smalls}$ (a buffer integer used to construct~$\smalls_{\max}$). As in Step~1, a larger $h_R$ is ideally preferable if $P_{\SSS}X_t$ is highly persistent, but its choice is not overly critical given that the primary purpose of this step is only to conjecture a reasonable upper bound for~$\nd$. Specifically, we recommend $h_R=[T^{1/5}]$, which is the same rate as the \citet{andrews1991} optimal bandwidth for stationary series (Remark~\ref{remLRV}). 

With $h_R$ selected, the BU procedure is implemented by sequentially testing $H_0: \nd=\smalls_0$ for $\smalls_0=0,1,\ldots$. For each~$\smalls_0$, we construct the slack extractor $\widehat{P}_{\KK} = \sum_{j=1}^{\KK} \hat{f}_j \otimes \hat{f}_j$ with $\KK=\smalls_0+m$ for some $m>0$. The choice of projection $\widehat{P}_{\KK}$ is discussed in Section~\ref{secprojec} and specifically Remark~\ref{rem:newremark}, and the choice of $m$ is discussed in Remark~\ref{remadd}. In our simulation experiments, we found that setting $m=2$ results in good finite sample properties. Upon obtaining the estimate $\widehat\smalls_{\rm BU}$ from this procedure, we set $\smalls_{\max} = \widehat\smalls_{\rm BU} + m_{\smalls}$. The subsequent VR(2,1)-based TD procedure is remarkably robust to the choice of $\smalls_{\max}$, so we recommend a slightly larger value $m_{\smalls}=5$ to reduce the risk of setting $\smalls_{\max}$ too small.

\subsubsection*{Step 3: Eigenvector estimation for the TD procedure}
We re-estimate the eigenvectors of $\widehat{\Lambda}_{1,P}$ in \eqref{eqlongrun2} specifically for the VR(2,1)-based TD procedure. As discussed in Remark~\ref{remvr21}, we implement the VR($2,1$)-based test with $h_L=h_R=0$. Consequently, a different set of eigenvectors is required to construct the slack extractor for the TD procedure with $a_R=0$; see Theorem~\ref{thmslack1} regarding the requirements depending on the value of~$a_R$. This step does not require the selection of any additional tuning parameters. We then construct $\widehat{\Lambda}_{1,P}$ in \eqref{eqlongrun2} with $a_R=0$ and compute its eigenvectors using standard~FPCA. To distinguish these eigenvectors from those in Step~1, we denote them~$\{\hat{g}_j\}_{j\geq 1}$.

\subsubsection*{Step 4: TD procedure for final estimation}
The final step is to implement the TD procedure based on the VR(2,1) tests. The VR($2,1$)-based test is implemented with $h_L=h_R=0$; see Remark~\ref{remvr21}. In this step, we test the hypothesis $H_0:\nd = \smalls_0$ against $H_1:\nd = \smalls_0-1$ for $\smalls_0 = \smalls_{\max}, \smalls_{\max},\ldots, 1$ to obtain the final estimate~$\widehat\smalls_{\rm TD}$. Thus, the only tuning parameter for this step is~$m$, which is used to determine~$\KK = \smalls_0 +m$. Once $\KK$ is specified for each~$\smalls_0$, the slack extractor is constructed as $\widehat{P}_{\KK}= \sum_{j=1}^{\KK} \hat{g}_j \otimes \hat{g}_j$ using the eigenvectors computed in Step~3. As in Step~2, based on our simulation evidence, we set $m=2$. 

\subsubsection*{Possible variants}
Variants of the procedure described above can be implemented with some adjustments.
\begin{enumerate}[label=(\roman*)]
\item In Step~2, $h_L > 0$ can be used for the BU procedure. However, our simulation results suggest that this does not substantially affect finite sample performance. Since any specific choice of $h_L>0$ might appear arbitrary, $h_L=0$ may generally be recommended in practice.
\item The TD procedure in Step~4 can be based on VR(1,0) or VR(2,0) tests. In such cases, Step~3 must be adjusted by employing $h_P > 0$ in accordance with Theorem~\ref{thmslack1}. In fact, this implies that the eigenvectors from Step~1 can be used also in Step~3 (which can therefore be skipped). Unlike the VR(2,1)-based tests in Step~4 above, these variants require additional tuning parameters: $h_L$ and~$h_R$. While $h_R$ can be selected using existing optimal bandwidth selection methods (Remark~\ref{remLRV}), our simulations indicate that such choices do not necessarily guarantee superior finite sample properties. We may let $h_L=0$ for simplicity, although our evidence suggests that $h_L > 0$ might yield better performance in finite samples. The difficulty in choosing $h_L$ and $h_R$ for the VR(1,0) or VR(2,0) is another advantage of the VR(2,1)-based tests.
\end{enumerate}

\subsection{Computational aspects}
\label{revremadd2}
This section describes computation of the proposed tests based on functional data~$X_t$. In cases where $X_t$ is not observed as a function but only partially as a vector-valued time series, the implementation of our procedure can be treated as a special case. We describe computation of the projection $\widehat{P}_{\KK}$, the generalized eigenvalues $\mu_j$ in \eqref{eqgev}, and their associated eigenvectors,~$\nu_j$. 

\subsubsection*{Computing the projection}
Let $\{ \hat{f}_j\}_{j=1}^{\KK}$ be the eigenvectors corresponding to the $\KK$ largest eigenvalues of $\widehat{\Lambda}_{1,P}$ defined in~\eqref{eqlongrun2}. These eigenvectors are computed by the standard FPCA method. We note that to obtain these eigenvectors using the FPCA, it is typically necessary to represent $X_t$ using basis functions (e.g., B-splines or Fourier bases), which is not required for the vector-valued case. Details of this representation procedure are well-established in the literature (e.g., \citealp{Ramsay2005}). Having obtained $\{ \hat{f}_j\}_{j=1}^{\KK}$, the projection is given by $\widehat{P}_{\KK} = \sum_{j=1}^{\KK} \hat{f}_j \otimes \hat{f}_j$.

\subsubsection*{Computing the eigenvalues}
As discussed in Remark~\ref{remisomorphism}, the eigenvalue problem in \eqref{eqgev} can be recast into the following $\KK$-dimensional eigenvalue problem:
\begin{equation}
\label{eqapp01}
\mu_j [ \widehat{P}_{\KK} \widehat{\Lambda }_{d_L, L} \widehat{P}_{\KK}] \nu_{j,\KK} =  [ \widehat{P}_{\KK} \widehat{\Lambda }_{d_R, R} \widehat{P}_{\KK}]\nu_{j,\KK}
\end{equation}
for some eigenvectors $\{\nu_{j,\KK}\}_{j=1}^\KK$ defined on~$\mathbb{R}^{\KK}$. From the isomorphism discussed in Remark~\ref{remisomorphism}, we find that the $\KK\times \KK$ matrices $[ \widehat{P}_{\KK} \widehat{\Lambda }_{d_L, L} \widehat{P}_{\KK}]$ and $[ \widehat{P}_{\KK} \widehat{\Lambda }_{d_R, R} \widehat{P}_{\KK}]$ can be defined explicitly as
\begin{equation}
\label{eqapp02}
[ \widehat{P}_{\KK} \widehat{\Lambda }_{d_m, m} \widehat{P}_{\KK}]_{ij}= \langle \widehat{f}_i,  \widehat{\Lambda }_{d_m, m}  \widehat{f}_j\rangle, \quad 1\leq i,j\leq \KK, \quad m \in \{L,R\}. 
\end{equation}
Since $\{\hat{f}_j\}_{j=1}^{\KK}$ and $\widehat{\Lambda}_{d_m,m}$ are computed from the given sample, these matrices can be readily constructed. Specifically, the inner product is calculated as
\begin{equation*}
\langle \widehat{f}_i,  \widehat{\Lambda }_{d_m,m}  \widehat{f}_j\rangle  = \sum_{s=-T+1} ^{T-1} \mathrm{k} (s/h)\langle \widehat f_i,\widehat{\Gamma}_{d_m,s} \widehat f_j\rangle,
\end{equation*}
where $\widehat{\Gamma}_{d_m,s}$ is defined in~\eqref{eqslrv}. Computing the eigenvalues and eigenvectors from \eqref{eqapp01} and \eqref{eqapp02} is a standard finite-dimensional eigenvalue problem. Consequently, the eigenvalues $\{\mu_j\}_{j=1}^{\KK}$ required to construct the test statistics can be easily obtained.

\subsubsection*{Computing the eigenvectors}
$\mu_j$ and $\nu_j$ in \eqref{eqgev} are expressed as eigenelements of two operators, but computing these eigenvalues is straightforward as discussed above. In fact, the generalized eigenvalue problem can be simplified into \eqref{eqapp01} involving two $\KK \times \KK$ sample (long-run) variance matrices. It is equally straightforward to compute $\nu_j$ from $\nu_{j,\KK}$: if $\nu_{j,\KK} = (c_{j,1}, \ldots , c_{j,\KK})' \in \mathbb{R}^\KK$ is an eigenvector associated with the generalized eigenvalue problem in~\eqref{eqapp01}, then $\nu_j = \sum_{s=1}^{\KK} c_{j,s} \hat{f}_s$. Note that this simple computation of the eigenelements is possible due to the presence of $\widehat{P}_{\KK}$ in~\eqref{eqgev}. Without $\widehat{P}_{\KK}$, computation of the eigenelements of the linear operators $\widehat{\Lambda}_{d_m,m}$ for $m \in \{ L, R \}$ requires a high-dimensional approximation that is potentially both inaccurate and numerically unstable. This further emphasizes the importance of a feasible choice of~$\widehat{P}_{\KK}$.

\section{S.2 \quad Relationships with existing tests}
\label{secexisting}

As discussed in Remark~\ref{remisomorphism}, the projected time series $\widehat{P}_{\KK}X_t$ based on our slack extractor may be understood as a finite-dimensional vector-valued time series, $x_{\KK,t} = (\langle X_t, \hat{f}_1 \rangle,\ldots,\langle X_t, \hat{f}_{\KK}\rangle)'$, where $\hat{f}_j$ is an eigenvector of $\widehat{\Lambda}_{1,P}$ as used in, e.g., Theorem~\ref{thmslack1}. This time series is expected to have $\nd$ stochastic trends (and cointegration rank~$\KK-\nd$). Of course, as discussed in Remarks~\ref{remslack} and~\ref{rem:newremark}, this requires a careful choice of~$\widehat{P}_{\KK}$. Thus, it may be tempting to apply any existing cointegration rank tests developed in a finite-dimensional setting to~$x_{\KK,t}$. However, from Remark~\ref{rem1} we already know that dimension reduction by a projection map is not guaranteed to preserve parametric assumptions such as the VAR assumption, and hence not all tests developed in a finite-dimensional setup are applicable. Nevertheless, in special cases, our variance ratio tests are related to some well-known tests developed in finite-dimensional and function-valued settings. This will be discussed in detail in this section.

\subsection{VR(2,1)-based tests}

\subsubsection{The test of \citet{Breitung2002}}
\label{sec_breitung}

Suppose we are in the conventional Euclidean space setting with $\mathcal H= \mathbb{R}^p$ and consider the VR(2,1)-based test using $\Ftr$ with $\KK = p \geq \smalls_0$ and $a_L=a_R=0$. Then the VR(2,1)-based test is identical to the test proposed by \citet{Breitung2002}. Note that, in this simple case, $\widehat{P}_{\KK} = I$, and thus Assumption~\ref{assumvr1} or its low-level counterparts are no longer needed. However, our VR(2,1)-based test can also be applied even when the number of nonzero eigenvalues of $\RO$ is smaller than $p-\nd$ using $\widehat{P}_{\KK}$ of rank $\KK < p$; see Section~\ref{secprojec} and note that we can distinguish nonzero eigenvalues as discussed in Theorem~\ref{thmslack2} and Remark~\ref{remslack2}. 

In the case where $\mathcal H$ can be a general Hilbert space, $\{\widehat{P}_{\KK}X_t\}_{t=1}^T$ can also be understood as the vector-valued time series~$\{x_{\KK,t}\}_{t=1}^T$. Moreover, with $a_L=a_R = 0$, the unnormalized sample variance of $\{\widehat{P}_{\KK}X_t\}_{t=1}^T$ can simply be written as $\widehat{P}_{\KK}\widehat{\Lambda}_{1,R}\widehat{P}_{\KK}$, so that the VR(2,1) eigenvalue problem reduces to the generalized eigenvalue problem associated with the sample variances of $\{\sum_{s=1}^t x_{\KK,s}\}_{t=1}^T$ and~$\{x_{\KK,t}\}_{t=1}^T$. Thus, $\mathcal F_{\tr}(\{\tilde{\mu}_j\}_{j=1}^{\nd})$ and its limits are identical to those of \citepos{Breitung2002} test for examining \eqref{eqhypo}, implemented assuming that $\{x_{\KK,t}\}_{t=1}^T$ are observed. From this result, it is clear that implementation of the VR(2,1)-based test with $a_L=a_R=0$ is particularly simple because it reduces to application of \citepos{Breitung2002} test to the $\KK$-dimensional time series~$\{x_{\KK,t}\}_{t=1}^T$.

\subsubsection{The test of \citet{NSS}}
\label{sec_NSS}

Consider the case where $\mathcal H$ is infinite-dimensional. Then the VR(2,1)-based test using $\Ftr$ with $\KK \geq \smalls_0$ and $a_L=a_R=0$ is equivalent to the $\mathcal T_{\mathcal C}$ test proposed by \citet{NSS}. However, \citet{NSS} required $\RO$ to be injective on $\mathcal H_{\SSS}$ (such that $\RO$ has infinitely many positive eigenvalues), but this condition is not needed for our VR(2,1)-based test. This is an important feature of our VR(2,1)-based test compared to that of \citet{NSS}. Because they consider infinite-dimensional curve-valued time series, the injectivity assumption does not seem very restrictive in their setup. However, we want to accommodate the case where the time series takes values in a possibly finite-dimensional subspace of~$\mathcal H$, and in such cases the injectivity condition may not hold.

\subsection{VR(1,0)-based tests}

\subsubsection{The test of \citet{shintani2001simple}}
\label{sec_shintani}

Consider the VR(1,0)-based test using $\Ftr$ with $\KK\geq \smalls_0$, $a_L\geq 0$, and $a_R=0$. Suppose that $\mathcal H = \mathbb{R}^p$, $\KK = p$, and $\RO$ has $p-\nd$ positive eigenvalues. Then the VR(1,0)-based test is equivalent to the test proposed by \citet{shintani2001simple}. However, as discussed in Appendix~\ref{sec_breitung}, unlike \citepos{shintani2001simple} test, our VR(1,0)-based test can also be applied even when $\RO$ has some zero eigenvalues. As in our discussion of \citepos{Breitung2002} test in Appendix~\ref{sec_breitung}, it may be deduced that the VR(1,0)-based test can be easily implemented by applying \citepos{shintani2001simple} test to the vector-valued time series~$\{x_{\KK,t}\}_{t=1}^T$.

\subsubsection{The test of \citet{Chang2016}}
\label{sec_chang}

Again, we consider the infinite-dimensional Hilbert space setting and the VR(1,0)-based test using $\Fmax$ with $\KK=\smalls_0$, $a_L=0$, and $a_R>0$. We may deduce from Theorem~\ref{thmvd0}(i) that
\begin{align}
	1/\Fmax(\{{\widetilde{\mu}_j}\}_{j=1}^{\smalls_0}) &\dto \min_{1\leq j\leq \smalls_0} \left\{\lambda_{j} \left\{\int W_{\smalls_0} W_{\smalls_0}'\right\}\right\} &\text{under $H_0$ of \eqref{eqhypo}},\\
	1/\Fmax(\{{\widetilde{\mu}_j}\}_{j=1}^{\smalls_0}) &\pto 0 &\text{under $H_1$ of \eqref{eqhypo}}. 
\end{align}
Based on these results, we can construct a consistent test by rejecting the null hypothesis if $1/\Fmax(\{{\widetilde{\mu}_j}\}_{j=1}^{\smalls_0})$ is smaller than the appropriate critical value corresponding to the employed significance level. This test is equivalent to that proposed by \citet{Chang2016}. Their test is constructed for $\KK = \smalls_0$ regardless of how many eigenvalues of $\RO$ are positive while our tests, in general, allow $\KK>\smalls_0$ as long as $\RO$ has enough positive eigenvalues; note that, in this functional setting, $\RO$ usually has infinitely many positive eigenvalues. As discussed and shown by \citet{NSS} in detail, letting $\KK=\smalls_0$ is generally disadvantageous because it requires us to extract information on the $\nd$-dimensional subspace $\mathcal H_{\NNN}$ with an $\smalls_0$-dimensional projection~$\widehat{P}_{\KK}$, so under the null $\nd = \smalls_0$ we could fail to successfully capture all the stochastic trends. On the other hand, if $\KK > \smalls_0$ and thus $\widehat{P}_{\KK}$ is allowed to be a larger dimensional projection, such failure is less likely.

\subsection{Inverse VR tests}	
\label{sec_nyblom}

\subsubsection{The tests of \citet{nyblom2000tests}}

We now consider the inverse VR test with~$\Ftr$. If $\mathcal H = \mathbb{R}^p$, $\KK=p$, and $\RO$ has $p-\nd$ positive eigenvalues, the inverse VR test with $h_L=0$ is equivalent to the test of \citet{nyblom2000tests} derived from the LBI principle. Again, our test can be applied even when $\RO$ contains zero eigenvalues, which sets it apart from their test. As in our discussion of \citepos{Breitung2002} test in Appendix~\ref{sec_breitung}, it may be deduced that the inverse VR test can be implemented in practice by applying the test of \citet{nyblom2000tests} to~$\{x_{\KK,t}\}_{t=1}^T$.

\subsubsection{The KPSS-type stationarity tests}

Consider still the inverse VR test with $\Ftr$, and specifically the case where $\smalls_0=0$. This inverse VR test examines the null hypothesis of stationarity against the alternative hypothesis of unit root nonstationarity. If $\mathcal H=\mathbb{R}$, then this test is identical to the standard KPSS stationarity test \citep{KPSS92}. If $\mathcal H$ is infinite-dimensional, then it can be shown without difficulties that the inverse VR test statistic is similar to the functional KPSS test statistic proposed by \citet{Horvath2014}, but it differs in the way that it pivotalizes the asymptotic null distribution. Furthermore, it is then a natural consequence of the works of \citet{Horvath2014} and \citet{kokoszka2016kpss} that our stationarity test based on the inverse VR test should have good power against the alternative of various types of nonstationarity such as structural breaks and/or unrecognized deterministic trends.

\subsection{VR(2,0)-based tests}\label{section_supp_revision} \revlab{ce_2nd_major3_supp}
To the best of the authors' knowledge, no existing test can be viewed as a VR(2,0)-based test in either finite or infinite-dimensional setups, which stands in contrast to the VR(2,1), VR(1,0), and inverse VR tests having clear counterparts. In fact, VR(2,0) can be viewed as an intermediate bridge between the VR(2,1) and VR(1,0) tests, and its theoretical justification is derived from a synthesis of the arguments used for both. While earlier studies, such as \citet{Chang2016} and \citet{NSS}, developed tests similar to VR(1,0) and VR(2,1) in isolation, we demonstrate that VR(2,0) serves as a link that integrates these different approaches into a unified framework.

\section{S.3 \quad Supplementary simulation results}
\label{appsim}
 
In this section, we study the performance of our ADI tests in a finite-dimensional VAR setup and compare with \citepos{Johansen1991} trace test. This serves two purposes. First, when \citepos{Johansen1991} test is correctly specified, it is expected to do very well and is a useful benchmark. Second, if we apply slack extraction to the time series before applying \citepos{Johansen1991} test, we can investigate the extent to which it is not ADI (see Remark~\ref{rem1}). In any case, this setup will demonstrate the broad applicability of our proposed testing procedures.

We simulate a time series $\{ X_t \}_{t=1}^T$ taking values in~$\mathbb{R}^{10}$. Specifically, $X_t = \mathbf D ( x_{\NNN,t}', x_{\SSS,t}')'$, where the time series $x_{\NNN,t} \in \mathbb R^{\nd}$ and $x_{\SSS,t} \in \mathbb R^{10-\nd}$ are generated as
\begin{align*}
x_{\NNN,t}  &=  x_{\NNN,t-1} + \mathbf A_{11} x_{\SSS, t-1} + \mathbf A_{12} x_{\SSS, t-2} + \epsilon_{\NNN,t},\\
x_{\SSS,t}  &= \mathbf A_{22} x_{\SSS,t-2} + \epsilon_{\SSS,t}.
\end{align*}
Here, $\epsilon_t=  ( \epsilon_{\NNN,t} ' ,\epsilon_{\SSS,t}')'$ is iid normal with mean zero and variance matrix $\diag (0.8, \ldots, 0.8)$, $\mathbf A_{11}$ and $\mathbf{A}_{12}$ are $\nd \times (10-\nd)$ coefficient matrices whose elements are randomly drawn from $\mathrm{Uni}[-1,1]$, and for the matrix~$\mathbf A_{22}$, we set its diagonal elements to~$0.3$, and for each row we randomly choose two non-diagonal entries and set them to~$-0.3$. Finally,
\begin{equation*}
\mathbf D = \left(\begin{matrix}
	\mathbf I_{\nd} &\mathbf 0_{\nd \times (10-\nd)} \\ 
	\widetilde{\mathbf D} & \mathbf I_{10-\nd}
\end{matrix}\right) ,
\end{equation*}
where $\mathbf I_{d}$ is the identity matrix of dimension $d$ and for each row of~$\widetilde{\mathbf{D}}$, two entries are randomly assigned to have values $-1$ and~$1$. The coefficients are fixed across simulations. Since the matrix $\mathbf D$ has full rank, $X_t$ has $\nd$ linearly independent common stochastic trends.

In this setup, $\Delta X_t = \Theta X_{t-1} + \Theta_1 \Delta X_{t-1} + \epsilon_t$ holds for some $\Theta, \Theta_1 \in \mathbb R^{10 \times 10}$, and from this representation one may apply \citepos{Johansen1991} trace test to identify the number of stochastic trends. In addition to the classic approach, we apply the trace test to projected time series from two slack extractors, $\widehat P_{\nd}$ and~$\widehat P_{\nd + 2}$, constructed from~$\widehat\Lambda_{1,P}(0)$. The number of stochastic trends determined by trace tests applied to $\{ X_t\}_{t \geq 1}$, $\{\widehat P_{\nd} X_t \}_{t\geq 1}$, and $\{ \widehat P_{\nd+2} X_t\}_{t \geq 1}$ are denoted $\widehat\smalls_{\rm{J}}$, $\widehat\smalls_{\rm{J} : \hat{P}_{\nd}}$, and~$\widehat\smalls_{\rm{J} : \hat{P}_{\nd+2}}$. We use one lagged difference in the VAR specification. We compare with our top-down and up-down tests based on VR(2,1), denoted $\widehat\smalls_{\rm TD }^{(2,1)}$ and~$\widehat\smalls_{\rm UD} ^{(2,1)}$.

The simulation results are summarized in Table~\ref{johansen:finite} for $\nd = 3$ and $\nd = 5$ and $T= 500,1000,2000,3000$. For each test procedure, we report the frequency of correctly determining $\nd$ at 5\% level. As expected, \citepos{Johansen1991} test applied to $\{X_t\}_{t \geq 1}$ performs well even with the smallest sample. The results for $\widehat\smalls_{\rm{J}: \hat{P}_{\nd}}$ and $\widehat\smalls_{\rm{J}: \hat{P}_{\nd+2}}$ exhibit the lowest accuracy, and the large distortions do not improve for the larger sample sizes. This shows very clearly that \citepos{Johansen1991} test is not ADI as discussed in Remark~\ref{rem1}. In contrast, our proposed procedures perform well, although there is some distortion for the smallest sample size. These simulation results hence confirm the broad applicability of our ADI tests, particularly as a complement to the \citepos{Johansen1991} test, even in a finite-dimensional setting.
 
\begin{table}[tbp]
\caption{Comparison with Johansen's trace test in finite-dimensional VAR setting}
\label{johansen:finite}
\vskip -8pt
\begin{tabular*}{\textwidth}{@{\extracolsep{\fill}}ccccccccccc}
\toprule 
& \multicolumn{5}{c}{$\nd=3$} & \multicolumn{5}{c}{$\nd=5$} \\
\cmidrule{2-6} \cmidrule{7-11}
$T$ &$\widehat\smalls_{\rm{J}}$ &$\widehat\smalls_{\rm{J} : \widehat P_{\nd}}$ &$\widehat\smalls_{\rm{J} : \widehat P_{\nd+2}}$&$\widehat\smalls_{\rm UD}^{(2,1)}$ &$\widehat\smalls_{\rm TD}^{(2,1)}$&$\widehat\smalls_{\rm{J}}$ &$\widehat\smalls_{\rm{J} : \widehat P_{\nd}}$ &$\widehat\smalls_{\rm{J} : \widehat P_{\nd+2}}$&$\widehat\smalls_{\rm UD}^{(2,1)}$ &$\widehat\smalls_{\rm TD}^{(2,1)}$ \\ 
\midrule  
\phantom{0}500 & 0.931 & 0.694 & 0.819 & 0.912 & 0.854 & 0.900 & 0.112 & 0.524 & 0.942 & 0.940 \\
  1000 & 0.938 & 0.691 & 0.835 & 0.962 & 0.956 & 0.921 & 0.114 & 0.534 & 0.963 & 0.967 \\ 
  2000 & 0.940 & 0.684 & 0.850 & 0.961 & 0.961 & 0.934 & 0.113 & 0.543 & 0.963 & 0.966 \\ 
  3000 & 0.939 & 0.683 & 0.857 & 0.961 & 0.962 & 0.936 & 0.115 & 0.557 & 0.958 & 0.960 \\ 
\bottomrule
\end{tabular*}
\vskip 4pt
\footnotesize{Notes: Frequencies of correct dimension determination based on 10,000 Monte Carlo replications with significance level~5\%. \citepos{Johansen1991} trace test applied to $\{ X_t\}_{t\geq 1}$, $\{\widehat{P}_{\nd} X_t\}_{t\geq 1}$, and $\{\widehat{P}_{\nd+2} X_t\}_{t\geq 1}$, is denoted $\widehat\smalls_{\rm{J}}$, $\widehat\smalls_{\rm{J} : \hat{P}_{\nd}}$, and $\widehat\smalls_{\rm{J} : \hat{P}_{\nd+2}}$. Our top-down and up-down tests based on VR(2,1) are denoted $\widehat\smalls_{\rm TD}^{(2,1)}$ and~$\widehat\smalls_{\rm UD}^{(2,1)}$.}
\end{table}

\section{S.4 \quad Proof of Lemma \ref{lem1}}
\label{sec:prooflemma}

We first prove part~(i). For any $v_1,v_2\in \mathcal H$, 
\begin{align*}
	\langle P_{\NNN} \widehat{\Lambda}_{d,m} P_{\NNN} v_1, v_2 \rangle 
	={}& \sum_{t=1}^T \langle P_{\NNN}X_{d,t},v_1 \rangle \langle P_{\NNN}X_{d,t},v_2 \rangle \\ 
	&+ \sum_{s=1}^{h_m} \km \Big(\frac{s}{h_m}\Big) \sum_{t=s+1}^T (\langle P_{\NNN}X_{d,t-s},v_1 \rangle \langle P_{\NNN}X_{d,t},v_2 \rangle 
	+ \langle P_{\NNN}X_{d,t},v_1 \rangle \langle P_{\NNN}X_{d,t-s},v_2 \rangle).
\end{align*}
As in our proof of Theorem~\ref{thmslack1}, we will use that $|\km(\cdot)| \leq 1$ and $\km(s/h_m) = 0$ if $|s|> h_m$. For $d \geq 1$ we thus find that $h_m^{-1} T^{-2d}| \langle P_{\NNN} \widehat{\Lambda}_{d,m} P_{\NNN} v_1, v_2 \rangle |$ is bounded from above by
\begin{equation}
	\frac{1}{h_m T^{2d}} \sum_{s=0}^{h_m} \bigg| \sum_{t=s+1}^T \langle P_{\NNN}X_{d,t-s},v_1 \rangle \langle P_{\NNN}X_{d,t},v_2 \rangle + \sum_{t=s+1}^T \langle P_{\NNN}X_{d,t},v_1 \rangle \langle P_{\NNN}X_{d,t-s},v_2 \rangle \bigg| .
	\label{eqlempf01}
\end{equation}
Because $\langle P_{\NNN}X_{1,t},v_j \rangle /T^{1/2}$, viewed as a partial sum process, converges weakly to Brownian motion, the continuous mapping theorem implies that \eqref{eqlempf01} is~$\bigOp (1)$, and the result for $d \geq 1$ follows directly. Next, $P_{\NNN} \widehat{\Lambda}_{0,m} P_{\NNN}$ is the sample long-run variance of $P_{\NNN} \Delta X_t$, where $P_{\NNN} \Delta X_t$ is $L^4$-$q$-approximable with $q\sum_{j=q+1}^{\infty} \| \Phi_j \|_{\op}=\smallo (1)$, so we find from \citet[Theorem~2]{horvath2013estimation} that $T^{-1} P_{\NNN} \widehat{\Lambda}_{0,m} P_{\NNN} = \bigOp (1)$, which proves the result for $d=0$.
	
Next, we prove part~(ii). Now $|\langle P_{\NNN} \widehat{\Lambda}_{d,m} P_{\SSS}^{\KK} v_1,v_2\rangle |$ is bounded from above by
\begin{equation}
\sum_{s=0}^{h_m} \bigg| \sum_{t=s+1}^T  \langle P_{\SSS}^{\KK}X_{d,t},v_1 \rangle \langle P_{\NNN}X_{d,t-s},v_2 \rangle+ \sum_{t=s+1}^T  \langle P_{\SSS}^{\KK}X_{d,t-s},v_1 \rangle\langle P_{\NNN}X_{d,t},v_2 \rangle \bigg| . 
\label{eqlempf02}
\end{equation}
Setting $d \geq 2$ in \eqref{eqlempf02} and multiplying by $h_m^{-1}T^{1-2d}$, this is $\bigOp (1)$ for the same reasons as for \eqref{eqlempf01} together with the fact that the partial sum process $\langle P_{\SSS}^{\KK}X_{2,t},v_1 \rangle/T^{1/2}$ converges weakly to Brownian motion. This proves part~(ii) for $d \geq 2$. 
For $d=1$, we note from \eqref{eqbn} that $\langle P_{\NNN}X_{1,t-s},v_j \rangle$ is nonstationary with stationary first difference $\langle P_{\NNN}X_{0,t-s},v_j \rangle$, while $\langle P_{\SSS}^{\KK}X_{1,t},v_j \rangle$ is stationary. It can then be shown, as in standard results \citep[e.g.,][]{phillips87}, that the summations over $t$ in~\eqref{eqlempf02}, multiplied by $T^{-1}$, converge weakly to stochastic integrals, uniformly in~$s$. This proves the result for $d=1$.
Finally, under our assumptions, we deduce from Lemma~8.1(b) of \citet{phillips1995fully} that both $T^{-1}|\langle P_{\NNN} \widehat{\Lambda}_{0,m} P_{\SSS}^{\KK} v_1,v_2 \rangle |$ and $T^{-1}|\langle P_{\SSS}^{\KK} \widehat{\Lambda}_{0,m} P_{\NNN}v_1,v_2\rangle |$ are $\bigOp (h_m^{-2}) + \bigOp ((Th_m)^{-1/2})$, and the result for $d=0$ follows.
	
Lastly, we prove part~(iii). Here, $h_m^{-1} T^{2-2d}|\langle P_{\SSS}^{\KK} \widehat{\Lambda}_{d,m} P_{\SSS}^{\KK} v_1,v_2\rangle|$ is bounded from above by
\begin{equation}
\frac{1}{h_m T^{2d-2}} \sum_{s=0}^{h_m} \bigg| \sum_{t=s+1}^T \langle P_{\SSS}^{\KK}X_{d,t-s},v_1 \rangle \langle P_{\SSS}^{\KK}X_{d,t},v_2 \rangle + \sum_{t=s+1}^T \langle P_{\SSS}^{\KK}X_{d,t},v_1 \rangle \langle P_{\SSS}^{\KK}X_{d,t-s},v_2 \rangle \bigg| .
\label{eqlempf03}
\end{equation}
For $d \geq 2$, this is $\bigOp (1)$ by the same arguments as in~(i) because $\langle P_{\SSS}^{\KK}X_{2,t},v_j \rangle/T^{1/2}$, viewed as a partial sum process, converges weakly to Brownian motion. Next, $T^{-1} P_{\SSS}^{\KK} \widehat{\Lambda}_{1,m} P_{\SSS}^{\KK}$ converges to the long-run variance of $\{ P_{\SSS}^{\KK} X_t\}_{t\geq 1}$ since $\{ P_{\SSS}^{\KK}X_{1,t} \}_{t\geq 1}$ is $L^4$-$q$-approximable with $q\sum_{j=q+1}^{\infty} \| \tilde\Phi_j \|_{\op} =\smallo (1)$ under our assumptions \citep[Theorem~2]{horvath2013estimation}. Finally, we deduce from Lemma~8.1(a) of \citet{phillips1995fully} that $h_m^2 T^{-1} | \langle P_{\SSS}^{\KK} \widehat{\Lambda}_{0,m} P_{\SSS}^{\KK} v_1,v_2 \rangle |$ is convergent, so that $P_{\SSS}^{\KK}\widehat{\Lambda}_{0,m} P_{\SSS}^{\KK} = \bigOp ( T / h_m^2 )$.

\pdfbookmark[1]{\refname}{unnumbered}
\bibliography{lit_dfvrtest}

\end{document}